\documentclass[11pt]{article}
\usepackage[a4paper,margin=2.4cm]{geometry}
\usepackage[T1]{fontenc}
\usepackage{lmodern,microtype}
\usepackage{amsmath,amssymb,amsthm,mathtools}
\usepackage{graphicx,booktabs,tabularx,enumitem,float}
\usepackage{xcolor,url}
\usepackage[colorlinks=true,linkcolor=blue!50!black,citecolor=blue!50!black,urlcolor=blue!50!black]{hyperref}

\hypersetup{
  pdftitle={How many states does an exact relaxation model need?},
  pdfauthor={Murad Aznagulov},
  pdfsubject={Sharp state complexity and geometric obstructions for reversible chains}
}

\newtheorem{theorem}{Theorem}
\newtheorem{proposition}[theorem]{Proposition}
\newtheorem{lemma}[theorem]{Lemma}
\newtheorem{corollary}[theorem]{Corollary}
\newtheorem{conjecture}{Conjecture}

\theoremstyle{definition}
\newtheorem{definition}[theorem]{Definition}
\newtheorem{example}[theorem]{Example}
\theoremstyle{remark}
\newtheorem{remark}[theorem]{Remark}
\numberwithin{theorem}{section}
\numberwithin{equation}{section}
\newcommand{\NCERT}{381}
\newcommand{\E}{\mathbb{E}}
\newcommand{\R}{\mathbb{R}}
\newcommand{\Q}{\mathbb{Q}}
\newcommand{\K}{\mathcal{K}}
\newcommand{\LS}{L_S}
\newcommand{\Lip}{\operatorname{Lip}}
\DeclareMathOperator{\tr}{tr}
\DeclareMathOperator{\conv}{conv}
\DeclareMathOperator{\supp}{supp}
\DeclareMathOperator{\diag}{diag}

\title{How many states does an exact relaxation model need?\\[2mm]
\large Sharp state complexity and geometric obstructions for reversible chains}
\author{Murad Aznagulov\\[1mm]
\small Faculty of Physics, Lomonosov Moscow State University, Moscow, Russia\\
\small \texttt{amurad12@icloud.com}}
\date{November 2026}

\begin{document}
\maketitle

\begin{abstract}
How many states does a reversible Markov model need to reproduce two
prescribed relaxation modes exactly? We maximize the smaller of the two
stationary coordinate variances over all finite reversible chains whose
states lie in a given convex region and whose coordinates relax at rates
$1$ and $\rho$. Detailed balance turns this into the equilibrium of a
tension-only cable network whose loads depend on the stationary masses.

For the moment curve of the Wright--Fisher diffusion no finite chain is
optimal: maximizing sequences converge to a unique beta law, and the best
deficit with $N$ states is of order $N^{-2}$. We find the exact leading
constant for states on the curve and for states in its convex hull,
together with the optimal cube-root density of states. The hull constant
is the curve constant times $c/(1+c)$, $c=(\rho-2)/\rho$; at $\rho=5/2$
the hull needs $\sqrt6$ times fewer states, although the limiting law is
the same. For $\rho>8/3$ any upward enlargement of the hull lowers the
constant again. All lower bounds hold for arbitrary transition graphs.

At every fixed number of states an optimizer exists. On a seven-gon the
smallest optimal chain jumps from three to four states while the optimal
value stays constant, and a family of polygons forces the state count to
grow linearly with the number of sides. Monotone vector fields give exact
upper certificates, which solve an asymmetric quadrilateral at all rates
and \NCERT{} polygon--rate instances in exact arithmetic. We formulate a
finite-realizability problem and a finite-dictionary conjecture for
polygons, and derive state budgets and exclusion tests for finite kinetic
models.
\end{abstract}

{\small\setcounter{tocdepth}{1}\tableofcontents}

\section{Introduction}

Finite Markov models are routinely used in place of continuous
dynamics: Markov state models of molecular kinetics~\cite{Prinz11,NoeNuske13,NoeReview14},
lattice discretizations of diffusions in population
genetics~\cite{FerrerAdmetlla16}, and moment-matching discretizations
in economics~\cite{FT17}. A model of this kind is usually judged by
whether it reproduces a few slow relaxation modes. This paper asks a
sharper version of the question. Suppose a reversible model must reproduce
two prescribed modes \emph{exactly}, and its states must stay in a
prescribed region of observable space. How much stationary fluctuation
can such a model carry, and how many states does it need to come close
to that limit?

The answer turns out to be far from a counting exercise. For the
moment curve of the Wright--Fisher diffusion, no finite model reaches
the optimum. The best $N$-state deficit decays like $N^{-2}$ with an
explicitly computable constant, and the constant depends on fine
geometric details that leave the limiting stationary law unchanged.
At the rate ratio $\rho=5/2$, allowing the states to move slightly off
the moment curve into its convex hull reduces the leading constant by a
factor of six. Raising the upper boundary of the hull by any small
amount reduces it further when $\rho>8/3$. On polygons the optimal value
can stay fixed while the smallest optimal model jumps from three to four
states, and a family of polygons forces the state count to grow linearly
with the number of sides. All lower bounds hold over the full class of
reversible transition graphs, not only for birth--death chains or other
structured families.

\subsection{The problem}

Let $Z_t=(u_t,v_t)$ be a finite reversible continuous-time Markov chain
whose states $x_i$ lie in a compact convex set $P\subset\R^2$ with
$0\in\operatorname{int}P$. We require the two coordinate observables to be
exact relaxation modes,
\begin{equation}\label{eq:linmean}
 \mathbb E[Z_t\mid Z_0=x]=e^{-t\Lambda}x,\qquad
 \Lambda=\operatorname{diag}(1,\rho),\quad \rho>1,
\end{equation}
and maximize the smaller stationary variance $\min(\E u^2,\E v^2)$. The
supremum over all finite chains is the \emph{capacity} $s_W(P,\rho)$, and
$s_N(P,\rho)$ is the supremum over chains with at most $N$ states. Only
the ratio $\rho$ of the two rates matters, because rescaling time
normalizes the slower rate to one.

Three questions follow. What fluctuation can two exact modes sustain in
a given geometry? Is the optimum attained by a finite chain? If not, how
fast does $s_N$ approach $s_W$, and what do the optimal $N$-state models
look like?

Reversibility turns the problem into mechanics. Multiply the generator
equations by the stationary masses $p_i$ and write $J_{ij}=p_iQ_{ij}$.
Then \eqref{eq:linmean} becomes the equilibrium of a tension-only cable
network: the cable $ij$ carries tension $J_{ij}$ per unit length and the
site $x_i$ carries the load $p_i\Lambda x_i$. Classical cable-web theory
characterizes the loads that such networks can support at fixed
nodes~\cite{Milton17,BMMS19}. In our problem the node positions and the
masses are free, and the masses enter both the load and the objective.
This coupling is what produces the phenomena below.

\begin{figure}[t]
 \centering
 \includegraphics[width=\textwidth]{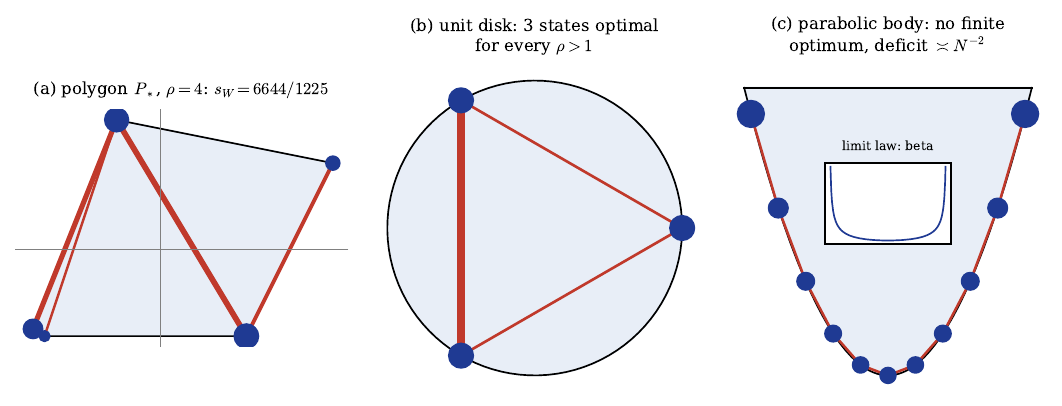}
 \caption{Three realizations of two exact modes at \(\rho=4\).
 (a) The asymmetric quadrilateral \(P_*\), where a finite
 certificate identifies the optimum. (b) The disk, with a three-state
 optimum. (c) A parabolic barrier, where finite designs approach a
 unique limiting mass distribution along the arc. Node sizes represent
 stationary masses and edge widths represent conductances.}
 \label{fig:three}
\end{figure}

\subsection{Main results}

\paragraph{Parabolic rigidity and the Wright--Fisher diffusion.}
For $\rho>2$, put $c=(\rho-2)/\rho$. If $P$ lies above the parabolic arc
$\Gamma$ on which the observables $u$ and $v\propto u^2-c$ of the
Wright--Fisher diffusion live, then $s_W(P,\rho)\le c$, with equality
exactly when $\Gamma\subset P$ (Theorem~\ref{thm:parab}). In the equality
case no finite chain is optimal. Every maximizing sequence has stationary
laws converging to a unique beta law on the arc and cable stresses
converging to a unique tangential stress. The deficit at $N$ states is of
exact order $N^{-2}$. The proof rests on a quartic energy identity,
\eqref{eq:defect}, which relates the variance deficit to the fourth
moment of the jumps and to the vertical distance of the states from the
arc. On the arc itself the optimal upper construction is the Moran model:
the symmetric Hahn birth--death chains are, up to time scaling, the
continuous-time Moran model with mutation
(Proposition~\ref{prop:moran}). Since the means of any frequency
distribution lie in the convex hull $\conv\Gamma$, a finite model whose
states represent \emph{distributions} of allele frequency, and which
reproduces the two slowest Wright--Fisher modes exactly, is precisely a
design in $\conv\Gamma$. Theorem~\ref{thm:microstate} turns this into an
explicit state budget: deficit $\delta$ requires at least
$1+c/\bigl(2\sqrt{(c+2)\delta}\bigr)$ states.

\paragraph{Sharp constants and the optimal placement of states.}
The $N^{-2}$ rate leaves the main practical question open: how small can
the constant be, and where should the states go? Let $j_\rho(x)$ be the
horizontal force across a vertical cut at $x$ of the limiting beta law,
and $I_\rho=\int_{-1}^1 j_\rho^{1/3}$. Theorem~\ref{thm:arcsharp} shows
\[
 (N-1)^2\bigl(c-s_N^\Gamma\bigr)\to C_\Gamma(\rho)=\frac{I_\rho^3}{4c},
 \qquad
 (N-1)^2\bigl(c-s_N(\conv\Gamma)\bigr)\to C_K(\rho)=\frac{c}{1+c}\,C_\Gamma(\rho),
\]
for states on the arc and in its convex hull, respectively. In both cases
optimal states are distributed along the arc with density proportional to
$(1-x^2)^{a_0/3}$, $a_0=1/(\rho-2)$. This is a cube-root allocation rule
of the Panter--Dite type known from high-resolution
quantization~\cite{PanterDite}, but here the weight is a mechanical cut
force, and the grid has to admit nonnegative rates with both generator
equations exact. The equispaced Moran chain has coefficient $1$, while
$C_\Gamma<1$ for every $\rho>2$. At $\rho=5/2$ one has
$C_\Gamma\approx0.757$ and $C_K\approx0.126$. Inverting these limits, the
number of states needed for deficit $\delta$ on the arc and in the hull
differs asymptotically by the factor $\sqrt{(1+c)/c}$, which equals
$\sqrt6$ at $\rho=5/2$. The interpretation is concrete. An arc state is a
single allele frequency, a hull state may represent a frequency
distribution through its first two moments, and the gain comes
from letting the states near the two ends of the arc carry such
spread. Corollary~\ref{cor:slacklocation} proves that optimal designs
place their vertical displacement only near the endpoints. The lower
bounds hold for arbitrary transition graphs. The matching constructions
are graded reversible paths with both modes exact.

\paragraph{Sensitivity to the confinement.}
The leading constant is not a function of the limiting stationary law.
If the hull $\{u^2\le t\le1\}$ is enlarged to $\{u^2\le t\le T\}$ with
$T>1$, the limit law is still the same beta law, yet for every
$\rho>8/3$ the scaled deficit strictly decreases, with a gain of first
order in $T-1$ (Theorem~\ref{thm:raisedcap}). The improvement comes from
a curve that coincides with the arc in the centre and bends upward near
the ends; its equilibrium trades horizontal for vertical variance at a
better rate than is available in $\conv\Gamma$. Two smooth bodies can
therefore share the same optimal continuum model and still require
different numbers of states.

\paragraph{Attainment at a fixed state cap and phase transitions in the state count.}
In contrast to the unrestricted problem, every fixed cap is attained.
Weak limits of $N$-state feasible laws remain feasible with at most $N$
states, even when conductances diverge as sites merge
(Theorem~\ref{thm:capatta}). For rational polygons the capped value is a
semialgebraic function of $\rho$ with finitely many analytic branches
(Corollary~\ref{cor:capalg}). On three sites the conductances are forced
by the sites and masses, through the explicit formula \eqref{eq:threeJ},
and a sharp rate threshold follows. This yields a seven-gon on which the
capacity equals $1$ for every rate, while the smallest optimal chain
changes from three to four states at $\rho=1+b^{-2}$
(Theorem~\ref{thm:sevenphase}). The two competing optimal laws have the
same means, variances and mixed moment. What separates them is whether
nonnegative conductances exist, a dynamical constraint that is invisible
to moments. A hexagon family has a complete phase diagram with jumps
$4\to6$, $3\to6$ and $5\to6$ (Theorem~\ref{thm:hexphase}). Finally, the
optimal number of states cannot be bounded uniformly in the number of
sides: for a family of $(k+2)$-gons circumscribed about the parabola,
every design beating the value $1/2$ at $\rho=4$ needs more than
$1+k/\sqrt{160}$ states (Theorem~\ref{thm:linear}).

\paragraph{Continuum relaxation and certificates.}
The closure of finite designs is described exactly by pairs of a mass
measure and a positive semidefinite stress measure in weak equilibrium
(Theorem~\ref{thm:closure}). Its dual consists of monotone vector fields.
Each such field gives a computable upper bound, and strong duality holds
with polynomial certificates (Theorems~\ref{thm:gap}
and~\ref{thm:strongdual}). A finitely supported optimal mass law is always
realized by a finite design on exactly its support
(Theorem~\ref{thm:finsupp}). These tools give a closed-form all-rate
capacity for an asymmetric quadrilateral, with three regimes and two
explicit breakpoints (Theorem~\ref{thm:pstar}), and \NCERT{} exact
polygon--rate solutions, each consisting of an algebraic design and a
matching monotone certificate checked in exact arithmetic
(Theorem~\ref{thm:computer}).

\subsection{Conjectures and what they would give}

The polygonal case is where the main open problems lie. A polygon inside
the parabolic region always has capacity below $c$, so the obstruction
of Theorem~\ref{thm:parab} disappears, and one expects a finite optimum.

\begin{itemize}[leftmargin=*]
\item \emph{Finite realizability (FRR).} For every convex polygon $P$ with
 $0\in\operatorname{int}P$ and every $\rho>1$, the capacity $s_W(P,\rho)$
 is attained by a finite reversible chain
 (Definition~\ref{def:frr}).
\item \emph{Finite-dictionary conjecture $\Sigma$.} Moreover, an optimal
 chain can be found on an explicit set $S(P,\rho)$ of $O(n^2)$
 candidate sites: vertices, axis crossings, the points where a side is
 tangent to the load field $\Lambda x$, and the roots of a leaf equation
 dictated by the local structure of optimal networks
 (Conjecture~\ref{conj:sigma}).
\end{itemize}

\noindent Conjecture~$\Sigma$ implies FRR with $O(n^2)$ states. It would
also reduce the capacity of any polygon to a single finite linear
program over an explicitly computable site set, so the capacity would be
an algebraic number for algebraic data and computable exactly instead
of through two-sided approximation (Theorem~\ref{thm:computecap}). The
paper collects substantial evidence. FRR holds with at most six states
for every body near $\rho=1$ (Theorem~\ref{thm:iso}), for every
reflection-symmetric body at every rate (Theorem~\ref{thm:sym}), and for
bodies with affine vertical midline (Theorem~\ref{thm:affinemidline}).
Conjecture~$\Sigma$ is asymptotically exact as $\rho\to\infty$ with only
eight dictionary sites (Theorem~\ref{thm:eight}), and it holds in all
\NCERT{} certified instances. The invariant-core theorem rules out
dynamically generated boundary arcs as carriers of diffuse limiting mass
(Theorems~\ref{thm:invcore} and~\ref{thm:barrier}). The linear lower
bound of Theorem~\ref{thm:linear} shows that no dictionary with a bounded
number of \emph{active} sites can work, so the gap between $\Omega(n)$ and
$O(n^2)$ is the remaining question about state complexity. If $\Sigma$
fails, it fails on an integral polygon with an exactly verifiable integral
witness (Theorem~\ref{thm:arith}). A search over rational data is
therefore a semi-decision procedure for a counterexample. What remains is
to show that mass on flat portions of the original sides can be replaced
by finitely many atoms without losing force feasibility. The seven-gon
shows that matching moments alone cannot settle this.

\subsection{Consequences for finite kinetic models}

Since every lower bound holds for all reversible chains, the results
can be used as obstructions. Theorems~\ref{thm:microstate}
and~\ref{thm:robustmicro} give state budgets for any reversible
discretization of the Wright--Fisher diffusion that reproduces its two
slowest modes, exactly or up to measured generator residuals. The sharp
constants show how far the Moran model is from optimal and where the
states of an optimal discretization should be placed. The three-state
formula~\eqref{eq:threeJ} excludes a proposed three-state mechanism from
its observable sites and weights alone. The geometry of an extreme state
bounds its exit probability, which gives a pointwise test of a prescribed
eigenmode that differs from the usual Chapman--Kolmogorov validation of
Markov state models (Section~\ref{sec:inference}).

\subsection{Related work}

Moment-matching and hidden-Markov discretizations of Markov
processes~\cite{FT17,JM26} and approximate Markov models of the Wright--Fisher
diffusion~\cite{FerrerAdmetlla16} construct finite chains with good
approximation properties. Our focus is complementary: we fix two
exact modes and a confinement, and bound what \emph{every} reversible
realization can do. The mechanical side connects to cable webs and
trusses under tension~\cite{Milton17,BMMS19}; the exactly solvable
birth--death chains behind our upper bounds are the Hahn chains
of~\cite{Sasaki}, and the beta stationary law of the Moran model and
its diffusion limit is classical~\cite{KM62,GR13}. Markov state model
validation and variational slow-mode estimation are developed
in~\cite{Prinz11,NoeNuske13,NoeReview14}. Fixed-cap attainment also
appears in the author's work on the thermodynamic realizability of hidden
Markov processes~\cite{Az1}, and finite temporal resolution in the
detection of nonequilibrium is quantified in~\cite{Az2}.

\paragraph{Organization.}
Section~\ref{sec:setting} sets up designs and their three equivalent
readings. Section~\ref{sec:parab} treats parabolic confinement, the sharp
asymptotics and the raised cap, Section~\ref{sec:moran} the
Wright--Fisher state budget, and Section~\ref{sec:linear} the linear
state complexity of polygons. Section~\ref{sec:capped} proves attainment
at fixed caps and the seven-gon transition. Section~\ref{sec:relax}
develops the stress relaxation and monotone certificates,
Section~\ref{sec:polygons} the dictionary conjecture and exact polygon
solutions, and Section~\ref{sec:inference} the consequences for kinetic
models. The appendices contain the structure theory for polygons and the
remaining proofs.

\section{The design problem}\label{sec:setting}
%=====================================================================

Let $P\subset\R^2$ be a compact convex body with $0\in\operatorname{int}P$, $\rho>1$ and $\Lambda=\diag(1,\rho)$; write $x=(u,v)$.

\begin{definition}\label{def:design}
A \emph{design} consists of distinct sites $x_1,\dots,x_N\in P$, masses $p_i>0$ with $\sum_ip_i=1$, and conductances $J_{ij}=J_{ji}\ge0$ satisfying
\begin{equation}\label{eq:eq}
\sum_{j\ne i}J_{ij}(x_i-x_j)=p_i\Lambda x_i,\qquad i=1,\dots,N.
\end{equation}
The number of states is the number of distinct sites with positive mass.
The graph is allowed to be reducible. Its moments are $A=\sum_ip_iu_i^2$ and $B=\sum_ip_iv_i^2$. The \emph{capacity} is $s_W(P,\rho)=\sup\min(A,B)$ over all designs, and $s_N$ denotes the supremum over designs with at most $N$ sites.
\end{definition}

\noindent Equation \eqref{eq:eq} has three equivalent readings.
\begin{itemize}[leftmargin=*]
\item \emph{Markov chain.} $Q_{ij}=J_{ij}/p_i$ ($i\ne j$) is a reversible generator with stationary law $p$, and \eqref{eq:eq} says $Qu=-u$, $Qv=-\rho v$, i.e.\ \eqref{eq:linmean}.
\item \emph{Cable network.} $J_{ij}$ are tensions per unit length of cables joining the sites; the cables balance the body load $p_i\Lambda x_i$.
\item \emph{Spectral drawing.} For the weighted Laplacian $L$ of $J$ and $\Pi=\diag(p)$, \eqref{eq:eq} reads $Lu=\Pi u$, $Lv=\rho\Pi v$.
\end{itemize}
Here reversibility means detailed balance \(p_iQ_{ij}=p_jQ_{ji}\),
as in the classical theory of time reversal for Markov
chains~\cite{Kolmogoroff1936}.
Designs can be mixed linearly. Scale masses and conductances of two designs by $\lambda$ and $1-\lambda$, take the union and merge coincident sites. Mixing is linear in $(A,B)$. For node functions set $\Delta f=f_i-f_j$ on an edge and $\mathcal E(f,g)=\sum_{i<j}J_{ij}\Delta f\Delta g=-\E[fQg]$.

\begin{lemma}[Basic identities]\label{lem:id}
Every design satisfies \(\E u=\E v=\E uv=0\). For every node function \(f\),
\(\mathcal E(u,f)=\E[uf]\) and \(\mathcal E(v,f)=\rho\,\E[vf]\).
In particular \(\mathcal E(u,u)=A\) and \(\mathcal E(v,v)=\rho B\).
\end{lemma}
\begin{proof}
Pair the $u$-component of \eqref{eq:eq} with $f_i$ and sum: by symmetry of $J$, $\mathcal E(u,f)=\sum_ip_iu_if_i$. Likewise for $v$. With $f\equiv1$ we get $\E u=\E v=0$; with $f=v$ and $f=u$ we get $\E uv=\mathcal E(u,v)=\rho\E uv$, so $\E uv=0$ because $\rho\ne1$.
\end{proof}

The moment bound \(R(P)\) is the maximum of \(\min(A,B)\) over
probability measures on \(P\) satisfying
\(\mathbb E u=\mathbb E v=\mathbb E uv=0\).
Every design satisfies these constraints, so \(s_W(P,\rho)\le R(P)\).

\paragraph{Operational meaning.}
The coordinate condition concerns a specific pair of observables, not the full
transition law. It implies, under stationarity,
\[
\E[u(X_t)u(X_0)]=Ae^{-t},\qquad
\E[v(X_t)v(X_0)]=Be^{-\rho t},\qquad
\E[u(X_t)v(X_0)]=0.
\]
Nonlinear observables of a target continuous process are not required to
evolve in the same way. The number of states always means the number of
distinct sites with positive stationary mass; massless relay vertices can be
eliminated and are not counted. If the prescribed rates are $\lambda_1>0$ and $\lambda_2>\lambda_1$,
replacing $Q$ by $Q/\lambda_1$ normalizes the problem to
$\rho=\lambda_2/\lambda_1$. All thresholds therefore depend only on this ratio.

\begin{lemma}[Monotonicity in the rate ratio]\label{lem:mono}
$s_W(P,\rho)$ is non-increasing in $\rho$.
\end{lemma}
\begin{proof}
Let a design at $\rho'$ be given and $1<\rho\le\rho'$. Since $\E x=0$, the complete \emph{reset} network $J^0_{ij}=p_ip_j$ satisfies $\sum_jp_ip_j(x_i-x_j)=p_ix_i$, i.e.\ it is a design for $\Lambda=I$. With $a=(\rho-1)/(\rho'-1)$ the conductances $aJ+(1-a)J^0\ge0$ satisfy \eqref{eq:eq} for $\diag(1,a\rho'+1-a)=\Lambda$, with the same sites and masses.
\end{proof}

\section{Parabolic confinement and sharp asymptotics}\label{sec:parab}
%=====================================================================

Fix $\rho>2$ and put
\[
c=\frac{\rho-2}{\rho},\qquad a_0=\frac1{\rho-2},\qquad b_0=\sqrt{\frac{3\rho-4}{4c}},\qquad t=c+\frac v{b_0}.
\]
In this section $P$ is a compact convex body with $0\in\operatorname{int}P$ lying in the parabolic region $\{t\ge u^2\}$. Let
\[
X(u)=\bigl(u,b_0(u^2-c)\bigr),\qquad \Gamma=X([-1,1]),\qquad \beta(du)=\frac{(1-u^2)^{a_0-1}\mathbf 1_{(-1,1)}(u)\,du}{\int_{-1}^1(1-y^2)^{a_0-1}dy}.
\]

\begin{theorem}[Parabolic rigidity]\label{thm:parab}
$s_W(P,\rho)\le c$, with equality iff $\Gamma\subset P$. In the equality case:
\begin{enumerate}[label=(\alph*)]
\item no finite design attains $c$;
\item every maximizing sequence has mass measures converging weakly to $X_\#\beta$ and stresses converging weakly to $X_\#\bigl[\tfrac{\rho-2}2X'(u)X'(u)^{\mathsf T}(1-u^2)\beta(du)\bigr]$;
\item there is $k_{P,\rho}>0$ such that for all $N\ge2$
\begin{equation}\label{eq:N2}
\frac{k_{P,\rho}}{(N-1)^2}\le c-s_N(P,\rho)\le\frac1{(N-1)(N-1+2a_0)}.
\end{equation}
\end{enumerate}
\end{theorem}

\paragraph{The defect identity.} Let $d=t-u^2\ge0$ on $P$, $\mathcal W_4=\sum_{i<j}J_{ij}(\Delta u)^4$ and $\delta=c-\min(A,B)$. Since $\E t=c$, $\E d=c-A$; hence $A\le c$ and $0\le\E d\le\delta$. We claim
\begin{equation}\label{eq:defect}
\mathcal W_4+3\mathcal E(d,d)=4c(c-B)+\E\Bigl[d\bigl((6\rho-8)t-6\rho c+4d\bigr)\Bigr].
\end{equation}
\begin{proof}
On every edge $4\Delta u\,\Delta(u^3)-3(\Delta(u^2))^2=(\Delta u)^4$, so $3\mathcal E(u^2,u^2)=4\mathcal E(u,u^3)-\mathcal W_4=4\E u^4-\mathcal W_4$ by Lemma~\ref{lem:id}. Also $\mathcal E(t,f)=\rho\E[(t-c)f]$ because $t-c=v/b_0$. Expand $3\mathcal E(t-u^2,t-u^2)=3\mathcal E(t,t)-6\mathcal E(t,u^2)+3\mathcal E(u^2,u^2)$, add $\mathcal W_4$, substitute $u^2=t-d$ and use $(3\rho-4)\operatorname{Var}(t)=4cB$, which is the definition of $b_0$.
\end{proof}
\noindent Consequently, with $M=4c+\max\bigl(0,\max_P((6\rho-8)t-6\rho c+4d)\bigr)$,
\begin{equation}\label{eq:defectbound}
\mathcal W_4+3\mathcal E(d,d)\le M\delta .
\end{equation}
Since $A=c-\E d\le c$, every design has $\min(A,B)\le c$, so $s_W\le c$. On the arc, $d=0$, $A=c$ and $B=c-\mathcal W_4/(4c)$; a finite design attaining $c$ would have $A=B=c$, hence $\E d=0$ (all mass on the arc) and $\mathcal W_4=0$, so $\Delta u=0$ on every cable and $A=\mathcal E(u,u)=0$, contradicting $c>0$. This proves (a).

\paragraph{Hahn chains (upper bound and sufficiency).} For $n\ge1$ use the sites $X(u_i)$, $u_i=h_n(i-n/2)$, $0\le i\le n$, $h_n=2/\sqrt{n(n+2a_0)}$, with jump rates
\[
q_i^+=\frac{(n-i)(i+a_0)}{2a_0},\qquad q_i^-=\frac{i(n-i+a_0)}{2a_0},
\]
and masses defined by detailed balance $p_{i+1}/p_i=q_i^+/q_{i+1}^-$. These are the symmetric Hahn birth--death chains~\cite{Sasaki} up to time scaling. Directly, $h_n(q_i^+-q_i^-)=-u_i$ and $h_n^2(q_i^++q_i^-)=(\rho-2)(1-u_i^2)$, which give $Qu=-u$ and $Q(u^2-c)=-\rho(u^2-c)$, i.e.\ \eqref{eq:eq}. On the arc $\mathcal W_4=h_n^2A$, and \eqref{eq:defect} with $d=0$ gives $A=c$, $B=c-1/(n(n+2a_0))$. With $N=n+1$ this proves sufficiency of $\Gamma\subset P$ and the upper bound in \eqref{eq:N2}.

\paragraph{Quartic energy (lower bound).} Order the distinct abscissae $r_1<\dots<r_m$ and let $g_k=r_{k+1}-r_k$. Summing horizontal equilibrium to the right of the $k$-th gap gives the cut force $C_k=\E[u\mathbf 1_{u>r_k}]=\sum_{\text{edges across }k}J_{ij}|\Delta u|\le\E u_+=\frac12\E|u|$. Telescoping gives $A=\sum_kC_kg_k$; since the cube of a sum of non-negative gaps dominates the sum of cubes, $\mathcal W_4\ge\sum_kC_kg_k^3$, and Jensen gives
\begin{equation}\label{eq:quartic}
\mathcal W_4\ge\frac{A^3}{(\sum_kC_k)^2}\ge\frac{4A^2}{(N-1)^2}.
\end{equation}
If $A\ge c/2$, \eqref{eq:defectbound} gives $\delta\ge c^2/(M(N-1)^2)$; otherwise $\delta>c/2$. Thus $k_{P,\rho}=\min(c/2,c^2/M)$ works.

\paragraph{Identification of limits and necessity.} Appendix~\ref{app:parab} proves (b) and the necessity of $\Gamma\subset P$ from a positive-semidefinite stress identity (Lemma~\ref{lem:universal-psd}). The proof works for arbitrary distributional stresses in equilibrium, without invoking the density direction of Theorem~\ref{thm:closure}: saturation forces tangential stress on the arc, and the equilibrium equations identify its unique beta-law mass.

\begin{corollary}[Smooth positively curved bodies without finite optimum]\label{cor:smooth}
For every $\rho>2$ there is a compact convex body with $C^\infty$ boundary of everywhere positive curvature on which the capacity is not attained by any finite chain, with deficit \eqref{eq:N2}.
\end{corollary}
\begin{proof}
In $(u,t)$ coordinates let $\phi(t)=0$ for $t\le1$ and $\phi(t)=(t-1)^2e^{-1/(t-1)}$ for $t>1$, and $P=\{u^2-t+\phi(t)\le0\}$. $\phi$ is smooth, non-negative, superlinear and strictly convex for $t>1$, so $P$ is compact, lies above the parabola, contains $\Gamma$ (where $t\le1$), and contains the origin in its interior because $0<c<1$. The defining function is convex with non-vanishing gradient on the zero level; its tangential Hessian is $2\xi^2>0$ for $t\le1$ (tangent vectors satisfy $\eta=2u\xi$) and its Hessian is positive definite for $t>1$. Hence the boundary is smooth with positive curvature, preserved by the affine change back to $(u,v)$. Apply Theorem~\ref{thm:parab}.
\end{proof}

\begin{proposition}[Three states are optimal on the disk]\label{prop:disk}
Let $P$ be the unit disk and $\rho>1$. The design on $x_1=(1,0)$, $x_{2,3}=(-\frac12,\pm\frac{\sqrt3}2)$ with $p_i=\frac13$, $J_{12}=J_{13}=\frac19$, $J_{23}=\frac{3\rho-1}{18}$ has $A=B=\frac12$, and $s_W(P,\rho)=\frac12$.
\end{proposition}
\begin{proof}
At $x_1$: $\frac19[(\frac32,-\frac{\sqrt3}2)+(\frac32,\frac{\sqrt3}2)]=(\frac13,0)=p_1\Lambda x_1$. At $x_2$: $\frac19(-\frac32,\frac{\sqrt3}2)+\frac{3\rho-1}{18}(0,\sqrt3)=(-\frac16,\frac{\rho\sqrt3}6)=p_2\Lambda x_2$, and $x_3$ is symmetric. $A=\frac13(1+\frac14+\frac14)=\frac12=B$, while $A+B=\E|x|^2\le1$ for every measure on the disk.
\end{proof}

\begin{remark}
Corollary~\ref{cor:smooth} and Proposition~\ref{prop:disk} show that curvature alone does not decide finiteness. On the disk the trivial bound $A+B\le1$ is saturated by a finite centred measure on the circle; at the parabolic barrier, saturation forces the mass to spread along the whole arc. Conversely, a polygon inside the parabolic region always has $s_W<c$, since a finite union of segments meets the parabola in finitely many points; the same holds for bodies with real-analytic boundary inside the region.
\end{remark}

\subsection{Sharp constants on the arc and in its convex hull}

Write \(K_T=\{(u,t):u^2\le t\le T\}\) for \(T\ge1\), identifying
the body with its image in the original \((u,v)\) coordinates.
Then \(K_1=\conv\Gamma\). Let \(s_N^\Gamma(\rho)\) denote the capped
value for sites on the arc itself. The horizontal force across a cut
of the limiting beta law is
\[
 j_\rho(x)=\int_x^1 u\,\beta(du)
 =\frac{(1-x^2)^{a_0}}{2a_0\int_{-1}^1(1-y^2)^{a_0-1}\,dy},
 \qquad I_\rho=\int_{-1}^1 j_\rho(x)^{1/3}\,dx .
\]

\begin{theorem}[Sharp state-count asymptotics]\label{thm:arcsharp}
For every \(\rho>2\),
\begin{align}
 \lim_{N\to\infty}(N-1)^2\bigl(c-s_N^\Gamma(\rho)\bigr)
 &= C_\Gamma(\rho):=\frac{I_\rho^3}{4c},
 \label{eq:arcsharp}\\
 \lim_{N\to\infty}(N-1)^2\bigl(c-s_N(K_1,\rho)\bigr)
 &= C_K(\rho):=\frac{I_\rho^3}{4(c+1)}
   =\frac{c}{c+1}C_\Gamma(\rho).
 \label{eq:hullsharp}
\end{align}
In either case asymptotically optimal designs have stationary horizontal
law converging to \(\beta\) and empirical horizontal site density
\begin{equation}\label{eq:arcdensity}
 \eta_*(x)=\frac{j_\rho(x)^{1/3}}{I_\rho}
          =\frac{(1-x^2)^{a_0/3}}{\int_{-1}^1(1-y^2)^{a_0/3}\,dy}.
\end{equation}
\end{theorem}

The site density counts all states equally. The stationary law counts
them with their invariant weights. The cube-root allocation has the
variational form of high-resolution scalar quantization~\cite{PanterDite}.
Here its weight is a cut force, and the grid must admit positive rates
with both generator equations exact. Fubini gives \(\int j_\rho=c\), so
\[
 C_\Gamma(\rho)=
 \frac{\bigl(\int_{-1}^1(1-x^2)^{a_0/3}\,dx\bigr)^3}
      {4\int_{-1}^1(1-x^2)^{a_0}\,dx}<1
\]
by strict H\"older. The equispaced Hahn chain has coefficient \(1\).
At \(\rho=5/2\) the two optimal coefficients are approximately
\(0.757254\) and \(0.126209\). The factor of six arises from allowing a
small vertical displacement of sites with exact rates and with the
same limiting stationary law.

\begin{proof}
\emph{Bounds for arbitrary graphs.}
Let \(\nu\) be the horizontal mass law, and set
\(j_\nu(x)=\int_{(x,\infty)}u\,\nu(du)\).
As in \eqref{eq:quartic}, \(j_\nu\) is constant on each gap, equal to
the cut force \(C_k\). It vanishes outside the extreme sites.
For \(n=N-1\), the cut inequality and H\"older give
\begin{equation}\label{eq:cutholder}
 n^2\mathcal W_4\ge n^2\sum_k C_kh_k^3
 \ge \left(\int j_\nu^{1/3}\right)^3,
\end{equation}
because there are at most \(n\) gaps. The first inequality is an
equality for a nearest-neighbour chain with distinct abscissae.
For sequences with deficit tending to zero, Theorem~\ref{thm:parab}
gives \(\nu\rightharpoonup\beta\). The limit is atomless, so
\(j_\nu\) converges pointwise to \(j_\rho\) extended by zero.
Dominated convergence on the compact projection gives
\(\int j_\nu^{1/3}\to I_\rho\).
On the arc \(A=c\) and \(\mathcal W_4=4c(c-B)\).

For an arbitrary design above the parabola write
\(S_i=\sum_{j\ne i}Q_{ij}(u_j-u_i)^2\).
The identities for \(Qu\) and \(Qt\) give
\[
 Qd=(\rho-2)(1-u^2)-\rho d-S=(\rho-2)(1-t)-2d-S .
\]
Substitute \(\mathcal E(d,d)=-\E[dQd]\) into \eqref{eq:defect}
and use \(c-A=\E d\). The result is
\begin{equation}\label{eq:slackidentity}
 \mathcal W_4 =
 4c(c-B)+4(c-A)
 -(3\rho-2)\E[d(1-t)]-2\E[d^2]-3\E[dS].
\end{equation}
The three subtracted terms are nonnegative in \(K_1\), hence
\(\mathcal W_4\le4(c+1)(c-\min(A,B))\).
Apply \eqref{eq:cutholder} to designs within \(o(n^{-2})\) of each
capped optimum. The Hahn chains guarantee that their deficits tend
to zero, so the two lower bounds follow.

\emph{Reversible paths with both modes exact.}
Choose a smooth, strictly positive even probability density
\(\eta\) on \([-1,1]\) that is nonincreasing on \([0,1]\), and let
\(y_i\) be its \(i/n\)-quantiles. With \(b=\rho-2\), take
\[
 r_n=\sqrt{\frac{b}{b+1-y_{n-1}}},\qquad
 u_i=r_ny_i,\qquad t_i=u_i^2+\frac{k}{n^2}g(u_i),\quad 0\le i\le n .
\]
For the arc use \(g=0\). For the hull take an even nonnegative smooth
bump \(g\) compactly supported in two strips near \(\pm1\).
The constant \(k\) is fixed below. The sites lie in \(K_1\) for all
large \(n\), since the support of the bump is separated from
the endpoints. At an interior site let \(h_\pm=|u_{i\pm1}-u_i|\)
and let \(s_\pm\) be the left and right secant slopes of the graph
of \(t\), indexed by their respective signs. The rates
\begin{equation}\label{eq:secant_rates}
 q_i^+=\frac{-\rho(t_i-c)+u_i s_-}{h_+(s_+-s_-)},\qquad
 q_i^-=\frac{-\rho(t_i-c)+u_i s_+}{h_-(s_+-s_-)}
\end{equation}
solve \(Qu=-u\), \(Qt=-\rho(t-c)\).
On the unperturbed arc the two numerators are
\(D(u_i)-u_i h_-\) and \(D(u_i)+u_i h_+\), where
\(D(u)=b(1-u^2)\). All gaps are at most the outer gap, since the
derivative of the quantile map increases away from its median.
At \(u_i>0\), \(D(u_i)/u_i>D(r_n)/r_n=u_n-u_{n-1}\).
This proves positivity there, and reflection and the centre
give the remaining cases. At each outer site the inward rate
\(r_n\) divided by the outer gap gives both mode equations by
the choice of \(r_n\).

For large \(n\) the bump vanishes at each outer site and its
neighbour. At the interior triples it affects, the unperturbed
numerators have a uniform positive margin because those triples
stay away from \(\pm1\) and the maximal gap is \(O(n^{-1})\).
The change in each numerator is \(O(n^{-2})\), using Lipschitz
continuity of the bump for the secants. Strict convexity of
\(u^2+kg(u)/n^2\) gives \(s_+>s_-\).
Thus all adjacent rates are positive, and detailed balance
\(p_{i+1}/p_i=q_i^+/q_{i+1}^-\) supplies the stationary weights.

\emph{Matching upper bounds.}
For fixed \(\eta\) and \(g\), \(c-A=\E d=(k/n^2)\E g\) and
\(\mathcal E(d,d)\le (k/n^2)^2\Lip(g)^2 A\).
Since \(\mathcal W_4=\sum_k C_kh_k^3=O(n^{-2})\),
the defect identity gives \(B\to c\). Parabolic rigidity yields
\(\nu\rightharpoonup\beta\).
On the scaled gaps the function \(n h_k\) is uniformly bounded
and converges pointwise to \(1/\eta(x)\) for \(x\in(-1,1)\).
Consequently
\[
 n^2\mathcal W_4=\sum_k C_kh_k(nh_k)^2
 \ \longrightarrow\ J_\eta:=\int_{-1}^1\frac{j_\rho(x)}{\eta(x)^2}\,dx .
\]
For \(g=0\) the scaled deficit converges to \(J_\eta/(4c)\).
For \(g\ne0\) put \(L(x)=4-(6\rho-8)(1-x^2)\) and
\(\overline L_g=\E_\beta(gL)/\E_\beta g\).
By \eqref{eq:defect} the scaled coordinate deficits tend to
\(k\E_\beta g\) and \([J_\eta-k\E_\beta(gL)]/(4c)\), respectively.
Choose the strips close enough to the endpoints that
\(\overline L_g>0\). With
\(k=J_\eta/[(4c+\overline L_g)\E_\beta g]\),
both limits equal \(J_\eta/(4c+\overline L_g)\).
Densities proportional to \((1-x^2+\varepsilon)^{a_0/3}\)
give \(J_\eta\to I_\rho^3\) as \(\varepsilon\downarrow0\)
by dominated convergence. The support strips can be placed
arbitrarily close to the endpoints, so \(\overline L_g\to4\).
For each accuracy one fixes the smooth density and the bump first.
This gives the required upper bounds at all sufficiently large caps.

\emph{Equality and horizontal site density.}
Both optimal limits force \(n^2\mathcal W_4\to I_\rho^3\).
Put \(\eta_n=1/(nh_k)\) on each gap and zero elsewhere.
Its mass is the number of gaps divided by \(n\), at most one.
Applying the cut bound and H\"older to the intervals
\(L_0=[-1,x]\) and \(R_0=(x,1]\), for \(-1<x<1\), gives
\[
 n^2\mathcal W_4\ge\int\frac{j_\nu}{\eta_n^2}
 \ge
 \frac{(\int_{L_0} j_\nu^{1/3})^3}{(\int_{L_0}\eta_n)^2}
 +\frac{(\int_{R_0} j_\nu^{1/3})^3}{(\int_{R_0}\eta_n)^2},
\]
where the integrand is taken as zero outside the gaps.
If the limiting numerators before cubing are \(a,b>0\), the
right side has the unique minimum \((a+b)^3\) over masses
\(l+r\le1\) at \(l=a/(a+b)\), \(r=b/(a+b)\).
Thus \(\eta_n\) has mass tending to one and its cumulative
distribution converges to that of \(\eta_*\).
Replacing the mass on each gap by \(1/n\) at the left endpoint
changes the cumulative distribution by at most \(1/n\).
The distinct abscissae account for a fraction tending to one
of all sites, so the empirical horizontal distribution has
the same limit.
\end{proof}

\begin{corollary}[Where optimal designs spend their vertical displacement]
\label{cor:slacklocation}
For any asymptotically optimal sequence in \(K_1\), both scaled
coordinate deficits tend to \(C_K(\rho)\), and
\[
 (N-1)^2\E[d(1-t)]\to0,\qquad (N-1)^2\E[d^2]\to0 .
\]
Consequently the probability law with weights \(p_i d_i/\E d\)
concentrates near the two ends of the arc.
\end{corollary}
\begin{proof}
Put \(\delta=c-\min(A,B)\). Subtract \(\mathcal W_4\) in
\eqref{eq:slackidentity} from \(4(c+1)\delta\). The difference
is the sum of the three nonnegative displacement terms and
\(4c(\delta-(c-B))+4(\delta-(c-A))\).
Every summand is \(o((N-1)^{-2})\) by
\eqref{eq:hullsharp} and \eqref{eq:cutholder}.
Since \((N-1)^2\E d\to C_K(\rho)>0\), both \(1-t\) and \(d\)
tend to zero in expectation under the normalized weighted law.
Their common zero set in \(K_1\) consists of the two endpoints.
\end{proof}

\begin{figure}[H]
 \centering
 \includegraphics[width=.91\textwidth]{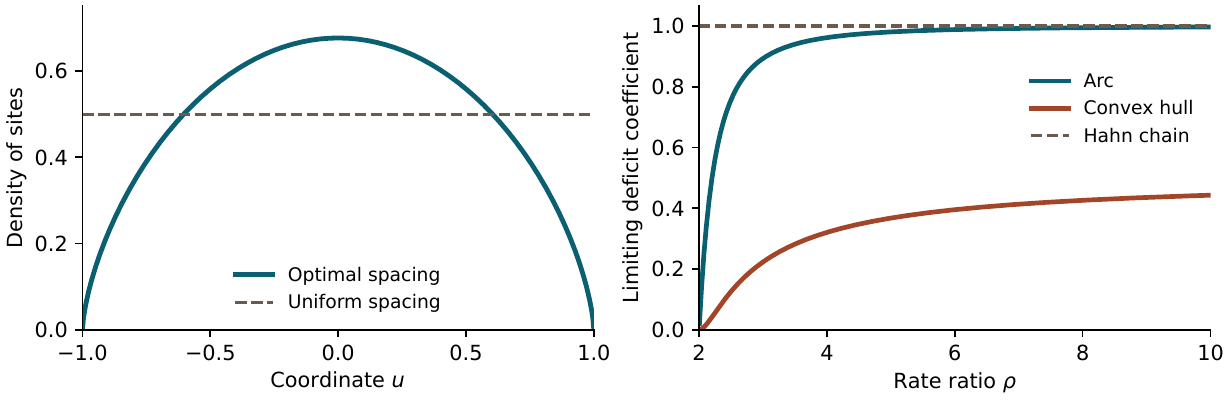}
 \caption{Sharp leading constants for two geometries.
 Left, both optimal geometries have the same limiting density of
 sites at \(\rho=5/2\), while the equispaced Hahn path has a uniform
 density. Right, the leading coefficients of the best finite-state
 deficits on the arc and in its convex hull, with the Hahn coefficient
 for reference. Curves are computed from
 \eqref{eq:arcsharp}--\eqref{eq:arcdensity}.}
 \label{fig:arcsharp}
\end{figure}

\subsection{Sensitivity to an expansion beyond the endpoints}

\begin{theorem}[Effect of raising the cap]\label{thm:raisedcap}
Let \(\rho>2\), \(T\ge1\) and
\(F_N(T)=(N-1)^2(c-s_N(K_T,\rho))\). Then
\begin{equation}\label{eq:cap_bounds}
 \frac{I_\rho^3}{4(c+1)+(3\rho-2)(T-1)}
 \le\liminf_{N\to\infty}F_N(T)
 \le\limsup_{N\to\infty}F_N(T)
 \le C_K(\rho).
\end{equation}
For every \(\rho>8/3\) and every \(T>1\), the last inequality
is strict. Moreover, for such \(\rho\) there are
\(m_\rho,\eta_\rho>0\) such that
\begin{equation}\label{eq:cap_strict}
 \limsup_{N\to\infty}F_N(T)
 \le\frac{I_\rho^3}{4(c+1)+m_\rho(T-1)}
 \qquad (1<T<1+\eta_\rho).
\end{equation}
\end{theorem}

The lower bound in \eqref{eq:cap_bounds} follows from
\eqref{eq:slackidentity} using
\(\E[d(t-1)]\le(T-1)(c-A)\) and then \eqref{eq:cutholder}.
The upper bound follows by inclusion. The strict improvement
comes from a curve that agrees with the arc in the centre and
bends upward near both endpoints. Its equilibrium admits an
improvement in vertical variance per unit horizontal variance
lost that is larger than the limiting ratio available in \(K_1\).
A finite reversible design with that ratio can be mixed in
at order \(N^{-2}\). Appendix~\ref{app:raisedcap} proves
\eqref{eq:cap_strict}. These bounds determine a change of order
\(T-1\) at \(T=1\) when \(\rho>8/3\), while the limiting
stationary law remains \(\beta\) by Theorem~\ref{thm:parab}.

\begin{corollary}[Bounds for general parabolic confinement]
\label{cor:arcconvex}
Under Theorem~\ref{thm:parab}, suppose that \(P\) contains
\(\Gamma\) and \(T=\max_P t\). Then the bounds
\eqref{eq:cap_bounds} hold with \(s_N(P,\rho)\) in place of
\(s_N(K_T,\rho)\).
\end{corollary}
\begin{proof}
Convexity gives \(K_1\subseteq P\subseteq K_T\), so the assertion
follows by monotonicity in the body.
\end{proof}

\section{Finite models of the Wright--Fisher diffusion}\label{sec:moran}
%=====================================================================

In the coordinate \(u=2x-1\) of the allele frequency \(x\in[0,1]\),
the Wright--Fisher diffusion with symmetric mutation has generator
\[
\mathcal Gf=\kappa(1-u^2)f''-uf',\qquad \kappa=\frac{\rho-2}2 ,
\]
with stationary law the beta law $\beta$ of Section~\ref{sec:parab}. Directly $\mathcal Gu=-u$ and $\mathcal G(u^2-c)=-\rho(u^2-c)$ with $c=(\rho-2)/\rho$: the mean frequency and the heterozygosity-type observable $v=b_0(u^2-c)$ relax exactly exponentially, with rates $1$ and $\rho$. The pair $(u,v)$ moves on the arc $\Gamma$, and every probability law over frequencies has its pair of means in the convex hull $\conv\Gamma=\{b_0(u^2-c)\le v\le b_0(1-c)\}$. Thus a finite kinetic model whose states represent frequency distributions, and which reproduces exactly the two slowest relaxation modes, is precisely a design in $\conv\Gamma$.

\begin{proposition}[The Moran model is the Hahn chain]\label{prop:moran}
The Hahn chain of Section~\ref{sec:parab} with $n$ steps is the continuous-time Moran model of $n$ individuals in which each ordered pair of individuals of different types resamples at rate $r=(\rho-2)/2$ and each individual mutates to the other type at rate $\mu=\frac12$: $q_i^+=(n-i)(ri+\mu)$, $q_i^-=i(r(n-i)+\mu)$.
\end{proposition}
\begin{proof}
$(n-i)(i+a_0)/(2a_0)=(n-i)\bigl(i/(2a_0)+\frac12\bigr)$ with $1/(2a_0)=(\rho-2)/2$, and similarly for $q_i^-$. The observable at count \(i\) is evaluated at
\(u_i=(2i-n)/\sqrt{n(n+2a_0)}\), a rescaled frequency.
For \(n\ge2\), the first two nonzero decay rates are
\(2\mu=1\) and \(2r+4\mu=\rho\). At \(n=1\), the quadratic
observable vanishes.
\end{proof}

\begin{theorem}[State-count bound for exact relaxation models]\label{thm:microstate}
Let $\rho>2$, $c=(\rho-2)/\rho$, and consider any reversible Markov chain with $N\ge2$ states and two observables $(u,v)$ with values in $\conv\Gamma$ satisfying $Qu=-u$, $Qv=-\rho v$. Then
\begin{equation}\label{eq:micro}
c-\min(\operatorname{Var}u,\operatorname{Var}v)\ \ge\ \frac{c^2}{4(c+2)(N-1)^2},
\end{equation}
while the Wright--Fisher diffusion has $\operatorname{Var}u=\operatorname{Var}v=c$ and the Moran model with $n=N-1$ individuals has deficit $1/((N-1)(N-1+2a_0))$. Consequently a model reproducing the two relaxation modes with fluctuation deficit $\delta$ must have
\[
N\ \ge\ 1+\frac{c}{2\sqrt{(c+2)\,\delta}}
\]
states. The Moran model reaches deficit at most \(\delta>0\) with
\(n_\delta=\max\{1,\lceil\sqrt{a_0^2+\delta^{-1}}-a_0\rceil\}\)
individuals, hence \(n_\delta+1\) states. As \(\delta\downarrow0\)
this is within a factor \(2\sqrt{c+2}/c\) of the minimum state count,
up to a relative error tending to zero.
\end{theorem}
\begin{proof}
On $\conv\Gamma$ one has $u^2\le t\le1$ and $0\le d=t-u^2\le1$, so the bracket in \eqref{eq:defect} satisfies $(6\rho-8)t-6\rho c+4d\le(6\rho-8)-6(\rho-2)+4=8$; hence $M=4c+8$ in \eqref{eq:defectbound} and $k=\min(c/2,c^2/M)=c^2/(4(c+2))$ in \eqref{eq:N2}. Means are zero by Lemma~\ref{lem:id}, so variances are $A$ and $B$. The Moran statement is Proposition~\ref{prop:moran} and the upper bound in \eqref{eq:N2}; solving \eqref{eq:micro} gives the lower state count. Solving
\(n(n+2a_0)\ge\delta^{-1}\) and rounding up gives \(n_\delta\).
The asymptotic comparison follows by dividing the two bounds.
\end{proof}

Let \(N_\delta^\Gamma\) and \(N_\delta^K\) be the least state caps
with deficit at most \(\delta\) on the arc and in its convex hull,
respectively. Theorem~\ref{thm:arcsharp} gives the sharp state
budgets
\[
 N_\delta^\Gamma-1\sim \sqrt{C_\Gamma(\rho)/\delta},\qquad
 N_\delta^K-1\sim \sqrt{C_K(\rho)/\delta}
 \qquad(\delta\downarrow0).
\]
They follow by inverting the two capped limits and using
monotonicity in \(N\). The two budgets have a direct modelling meaning.
An arc state represents a single allele frequency, while a hull state
can represent a distribution of frequencies through its first two
moments. At \(\rho=5/2\) the second kind of model needs asymptotically
\(\sqrt6\) times fewer states for the same accuracy.
Appendix~\ref{app:robust} gives a bound when the generator
eigenfunction equations hold only up to measured residuals.

\section{Linear state complexity}\label{sec:linear}
%=====================================================================

\begin{figure}[t]
\centering
\includegraphics[width=\textwidth]{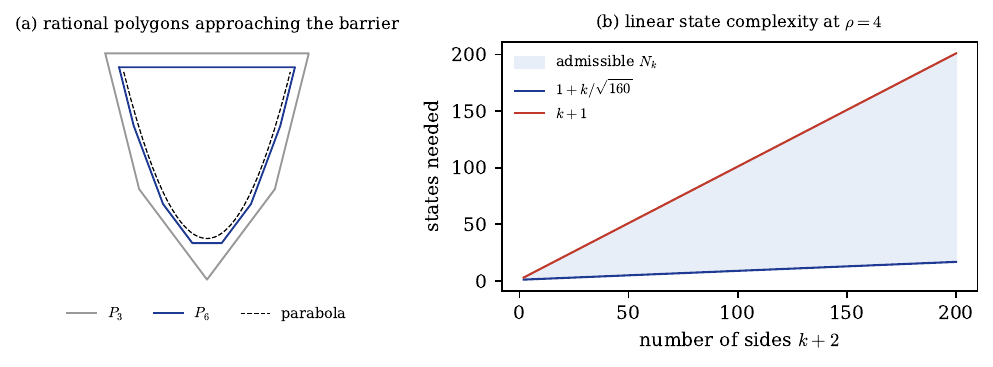}
\caption{(a) Rational polygons $P_k$ circumscribed about the parabolic barrier at $\rho=4$. (b) The least number of states needed to exceed $1/2$ on $P_k$ lies in the shaded band (Theorem~\ref{thm:linear}).}
\label{fig:linear}
\end{figure}

Take $\rho=4$, so $c=\frac12$, $a_0=\frac12$, $b_0=2$ and the barrier region is $\{2u^2-1\le v\}$. For $k\ge2$ let $a_j=-1+2j/k$ ($0\le j\le k$) and
\[
H_k=\{v\le1,\ v\ge4a_ju-2a_j^2-1\ (0\le j\le k)\},\qquad P_k=(1+2/k^2)H_k,
\]
rational $(k+2)$-gons circumscribed about the parabola $v=2u^2-1$ (Figure~\ref{fig:linear}a).

\begin{theorem}[Linear state complexity]\label{thm:linear}
$s_W(P_k,4)-\frac12=\Theta(k^{-2})$. If $N_k$ is the least number of sites of a design in $P_k$ with $\min(A,B)>\frac12$, then
\[
1+\frac{k}{\sqrt{160}}<N_k\le k+1 .
\]
In particular no bound $N(n)=o(n)$ on the size of near-optimal designs holds uniformly over $n$-gons at a fixed rate ratio.
\end{theorem}
\begin{proof}
Let $K=\{2u^2-1\le v\le1\}$ and $r=1+2/k^2$. Each tangent is active on an interval around its contact abscissa, so together with the lid these are exactly the $k+2$ sides; the end tangents force $|u|\le1$ below the lid, so every $u$ is within $1/k$ of a contact abscissa, and the tangent at $a_j$ errs by $2(u-a_j)^2\le2/k^2$. Hence $K\subset H_k\subset rK$ and $rK\subset P_k\subset r^2K$. Dilation by $r$ multiplies both moments by $r^2$.

Upper bound: the Hahn chain with $n=k$ fits in $K$, hence after dilation by $r$ in $rK\subset P_k$; its smaller moment is $r^2(\frac12-\frac1{k(k+1)})>\frac12$ for $k\ge2$, since $\frac1{k(k+1)}<\frac1{k^2}$ and $(1+2/k^2)^2(1-2/k^2)\ge1$. Its excess times $k^2$ tends to $1$, and $s_W(P_k,4)\le r^4/2$ by the outer inclusion; this gives the order $k^{-2}$.

Lower bound: on $K$ we have $0\le d\le t\le1$, so the bracket in \eqref{eq:defect} is $16t-12+4d\le8$ and $M\le10$, $k_{K,4}=\frac1{40}$. An $N$-site design in $P_k$ with $\min(A,B)>\frac12$, shrunk by $r^2$, is a design in $K$ with deficit $<\frac12(1-r^{-4})\le2(r-1)=4/k^2$. By \eqref{eq:N2}, $\frac1{40(N-1)^2}<\frac4{k^2}$.
\end{proof}

\section{Fixed state caps and jumps in the optimal state count}
\label{sec:capped}

Section~\ref{sec:parab} shows that the supremum over all finite chains
need not be attained. With a fixed bound on the number of states, it
always is. The proof uses only finite networks and does not rely on the
continuum relaxation of Section~\ref{sec:relax}.

\begin{lemma}[Finite realization from monotone work]\label{lem:finitework}
Let \(\mu=\sum_a p_a\delta_{z_a}\) be a probability law on a finite
set. If
\[
 \int \Lambda x\cdot V(x)\,d\mu(x)\ge0
 \quad\text{for every globally Lipschitz monotone field \(V\),}
\]
then \(\mu\) is the mass law of a nonnegative cable design on its
support.
\end{lemma}
\begin{proof}
Farkas' lemma reduces the equilibrium equations to the condition
\begin{equation}\label{eq:cappedfarkas}
 \sum_a p_a\Lambda z_a\cdot w_a\ge0\qquad\text{whenever}\qquad
 (w_a-w_b)\cdot(z_a-z_b)\ge0\quad(a,b=1,\dots,M).
\end{equation}
Fix one such assignment \(w\) and \(\varepsilon>0\). For sufficiently
small \(t>0\), the map \(z_a\mapsto z_a-t(w_a+\varepsilon z_a)\)
is \(1\)-Lipschitz. Squaring each distance inequality gives a
strict linear gain \(2t\varepsilon|z_a-z_b|^2\) against a quadratic
error in \(t\). Extend the map by Kirszbraun's theorem
\cite{Kirszbraun} to a \(1\)-Lipschitz map \(H\) of the plane.
Then \(V=(\operatorname{id}-H)/t\) is globally Lipschitz and monotone,
because \((H(x)-H(y))\cdot(x-y)\le|x-y|^2\).
Its work inequality gives
\(\sum_a p_a\Lambda z_a\cdot(w_a+\varepsilon z_a)\ge0\).
Let \(\varepsilon\) decrease to zero and apply Farkas' lemma.
\end{proof}

\begin{theorem}[Attainment at a fixed state cap]\label{thm:capatta}
Let \(P\) be compact and contain the origin. For any \(\rho>1\)
and any \(N\ge1\), \(s_N(P,\rho)\) is attained. More precisely,
any weak limit of feasible laws with at most \(N\) sites is
feasible with at most \(N\) sites. The latter statement also
holds along rates \(\rho_m\to\rho>1\).
\end{theorem}
\begin{proof}
A weak limit of probability laws on a compact set with at most
\(N\) atoms still has at most \(N\) atoms. This follows by representing
each law with \(N\) sites and \(N\) weights, padding by zero weights,
and passing to a subsequence in the product of \(P^N\) and the
probability simplex. For any globally Lipschitz monotone field \(V\),
the equilibrium equations give
\[
 \int \Lambda_m x\cdot V(x)\,d\mu_m(x)
 =\sum_{i<j}J_{ij}^{(m)}(x_i^{(m)}-x_j^{(m)})
                  \cdot(V(x_i^{(m)})-V(x_j^{(m)}))\ge0 .
\]
The left side converges to the corresponding work of the limiting
law at \(\Lambda\). Lemma~\ref{lem:finitework} realizes that law.
Apply the result to a maximizing sequence and use continuity of
the two moments on the compact set.
\end{proof}

\begin{corollary}[Exact algebraic capped optimization]\label{cor:capalg}
Let $P$ be a polygon defined by rational linear inequalities and $N\ge1$.
Then $(\rho,c)\mapsto[s_N(P,\rho)\ge c]$ is a semialgebraic relation,
and $s_N(P,\rho)$ is a semialgebraic function of $\rho>1$.
For algebraic $\rho>1$ its value is a real algebraic number, and at
least one maximizing design can be chosen with algebraic sites,
masses and conductances.  In principle, $s_N\ge c$ is decidable for
algebraic input $(P,\rho,c)$ by real quantifier elimination.
Moreover $s_N(P,\rho)$ is non-increasing in $\rho$, and has only
finitely many analytic branches as \(\rho\) varies over \((1,\infty)\).
The value is upper semicontinuous and left-continuous in \(\rho\).
\end{corollary}
\begin{proof}
For each $1\le m\le N$ introduce real variables for $m$ site
coordinates, $m$ strictly positive masses and $m(m-1)/2$
nonnegative conductances.  The constraints $x_i\in P$,
$\sum p_i=1$, \eqref{eq:eq} and $A\ge c$, $B\ge c$ are a finite
system of polynomial inequalities and equations with rational
coefficients in those variables and $\rho,c$; distinctness is expressed
by the strict polynomial inequalities $|x_i-x_j|^2>0$.  Their finite
disjunction is an existential formula over the real closed field.
Tarski--Seidenberg quantifier elimination gives the first assertion;
Theorem~\ref{thm:capatta} ensures the upper endpoint is an attained
value.  When the parameters are real algebraic, the resulting
nonempty semialgebraic feasibility set has a point over the real
algebraic numbers, giving an algebraic optimizer.  The graph of the
optimal-value function is semialgebraic by the same elimination, so
it has finitely many real-analytic semialgebraic pieces on the real
line.  Finally, mixing the conductances of a design at rate $\rho'$
with the complete reset conductances $p_ip_j$ preserves its sites
and masses and realizes every \(1<\rho\le\rho'\).
The compactness argument at varying rates in
Theorem~\ref{thm:capatta} gives upper semicontinuity.
Together with monotonicity it gives left-continuity.
\end{proof}

\subsection{Three states and a jump on a seven-gon}

\begin{theorem}[Unique conductances on three states]\label{thm:threeexact}
Let $x_i=(u_i,v_i)$, $i=1,2,3$, be distinct sites with masses
$p_i>0$, $\sum p_i=1$, satisfying
\[
 \E u=\E v=\E uv=0,\qquad A=\E u^2>0,\quad B=\E v^2>0.
\]
At a prescribed rate ratio $\rho>1$ there exists a reversible design
on \emph{exactly these sites and masses} if and only if
\begin{equation}\label{eq:threecrit}
 (\rho-1)v_iv_j\le B\qquad(1\le i<j\le3).
\end{equation}
When it exists, its three conductances are uniquely determined by
\begin{equation}\label{eq:threeJ}
 \boxed{J_{ij}=p_ip_j\left[1-(\rho-1)\frac{v_iv_j}{B}\right]}
 \qquad(i\ne j).
\end{equation}
Writing $m_v=\max_{i<j}v_iv_j$, the precise admissible rate interval
is $1<\rho\le1+B/m_v$ if $m_v>0$; if $m_v\le0$, every $\rho>1$
is admissible.  In the latter case one site lies on $v=0$ and
the other two lie on a common vertical line, on opposite sides of
that axis.  Conversely, every such nondegenerate centred and
orthogonal three-site law admits all $\rho>1$.
\end{theorem}
\begin{proof}
In the three-dimensional Hilbert space of node functions with
inner product $\langle f,g\rangle=\sum_i p_if_ig_i$, the three
vectors $\mathbf1$, $u/\sqrt A$ and $v/\sqrt B$ are orthonormal,
so form a complete basis.  For any reversible generator satisfying
$Q\mathbf1=0$, $Qu=-u$, $Qv=-\rho v$, its spectral
representation is forced:
\[
 Qf=-u\frac{\langle u,f\rangle}{A}
      -\rho v\frac{\langle v,f\rangle}{B}.
\]
Completeness also gives, for distinct $i,j$,
$u_iu_j/A+v_iv_j/B=-1$.  Therefore
\[
 Q_{ij}=-p_j\left(\frac{u_iu_j}{A}+
                      \rho\frac{v_iv_j}{B}\right)
 =p_j\left[1-(\rho-1)\frac{v_iv_j}{B}\right],
\]
and $J_{ij}=p_iQ_{ij}$ yields \eqref{eq:threeJ}.  The Markov
requirement $Q_{ij}\ge0$ is precisely \eqref{eq:threecrit}.
Conversely, if these inequalities hold, the displayed symmetric
conductances produce a genuine generator.  The centring and
orthogonality conditions give $Qu=-u$ and $Qv=-\rho v$ by direct
substitution; uniqueness has already been established.

Finally, $m_v\le0$ means that no two of the three vertical
coordinates have the same nonzero sign.  Since $\E v=0$ and $B>0$,
exactly one vertical coordinate is zero and the other two have
opposite signs.  If these occur at sites $i,j$, then
$0=\E uv-u_i\E v=p_jv_j(u_j-u_i)$, so $u_i=u_j$.
The converse follows directly from \eqref{eq:threecrit}.
\end{proof}

\begin{example}[A sharp three-state rate threshold]\label{ex:threecutoff}
For equal masses $p_i=1/3$ at sites
$(-1,1),(1,1),(0,-2)$ we have $A=2/3$, $B=2$ and
\[
 J_{12}=\frac{3-\rho}{18},\qquad
 J_{13}=J_{23}=\frac{\rho}{9}.
\]
Thus \emph{these three states} realize exactly the rate ratios
$1<\rho\le3$; at $\rho=3$ one cable disappears, and for larger
$\rho$ uniqueness forces a forbidden negative cable.  Applied to
the critical hexagon of Theorem~\ref{thm:hexphase},
\eqref{eq:threecrit} gives directly the exceptional threshold
$\rho\le1+1/b_*$ for its three-site optimum.
\end{example}

\begin{figure}[htbp]
\centering
\includegraphics[width=.74\textwidth]{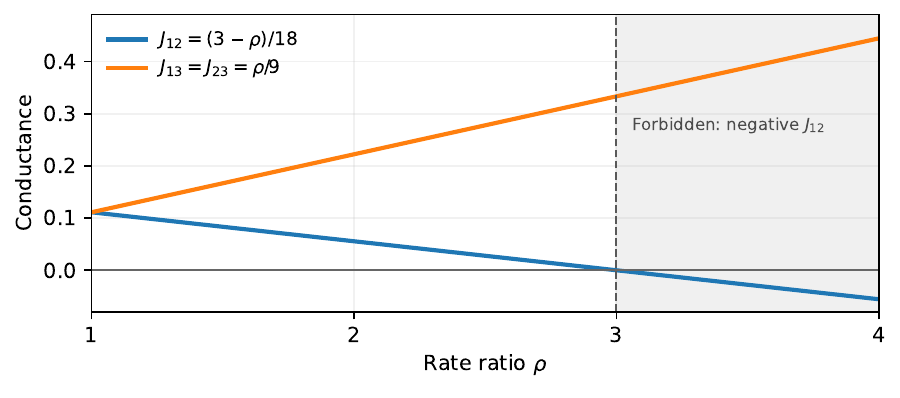}
\caption{Forced conductances for the three-site geometry
$(-1,1),(1,1),(0,-2)$ with equal masses. The vertical line is the
exact feasibility threshold $\rho=3$. The continuation $J_{12}<0$
to its right is \emph{not} a valid Markov chain; on three sites no
alternative choice of cables exists.}
\label{fig:threecutoff}
\end{figure}

\begin{theorem}[A tunable $3\to4$ phase transition on a seven-gon]
\label{thm:sevenphase}
For any $0<b<1$, put $a=\sqrt{1+b^2}$ and let $P_b$ be the
convex hull of the seven sites
\[
 \begin{gathered}
 A=(-a,b),\quad B=(a,b),\quad C=(0,-1/b),\\
 D=(-1,-1),\quad E=(-1,1),\quad F=(1,-1),\quad G=(1,1).
 \end{gathered}
\]
All seven sites are extreme vertices of $P_b$.  For every $\rho>1$,
\begin{equation}\label{eq:sevenphase}
 \boxed{\quad s_W(P_b,\rho)=1,\qquad
 N_{\rm opt}(P_b,\rho)=
 \begin{cases}3,&1<\rho\le1+b^{-2},\\
               4,&\rho>1+b^{-2}.\end{cases}\quad}
\end{equation}
The two optimal laws constructed below have the same means, mixed
moment and variances. Above $1+b^{-2}$ only one of them carries
nonnegative conductances.
For every rational $t\in(0,\sqrt2-1)$, choosing
\[
 b=\frac{2t}{1-t^2},\qquad a=\frac{1+t^2}{1-t^2}
\]
produces a \emph{rational} seven-gon.  Such rational seven-gons
have transition rates dense in $(2,\infty)$.
In particular, at $b=3/4$ the vertices are rational and
$N_{\rm opt}$ jumps from three to four at $\rho=25/9$.
\end{theorem}
\begin{proof}
Set $\alpha=1-b^2\in(0,1)$.  The strictly convex quadratic
$q_\alpha(u,v)=\alpha u^2+(1-\alpha)v^2$ equals one at all seven
listed points.  Hence each is a vertex of their convex hull,
$q_\alpha\le1$ on $P_b$, and equality is possible only at the
seven vertices.  Every design consequently satisfies
$\min(A,B)\le\alpha A+(1-\alpha)B\le1$.
Equality requires $A=B=1$ and the support to lie among the vertices.

On $A,B,C$ take
\[
 p_A=p_B=\frac1{2(1+b^2)},\qquad
 p_C=\frac{b^2}{1+b^2}.
\]
These weights give $\E u=\E v=\E uv=0$ and $A=B=1$.
The forced conductances of Theorem~\ref{thm:threeexact} are
\begin{equation}\label{eq:sevenJ3}
 J_{AB}=\frac{1-(\rho-1)b^2}{4(1+b^2)^2},\qquad
 J_{AC}=J_{BC}=\frac{\rho b^2}{2(1+b^2)^2}.
\end{equation}
They are nonnegative if and only if $1<\rho\le1+b^{-2}$.
For every $\rho>1$, use instead $D,E,F,G=(\pm1,\pm1)$
with masses $1/4$ and conductances
\begin{equation}\label{eq:sevenJ4}
 J_{ij}=\frac1{16}\left[1+2(\rho-1)
 \mathbf1_{\{u_i=u_j\}}\right]\quad(i\ne j).
\end{equation}
Lemma~\ref{lem:vertical} gives all equilibrium identities and
all $J_{ij}>0$; again $A=B=1$.  Thus $s_W=1$ at every rate.

To show that no alternative three-site optimizer exists after the
cutoff, recall that for a three-site law with
$\E u=\E v=\E uv=0$ and $A=B=1$, the orthonormal basis
$\{\mathbf1,u,v\}$ of $L^2(p)$ forces
\begin{equation}\label{eq:sevenorthogonal}
 u_i u_j+v_i v_j=-1\qquad(i\ne j)
\end{equation}
by completeness (as in Theorem~\ref{thm:threeexact}).
Among the seven vertices, the \emph{only} pairs with Euclidean
dot product $-1$ are $AB$, $AC$, and $BC$.
Indeed $a^2-b^2=1$, $b(1/b)=1$;
$0<a-b<1$, $a+b>1$, $1/b>1$;
the four square vertices have mutual dot products $0$ or $-2$.
These facts exclude every other pair.  Thus the only possible
three-site optimal support is $ABC$.  Its masses are uniquely fixed
by centring, and its conductance $J_{AB}$ in
\eqref{eq:sevenJ3} becomes negative for $\rho>1+b^{-2}$.
Two or fewer states cannot have the rank-two covariance matrix
$\operatorname{diag}(1,1)$.  This proves \eqref{eq:sevenphase}.

For rational $t\in(0,\sqrt2-1)$, the displayed parametrization
satisfies $a^2=1+b^2$ with $0<b<1$, so all seven vertices are
rational.  The map $b\mapsto1+b^{-2}$ is continuous and maps
$(0,1)$ onto $(2,\infty)$, while rational $t$ are dense in
$(0,\sqrt2-1)$. Therefore the corresponding rational transition
rates are dense in $(2,\infty)$.  The specific case $b=3/4$ and its
force balances are checked in exact rational arithmetic by
\texttt{anc/verify\_181\_heptagon.py}.
\end{proof}

\begin{figure}[t]
\centering
\includegraphics[width=.88\textwidth]{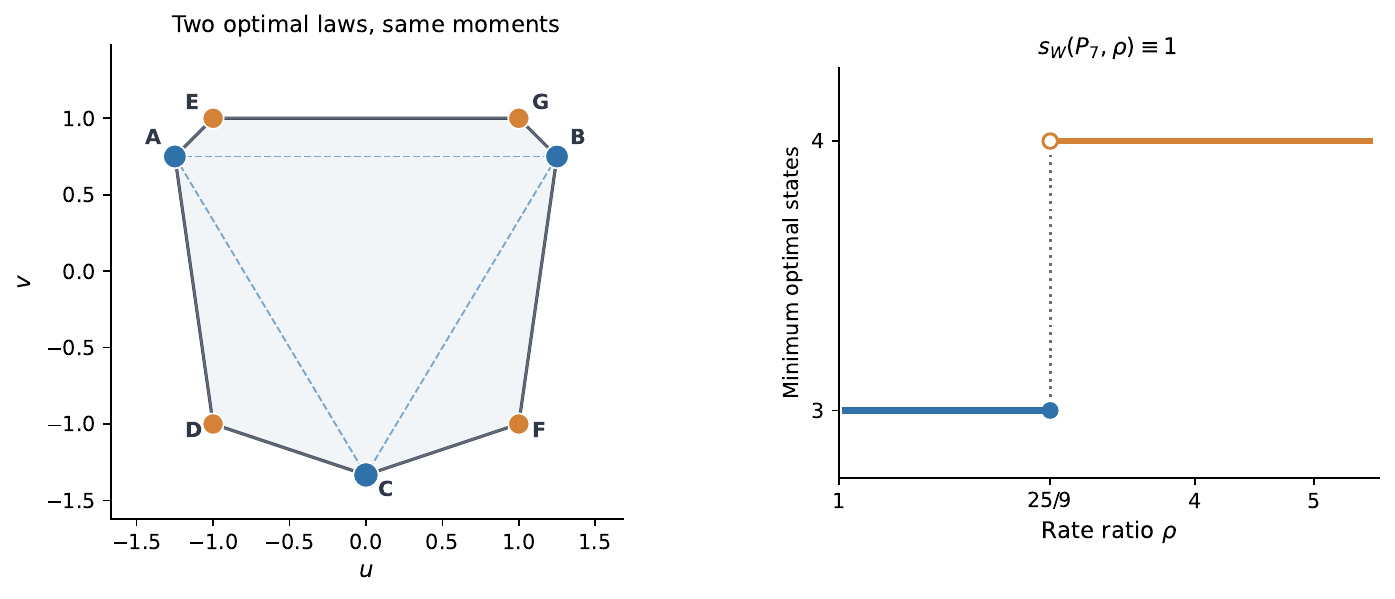}
\caption{The $b=3/4$ rational seven-gon in Theorem~\ref{thm:sevenphase}.
Left: two different optimal laws with identical five moments: the
three-site law $ABC$ and the four-site law $DEFG$.  Right: the
minimal optimal state count jumps from three to four at $\rho=25/9$,
while $s_W\equiv1$.  The parameter $b$ tunes this cutoff over the
entire interval $(2,\infty)$.}
\label{fig:sevenphase}
\end{figure}

\begin{remark}[Fixed caps and the polygon problem]
If FRR (Definition~\ref{def:frr}) fails at $(P,\rho)$, then
$s_N(P,\rho)<s_W(P,\rho)$ for every $N$. For rational polygons and
algebraic rates each fixed-$N$ value can be computed exactly, but no
finite list of such computations decides FRR without a uniform bound on
the optimal number of states.
\end{remark}

\section{Continuum stresses and upper certificates}\label{sec:relax}
%=====================================================================

To a design $W$ attach its mass measure and its \emph{chord stress}
\[
\mu_W=\sum_ip_i\delta_{x_i},\qquad T_W=\sum_{i<j}J_{ij}(x_i-x_j)(x_i-x_j)^{\mathsf T}\int_0^1\delta_{(1-t)x_i+tx_j}\,dt .
\]
$T_W$ is a positive semidefinite matrix measure, and integrating $\nabla\varphi$ along each chord in \eqref{eq:eq} gives
\begin{equation}\label{eq:weak}
\int\nabla\varphi:dT=\int\Lambda x\cdot\varphi(x)\,d\mu(x)\qquad(\varphi\in C^1(\R^2;\R^2)).
\end{equation}

\begin{theorem}[Exact stress relaxation]\label{thm:closure}
Let $\mu$ be a probability measure and $T$ a finite positive semidefinite matrix measure, both supported in $P$, satisfying \eqref{eq:weak}. Then there are designs at the same $\rho$ with sites in $\operatorname{int}P$ such that $(\mu_W,T_W)\rightharpoonup(\mu,T)$. Conversely every weak limit of designs satisfies \eqref{eq:weak}. Hence the set $\K$ of such pairs is compact and convex and
\begin{equation}\label{eq:relaxed}
s_W(P,\rho)=\max_{(\mu,T)\in\K}\min\Bigl(\int u^2d\mu,\int v^2d\mu\Bigr).
\end{equation}
\end{theorem}
\noindent The proof (Appendix~\ref{app:closure}) shrinks and smooths the pair, discretizes the stress on a lattice, and absorbs the force errors into a small reversible background network with uniformly positive conductances to four anchor points. Compactness follows from $\tr T(P)=\int x\cdot\Lambda x\,d\mu$, obtained by testing \eqref{eq:weak} with $\varphi(x)=x$.

A vector field $U$ is \emph{monotone} on a set if $(U(x)-U(y))\cdot(x-y)\ge0$ for all $x,y$ in it. Write $q_\alpha(x)=\alpha u^2+(1-\alpha)v^2$.

\begin{theorem}[Monotone certificate duality]\label{thm:gap}
Let \(\alpha\in[0,1]\), let \(U\) be monotone on \(P\), and assume \(F=q_\alpha+\Lambda x\cdot U\le C\) on \(P\). Then for every design
\begin{equation}\label{eq:gap}
C-\bigl(\alpha A+(1-\alpha)B\bigr)=\sum_ip_i\bigl(C-F(x_i)\bigr)+\sum_{i<j}J_{ij}(x_i-x_j)\cdot\bigl(U(x_i)-U(x_j)\bigr)\ \ge\ 0 .
\end{equation}
In particular $s_W\le C$. Equality for a design forces all mass onto $\{F=C\}$ and zero work on every cable.
\end{theorem}
\begin{proof}
$\sum_ip_iF(x_i)=\alpha A+(1-\alpha)B+\sum_ip_i\Lambda x_i\cdot U(x_i)$, and by \eqref{eq:eq} and symmetry, $\sum_ip_i\Lambda x_i\cdot U(x_i)=\sum_{i,j}J_{ij}(x_i-x_j)\cdot U(x_i)=\sum_{i<j}J_{ij}(x_i-x_j)\cdot(U(x_i)-U(x_j))$.
\end{proof}

\begin{theorem}[Strong duality and exact certificates]\label{thm:strongdual}
\[
s_W(P,\rho)=\inf_{\substack{0\le\alpha\le1,\ U\in C^1\\ e(U)\succeq0\text{ on }P}}\ \max_{x\in P}\bigl(q_\alpha(x)+\Lambda x\cdot U(x)\bigr),\qquad e(U)=\tfrac12(\nabla U+\nabla U^{\mathsf T}).
\]
For every $\eta>0$ there are a rational $\alpha$, a polynomial field $U$ with rational coefficients and $e(U)\succ0$ on $P$, and a rational $C<s_W+\eta$ with $q_\alpha+\Lambda x\cdot U<C$ on $P$. For rational polygons and rational $\rho$ these are finitely many real polynomial inequalities.
\end{theorem}
\begin{proof}
A $C^1$ field with $e(U)\succeq0$ on a convex set is monotone there (integrate along segments), so ``$\ge$'' is Theorem~\ref{thm:gap}. Fix $\alpha$, let $M=\max_Px\cdot\Lambda x$ and $L>M$. On the compact convex set of probability measures on $P$ and positive semidefinite stresses of trace at most $L$, apply the affine minimax theorem to $\int(q_\alpha+\Lambda x\cdot U)d\mu-\int e(U):dT$ with $U$ ranging over $C^1$ fields. The infimum over $U$ is $-\infty$ unless \eqref{eq:weak} holds, in which case it equals $\int q_\alpha d\mu$; the trace budget is not active on $\K$ by the trace identity. In the other order the maximum equals $\max_P(q_\alpha+\Lambda x\cdot U)+L\,\varepsilon(U)$ with $\varepsilon(U)=\max(0,-\min_P\lambda_{\min}(e(U)))$, and replacing $U$ by $U+\varepsilon(U)x$ gives a field with $e\succeq0$ and maximum at most that value. Minimizing over $\alpha$ (affine minimax on $\K$, Theorem~\ref{thm:closure}) proves the formula. For the certificates, take a near-optimal field, add $\delta x$ to obtain a uniform margin $e(U)\succeq\delta I$, approximate it in $C^1$ on a box containing $P$ by polynomials (Bernstein), and round coefficients and $\alpha$ to rationals.
\end{proof}

\begin{theorem}[Finite-support principle]\label{thm:finsupp}
Let $(\mu,T)\in\K$. If $\mu$ has finite support $X$, then there is a design on exactly the sites $X$ with mass law $\mu$. Consequently finite attainment $s_W=\max_N s_N$ holds if and only if some optimal pair in \eqref{eq:relaxed} has finitely supported mass.
\end{theorem}
\begin{proof}
Convolve any globally Lipschitz monotone field \(V\) with a smooth
nonnegative approximate identity. The smoothed fields \(V_\varepsilon\)
are monotone and converge to \(V\) uniformly on \(P\).
By \eqref{eq:weak} their work equals
\(\int e(V_\varepsilon):dT\ge0\).
Pass to the limit and apply Lemma~\ref{lem:finitework}.
The assertion about attainment follows from \eqref{eq:relaxed}.

\end{proof}

\begin{corollary}[Mass laws of the closure]\label{cor:masslaws}
A probability measure $\mu$ on $P$ is the mass law of some $(\mu,T)\in\K$ iff $\int\Lambda x\cdot U\,d\mu\ge0$ for every globally Lipschitz monotone field $U$ on $\R^2$.
\end{corollary}
\noindent The proof (Appendix~\ref{app:closure}) extends the functional $e(\varphi)\mapsto\int\Lambda x\cdot\varphi\,d\mu$ positively by Hahn--Banach and represents it by a stress measure.

\section{Polygonal confinement and exact certificates}\label{sec:polygons}
%=====================================================================

Let $P$ be a convex $n$-gon with $0\in\operatorname{int}P$.

\begin{definition}[Dictionary]
$S(P,\rho)$ consists of: the vertices; the intersections of the sides with the coordinate axes; for each side $E$ with outer normal $n_E$, the points of $E$ with $n_E\cdot\Lambda x=0$; and for each vertex $h$ and side $E$, the roots on $E$ of the leaf equation \eqref{eq:leaf} (if $K_h\equiv0$ on $E$, only its endpoints). $|S|=O(n^2)$. Let $\LS(P,\rho)$ be the value of $\max\min(A,B)$ over designs on $S$ --- a finite linear program.
\end{definition}

Each element of the dictionary has a structural origin. Leaves of an
optimal network hang on hubs along the curve \eqref{eq:leaf}
(Proposition~\ref{prop:cut}), axis crossings come from two-site
components, and the points with $n_E\cdot\Lambda x=0$ are where a side is
tangent to the load field. Only sites in the invariant core
(Theorem~\ref{thm:invcore}) can carry mass, and the equilibrium
equations automatically assign zero mass to the others, so they do no
harm in the linear program. Theorem~\ref{thm:barrier} rules out limiting
mass on curved boundary arcs of the core, but does not by itself imply
the conjecture.

\begin{definition}[Finite realizability]\label{def:frr}
We say that \emph{FRR holds at $(P,\rho)$} if the capacity is attained by
a finite design, that is, $s_W(P,\rho)=s_N(P,\rho)$ for some $N$. The
\emph{FRR problem} asks whether FRR holds for every convex polygon $P$
with $0\in\operatorname{int}P$ and every $\rho>1$.
\end{definition}

\begin{conjecture}[Finite-Dictionary Conjecture]\label{conj:sigma}
For every convex polygon $P$ with $0\in\operatorname{int}P$ and every $\rho>1$, $s_W(P,\rho)=\LS(P,\rho)$.
\end{conjecture}
\noindent Conjecture~\ref{conj:sigma} implies FRR on polygons, with $O(n^2)$ states. By Theorem~\ref{thm:eight} it is asymptotically exact as $\rho\to\infty$, and by Theorem~\ref{thm:linear} no dictionary with a uniformly bounded number of \emph{active} sites can work.

\subsection{An asymmetric quadrilateral at every rate}

Consider \(P_*=\conv\{(-3,-2),(2,-2),(4,2),(-1,3)\}\).
Its capacity is constant at small rate ratios, then departs from
the moment bound in two explicit regimes.

\begin{theorem}[All-rate capacity of \(P_*\)]\label{thm:pstar}
Put \(\rho_*=455/118\). For every \(\rho>1\),
\begin{equation}\label{eq:pstar-all}
 s_W(P_*,\rho)=
 \begin{cases}
 11/2,&1<\rho\le\rho_*,\\[1mm]
 \displaystyle\frac{22(103\rho-110)}{5(106\rho-179)},
       &\rho_*\le\rho\le4,\\[2mm]
 \displaystyle\frac{33\rho(52\rho-57)}{468\rho^2-1097\rho+575},
       &4\le\rho.
 \end{cases}
\end{equation}
A design with at most five sites attains this value. For
\(\rho\ge\rho_*\) the sites can be chosen in \(S(P_*,\rho)\).
\end{theorem}

The proof in Appendix~\ref{app:pstar} gives the primal weights
and the monotone certificates over the full rate intervals.
The two breakpoints mark changes in the explicit certificate
construction. In the low-rate regime, a single four-site law
remains feasible by interpolation of its conductances.
The certificate database additionally verifies dictionary attainment
at the particular rates \(3/2\) and \(2\).

\begin{figure}[htbp]
\centering
\includegraphics[width=.55\textwidth]{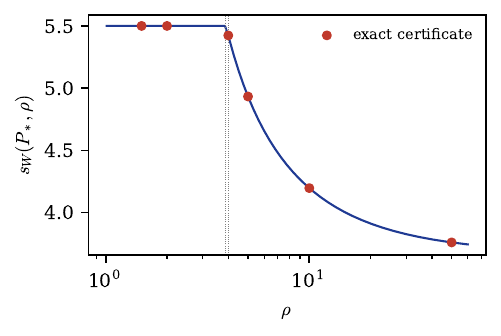}
\caption{Capacity of \(P_*\) from \eqref{eq:pstar-all}, with
exact dictionary certificates at six sampled rate ratios. The formula
holds on the full rate intervals.}
\label{fig:pstar}
\end{figure}

\subsection{Computer-assisted proofs}\label{sec:computer}

\begin{theorem}[Certified polygon--rate pairs]\label{thm:computer}
For each of the \NCERT{} pairs $(P,\rho)$ listed in the ancillary file \texttt{certificates.jsonl} --- rational polygons with $3$ to $7$ sides, rational $\rho\in[\frac32,100]$ --- Conjecture~\ref{conj:sigma} holds, and $s_W(P,\rho)$ is the exact algebraic number listed there.
\end{theorem}
\begin{proof}
For each instance the file contains (i) a design on dictionary sites, with sites, masses and conductances in $\Q(\sqrt{d_1},\dots,\sqrt{d_m})$, possibly with massless relay nodes (which are removed by Kron reduction as in the proof of Theorem~\ref{thm:sparse}, without changing masses, sites or moments), and (ii) $\alpha$, a rigid field $c+\omega Rx$ and boundary hinges $(h_k,d_k,b_k)$. The program \texttt{check\_certificates.py} recomputes $S(P,\rho)$ and verifies exactly: every site belongs to $S(P,\rho)$; masses are positive and sum to one; conductances are non-negative; \eqref{eq:eq} holds at every site and force balance holds at every relay node; $\alpha\in[0,1]$, $b_k\ge0$ and every hinge direction points into $P$. Then $C=\min(A,B)$ of the design is a lower bound for $\LS\le s_W$. By Lemma~\ref{lem:hinge} the field is monotone, and the program checks $F\le C$ on every piece of every polygon edge and hinge chord, cut at all chord intersections and for both one-sided traces, by exact maximization of a quadratic; by Proposition~\ref{prop:pwr} this gives $F\le C$ on $P$, hence $s_W\le C$ by Theorem~\ref{thm:gap}. Thus $s_W=\LS=C$. Zero tests are symbolic; strict signs of non-zero algebraic expressions are proved using rational interval enclosures of radicals, with increasing precision until zero is excluded (the verifier rejects undecided signs).
\end{proof}

The instances are rational triangles through heptagons at rates
between $3/2$ and $100$. Their optimizers have between three and nine
states, with sites and masses in multiquadratic fields in some cases.
Every instance comes with both a feasible design and a certificate that
bounds the objective on the whole polygon.

\section{Consequences for finite kinetic models}\label{sec:inference}

The results above are statements about every reversible chain that
reproduces two prescribed modes. They can therefore be read as
obstructions to proposed finite models.

\paragraph{State budgets for discretized diffusions.}
Theorem~\ref{thm:microstate} gives a lower bound on the number of states of
any reversible model that reproduces the two slowest Wright--Fisher modes
exactly. Theorem~\ref{thm:robustmicro} in Appendix~\ref{app:robust} keeps a
bound of the same order when the eigenfunction equations hold only up to
generator residuals $\varepsilon_u,\varepsilon_t$:
\[
 4(c+2)\delta+4\varepsilon_u+9\varepsilon_t\ \ge\ \frac{c^2}{18(N-1)^2}.
\]
A model that is both accurate in variance and small in residuals must
therefore have many states. These residuals are properties of the
generator, not of fitted correlation curves.

\paragraph{Three-state mechanisms.}
By Theorem~\ref{thm:threeexact}, the conductances of a three-state model
are determined by its observable sites and stationary weights. A proposed
three-state explanation of two observed modes is excluded as soon as
$(\rho-1)v_iv_j>B$ for some pair of states. No search over kinetic
parameters is needed, and the threshold rate is explicit.

\paragraph{Extreme states and exit probabilities.}
For a one-step kernel $K$ and an observable $u$ with a unique minimum at
state $0$, put $g=\min_{j\ne0}(u_j-u_0)$ and $q=1-K_{00}$. Since
$(Ku)_0-u_0=\sum_{j\ne0}K_{0j}(u_j-u_0)\ge gq$,
\begin{equation}\label{eq:introexit}
 |(Ku)_0-du_0|\ \ge\ \bigl[gq-(1-d)(-u_0)\bigr]_+ .
\end{equation}
If $u$ is required to satisfy $Ku=du$, the exit probability from the
extreme state is bounded by $(1-d)(-u_0)/g$. This pointwise ceiling can be
tested directly on trajectories, and it addresses a different question
from Chapman--Kolmogorov prediction~\cite{Chapman1928,Kolmogoroff1931}.
For a reversible stationary kernel and a centred observable $f$ with
normalized autocorrelations $r_j$, self-adjointness gives
\[
 \frac{\|Kf-df\|_p^2}{\operatorname{Var}f}=(r_2-r_1^2)+(d-r_1)^2 ,
\]
so exact agreement of two lags forces an eigenfunction. Approximate
agreement of estimated transition matrices at two lags, which is the
usual validation of Markov state models~\cite{Prinz11,NoeNuske13}, gives
no pointwise residual bound by itself, while \eqref{eq:introexit} does.

\section{Conclusion}\label{sec:discussion}

The main message of the paper is that the cost of realizing a continuum
equilibrium with finitely many states is a separate quantity from the
equilibrium itself. Under parabolic confinement the optimal continuum
model is the Wright--Fisher beta law in every admissible geometry, yet the
number of states needed to approach it depends on details that this law
does not see. Moving from the arc to its convex hull multiplies the
leading constant by $c/(1+c)$; enlarging the hull upward lowers it again
when $\rho>8/3$. In each case the horizontal cut force of the limiting
law fixes the optimal cube-root density of states, and the gain from the
hull comes entirely from vertical displacement near the two ends of the
arc.

On polygons, finite chains are expected to be optimal, but the minimal
state count is not a function of moments. The seven-gon has two optimal
laws with identical first and second moments, and positivity of the
conductances decides which one survives as the rate ratio grows.

Several problems remain open.
\begin{itemize}[leftmargin=*]
\item \emph{The constant above the hull.} Theorem~\ref{thm:raisedcap}
 brackets the leading constant for caps $T>1$ but does not determine it.
\item \emph{Finite realizability.} The invariant-core theorem excludes
 dynamically generated boundary arcs as carriers of diffuse mass. What
 remains is to show that mass on flat portions of the original sides can
 be replaced by finitely many atoms while keeping force feasibility.
 The seven-gon shows that moment matching alone cannot do this.
\item \emph{The dictionary conjecture.} Conjecture~$\Sigma$ would give
 FRR with $O(n^2)$ states and reduce the capacity of every polygon to a
 finite linear program. The linear lower bound of
 Theorem~\ref{thm:linear} leaves a gap between $\Omega(n)$ and $O(n^2)$
 states, and it would be interesting to close it even under
 Conjecture~$\Sigma$.
\item \emph{More modes and higher dimensions.} The invariant-core results
 are stated in $\R^d$, while the sharp asymptotics
 use the planar quartic identity. Whether a similar identity controls
 three or more exact modes is open.
\end{itemize}

\paragraph{Data and reproducibility.}
The sources include the polygon certificates and their exact
verifier, as well as \path{anc/verify_pstar_all_rates.py} for the
all-rate quadrilateral solution.
The script \path{anc/verify_181_heptagon.py} checks the rational
seven-gon of Theorem~\ref{thm:sevenphase} in exact arithmetic. The scripts
\path{anc/verify_arc_asymptotics.py} and
\path{anc/verify_convex_confinement.py} reproduce the sharp
coefficient plot, the graded-chain checks and the raised-cap
calculations. 

\clearpage
\appendix

\section{Optimal faces, sparsity and finite contacts}\label{sec:structure}
%=====================================================================

\begin{theorem}[Sparse skeleton]\label{thm:sparse}
\begin{enumerate}[label=(\alph*)]
\item Every design with $N\ge2$ has a design with the same sites and masses, hence the same $(A,B)$, with at most $2N-3$ positive conductances; for $N=1$ the sole feasible design is at the origin and has no cables.
\item Let $X$ be a finite set, let $\sigma$ be the optimal value of $\alpha A+(1-\alpha)B$ (resp.\ of $\min(A,B)$) over designs with sites in $X$, and assume $X$ is inclusion-minimal among sets carrying a design of value $\sigma>0$; let $N=|X|$. Then some design on $X$ of value $\sigma$ has at most $N+1$ (resp.\ $N+2$) cables. If it is connected, it is a spanning tree plus at most two (resp.\ three) cables.
\end{enumerate}
\end{theorem}
\begin{proof}
(a) The case $N=1$ is immediate. For $N\ge2$, with sites and masses fixed, $\{J\ge0:\eqref{eq:eq}\}$ is a non-empty pointed polyhedron; a vertex has at most $\operatorname{rank}$ positive entries. The rank of the cable matrix is at most $2N-3$, because the $2N$ equations are annihilated by the three infinitesimal rigid motions (total force and torque of the cable forces vanish identically).

(b) For fixed $X$, the constraints \eqref{eq:eq} and $\sum p_i=1$ are $2N+1$ linear equations in $(p,J)\ge0$, and $A,B$ are linear in $p$. The feasible polyhedron is bounded because $\sum_{i<j}J_{ij}|x_i-x_j|^2=A+\rho B$ (Lemma~\ref{lem:id}). An optimal vertex has at most $2N+1$ positive coordinates. For $\min(A,B)$ add a variable $t\ge0$ and constraints $t\le A$, $t\le B$ with slacks: $2N+3$ equations, and $t=\sigma>0$ at the optimum. It remains to see that all $N$ masses are positive at an optimal vertex. If $p_i=0$ and site $i$ carries no cable, the vertex is a design of value $\sigma$ on $X\setminus\{i\}$. If $p_i=0$ and $D_i=\sum_jJ_{ij}>0$, then \eqref{eq:eq} at $i$ gives $x_i=\sum_jJ_{ij}x_j/D_i$, and Kron reduction --- delete $i$ and add $J_{ij}J_{ik}/D_i$ between its neighbours $j,k$ --- preserves \eqref{eq:eq} at all other sites, since $J_{ij}(x_j-x_i)=\sum_k(J_{ij}J_{ik}/D_i)(x_j-x_k)$. Both cases contradict minimality of $X$. A connected graph on $N$ vertices has at least $N-1$ edges.
\end{proof}

\subsection*{Finite-contact dual certificates and effective bounds}
%=====================================================================

The unrestricted finite-attainment question is not settled by fixed-cap
compactness alone.  The next result isolates a sufficient condition which,
unlike the global PSD-stress representation theorem, is proved directly
from actual networks and their mass laws.  It also gives a quantitative
localization bound for nearly optimal designs.

\begin{theorem}[Finite-contact criterion without stress-density assumptions]
\label{thm:finitecontacts}
Let $P\subset\R^2$ be compact and convex, let $\rho>1$, and put
$s=s_W(P,\rho)$.  Suppose a continuous monotone field
$U:P\to\R^2$, $\alpha\in[0,1]$, and $C=s$ satisfy
\[
 F(x):=q_\alpha(x)+\Lambda x\cdot U(x)\le C\quad(x\in P).
\]
If the contact set $Z=\{x\in P:F(x)=C\}$ is finite, then an
optimal finite design exists with mass supported on $Z$ and with
at most $|Z|$ states.  This conclusion does not require
Theorem~\ref{thm:closure}.

More generally, if $C\ge s$ is any valid monotone-field upper bound,
$W$ is a design with $\min(A(W),B(W))\ge s-\varepsilon$, and
$\tau>0$, then
\begin{equation}\label{eq:contactlocalization}
\mu_W\{x\in P:C-F(x)\ge\tau\}
\le\frac{C-s+\varepsilon}{\tau}.
\end{equation}
The total monotone edge work of $W$ is at most $C-s+\varepsilon$.
\end{theorem}
\begin{proof}
The exact gap identity (Theorem~\ref{thm:gap}) gives
\[
 0\le\int_P(C-F)\,d\mu_W
 \le C-[\alpha A(W)+(1-\alpha)B(W)]
 \le C-\min(A(W),B(W)).
\]
The same estimate applies to the nonnegative edge-work term;
Markov's inequality yields \eqref{eq:contactlocalization}.

For finite contacts, take a weak limit of a maximizing sequence.
The gap bound and continuity give support in \(Z\).
The limiting law satisfies the monotone-work inequalities by weak
convergence, as in Theorem~\ref{thm:capatta}. Lemma~\ref{lem:finitework}
realizes it on its support, so the capacity is attained.

\end{proof}

\begin{figure}[htbp]
\centering
\includegraphics[width=.72\textwidth]{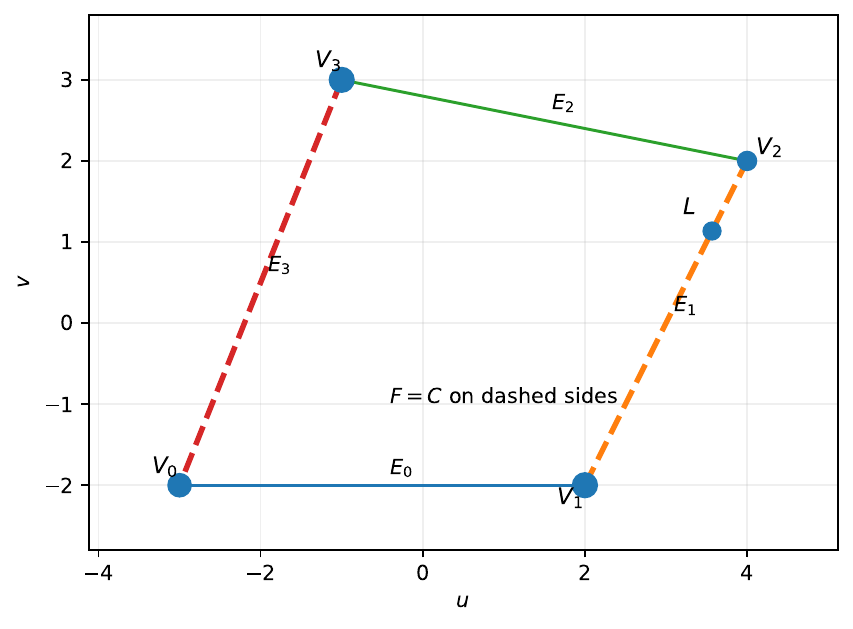}
\caption{The exact rigid calibration for $P_*$ at $\rho=3/2$:
dashed sides $E_1,E_3$ are entirely contained in the contact set
$\{F=s_W\}$, although an optimal network has only five states.
Marker size indicates stationary mass (not tension).  The contact
identities and equilibrium are exact, not merely graphical.}
\label{fig:flat178}
\end{figure}

\begin{theorem}[Effective two-sided approximation of the capacity]
\label{thm:computecap}
Given a convex polygon $P$ with rational vertices, a rational
$\rho>1$, and a rational tolerance $\varepsilon>0$, there is a
terminating algorithm which outputs an algebraic finite design $W$
and a rational polynomial monotone-field certificate with rational
upper bound $C$ such that
\[
 \min(A(W),B(W))\le s_W(P,\rho)\le C,
 \qquad C-\min(A(W),B(W))<\varepsilon.
\]
The termination assertion uses the strong duality and rational
polynomial-approximation conclusions of
Theorem~\ref{thm:strongdual}; it is \emph{not} an algorithm for
deciding exact finite attainment or Conjecture~\ref{conj:sigma}.
No efficiency claim is made.
\end{theorem}
\begin{proof}
For each $N$, the finite-cap value $s_N(P,\rho)$ and an algebraic
maximizing design are computable in principle by the real
quantifier-elimination formulation of
Corollary~\ref{cor:capalg}.  The values $s_N$ increase to $s_W$
by the definition of the unrestricted supremum.  In parallel
enumerate rational $\alpha\in[0,1]$, rational polynomial vector
fields $U$, and rational $C$, accepting them only if exact real
quantifier elimination verifies
$e(U)\succ0$ and $q_\alpha+\Lambda x\cdot U(x)<C$ throughout
$P$.  Every accepted candidate is a valid upper bound by
Theorem~\ref{thm:gap}.  By the rational approximation clause of
Theorem~\ref{thm:strongdual}, accepted bounds approach $s_W$
from above.  Dovetail the two computations and stop once an upper
bound minus an attained $s_N$ is less than $\varepsilon$.
This must happen after finitely many steps.  The output can be
checked by finite algebraic calculations and quantified polynomial
inequalities.  Nothing in the argument ensures that
$s_N=s_W$ for any fixed $N$.
\end{proof}

\begin{proposition}[Cut law and leaves]\label{prop:cut}
If a sub-network $S$ is joined to the rest by a single cable $ab$ with $a\in S$, then $m_S\Lambda\bar x_S=J_{ab}(x_a-x_b)$, where $m_S$ and $\bar x_S$ are the mass and barycentre of $S$. A leaf $x$ attached to $h$ satisfies $J_{xh}(x-h)=p_x\Lambda x$; hence $h=x-s\Lambda x$ with $s>0$ and
\begin{equation}\label{eq:leaf}
K_h(x):=\det(x-h,\Lambda x)=\rho(u-h_u)v-u(v-h_v)=0 .
\end{equation}
\end{proposition}
\begin{proof}
Sum \eqref{eq:eq} over $S$; internal forces cancel in pairs.
\end{proof}

\section{Invariant confinement and boundary exclusion}\label{sec:dyn}
%=====================================================================

The prescribed coordinate eigenmodes constrain not only the stationary
moments, but the entire convex path of conditional expectations.  This
produces a geometric reduction which is exact at \emph{every} state cap,
and a nonlinear inequality on all possible limiting mass laws.  Unlike
the stress-density theorem, these results use only finite reversible
chains, positivity of their semigroups and weak compactness of
probability measures.

It is useful to formulate the statements in $\R^d$, with a positive
diagonal rate matrix $\Lambda=\diag(\lambda_1,\ldots,\lambda_d)$.
A finite design is defined by the same force-balance equations
\eqref{eq:eq}, now in $d$ components.  Let $D_t=e^{-t\Lambda}$.

\begin{theorem}[Invariant confinement core]\label{thm:invcore}
Let $P\subset\R^d$ be compact and convex, with $0\in\operatorname{int}P$,
and define
\begin{equation}\label{eq:invcore}
 K_\Lambda(P)=\{x\in P:D_tx\in P\text{ for all }t\ge0\}
             =\bigcap_{t\ge0}e^{t\Lambda}P.
\end{equation}
Every site of every finite admissible design in $P$ belongs to
$K_\Lambda(P)$.  Conversely, every design with sites in
$K_\Lambda(P)$ is admissible in $P$.  Thus, in the planar setting,
\[
 s_N(P,\rho)=s_N(K_\Lambda(P),\rho)\quad(N\ge1),\qquad
 s_W(P,\rho)=s_W(K_\Lambda(P),\rho).
\]
Moreover $K_\Lambda(P)$ is the maximal forward-invariant subset of
$P$ under $D_t$, and is compact and convex.
\end{theorem}
\begin{proof}
For a finite design, put $Q_{ij}=J_{ij}/p_i$ for $i\ne j$
and $Q_{ii}=-\sum_{j\ne i}Q_{ij}$.  The coordinate eigenfunction
equations imply
\[
 \sum_j(e^{tQ})_{ij}x_j=D_tx_i\qquad(t\ge0).
\]
The matrix $e^{tQ}$ is stochastic; hence the left side is a convex
combination of sites in $P$.  Thus $D_tx_i\in P$ for all $t$,
so every site lies in the stated core.  The converse is inclusion.
The identities for the fixed-cap and unrestricted values follow.
Compactness and convexity follow from the intersection formula;
$D_sK_\Lambda(P)\subset K_\Lambda(P)$ follows from the semigroup
property, proving maximality.
\end{proof}

For a polytope $P=\{x:n_k\cdot x\le c_k,\ 1\le k\le m\}$,
with $c_k>0$, the core has the exact description
\begin{equation}\label{eq:corefacets}
 K_\Lambda(P)=\{x:n_k\cdot D_tx\le c_k
       \text{ for all }k,t\ge0\}.
\end{equation}
In the planar case, writing $n_k=(a_k,b_k)$ and
$\Lambda=\diag(1,\rho)$, this is equivalent to
\[
 \max_{0\le z\le1}\{a_ku z+b_kv z^\rho\}\le c_k
 \qquad(1\le k\le m).
\]
For rational $\rho=p/q>1$, the substitution $z=w^q$ reduces
membership in the core to finitely many univariate polynomial
nonnegativity tests.  Hence the core is semialgebraic for rational
polygonal data and rational $\rho$, although it need not be a polygon.

\begin{lemma}[Semigroup mass barrier]\label{lem:specmass}
Let $P\subset\R^d$ be a compact convex body with
$0\in\operatorname{int}P$, $\Lambda$ positive diagonal,
and $\mu$ any weak limit of stationary measures of finite
reversible designs with these fixed parameters.  For any affine slack
$h(x)=c-n\cdot x\ge0$ on $P$, put $h_t(x)=c-n\cdot D_tx$.
For $t\ge0$, $\delta>0$, and every Borel set
$E\subset\{x:h(x)\ge\delta\}$ one has
\begin{equation}\label{eq:specmass}
 \delta\,\mu(E)^2\ \le\ \int_E h_t(x)\,d\mu(x).
\end{equation}
The same inequality holds for the mass law of each individual design.
More generally, for any continuous $0\le f\le1$ satisfying
$h\ge\delta f$ on $P$,
\begin{equation}\label{eq:smoothspecmass}
\int_P f h_t\,d\mu\ge\delta\left(\int_Pf\,d\mu\right)^2.
\end{equation}
\end{lemma}
\begin{proof}
On a finite chain, write $T_t=e^{tQ}$.  Since
$Qh=-\sum_j\lambda_jn_jx_j$ on the node coordinates,
the exact coordinate modes give $T_th=h_t$.
If $h\ge\delta f\ge0$, positivity of the Markov semigroup gives
$T_th\ge\delta T_tf$.  By stationarity and reversibility,
\[
 \int f h_t\,dp=\langle f,T_th\rangle_p
 \ge\delta\langle f,T_tf\rangle_p
 \ge\delta\big(\textstyle\int f\,dp\big)^2.
\]
Indeed, $Q$ is self-adjoint and nonpositive on $L^2(p)$, whence
\[
\langle f,T_tf\rangle_p=
\left(\int f\,dp\right)^2+
\left\|T_{t/2}\left(f-\textstyle\int f\,dp\right)\right\|_{L^2(p)}^2.
\]
The continuous-test inequality passes to any weak mass limit.
For a Borel $E\subset\{h\ge\delta\}$, fix $\delta'<\delta$.
By regularity of the limiting measure, approximate $\mathbf1_E$
in $L^1(\mu)$ by continuous functions $0\le f_\ell\le1$
with support in the open set $\{h>\delta'\}$.  On this support
$h\ge\delta'f_\ell$, and the continuous-test inequality
applies.  Let $\ell\to\infty$, then $\delta'\uparrow\delta$.
For a finite chain one may take $f=\mathbf1_E$ directly.
\end{proof}

\begin{theorem}[No limiting mass on dynamically generated boundaries]
\label{thm:barrier}
Let $P\subset\R^d$ be a compact convex polytope with
$0\in\operatorname{int}P$, and let $\Lambda$ be positive diagonal.
For the invariant core $K=K_\Lambda(P)$, any weak limit $\mu$ of
stationary laws of finite reversible designs in $P$ satisfies
\begin{equation}\label{eq:nodynmass}
 \boxed{\ \mu(\partial K\setminus\partial P)=0.\ }
\end{equation}
No optimality hypothesis, uniform state cap or stress-density
representation is needed.
\end{theorem}
\begin{proof}
Choose one facet slack $h=c-n\cdot x$, $\delta>0$ and
$0<a<b<\infty$.  Let
\[
 D=\{x\in K:h(x)\ge\delta,\ \text{some }\tau\in[a,b]
                       \text{ satisfies }h_\tau(x)=0\}.
\]
All sets below are Borel, as they are projections of compact subsets
of $K\times[a,b]$.  Since $h_t(x)\ge0$ for all $t\ge0$ and
$x\in K$, every contact at $\tau>0$ obeys
$\partial_t h_t(x)|_{t=\tau}=0$.  The second derivative has a
uniform finite bound
\[
 |\partial_{tt}h_t(x)|\le M:=
 \max_{x\in P}\sum_{j=1}^d|n_j|\lambda_j^2|x_j|.
\]
Partition $[a,b]$ into at most $(b-a)/r+1$ intervals of
length at most $r$, with midpoints $t_\ell$.  Let $E_\ell$
consist of those $x\in D$ having a contact within the
$\ell$th interval.  Taylor's theorem gives
$0\le h_{t_\ell}(x)\le Mr^2/8$ on $E_\ell$.
By Lemma~\ref{lem:specmass},
\[
 \delta\mu(E_\ell)^2\le
    \int_{E_\ell}h_{t_\ell}\,d\mu
    \le (Mr^2/8)\mu(E_\ell),
 \qquad \mu(E_\ell)\le Mr^2/(8\delta).
\]
The covering therefore yields
$\mu(D)\le M((b-a)r+r^2)/(8\delta)\to0$ as $r\downarrow0$.

If $x\in\partial K\setminus\partial P$, then $x\in\operatorname{int}P$.
Some facet inequality must be active along its contracted orbit
at a finite $\tau>0$: otherwise $x$ has a uniform positive margin
in all the finitely many inequalities of \eqref{eq:corefacets}.
To see the uniformity for large times, use $D_tx\to0$ uniformly
for $x\in P$ and $c_k>0$; times in a bounded interval form a
compact set.  Such an $x$ belongs to one of countably many sets
$D$ obtained by taking $\delta=1/j$, $a=1/q$, $b=q$ for positive
integers $j,q$ and one of the finitely many facets.
Each $D$ has zero mass, proving \eqref{eq:nodynmass}.
\end{proof}

\begin{figure}[t]
\centering
\includegraphics[width=.73\textwidth]{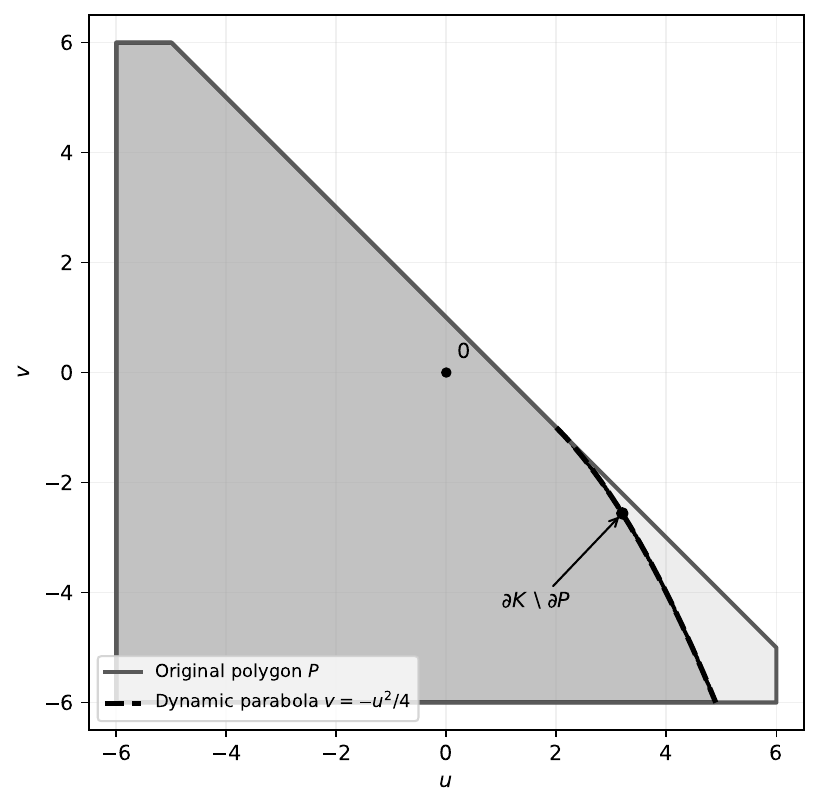}
\caption{Dynamically invariant core for $\rho=2$ inside the polygon
$|u|,|v|\le6$, $u+v\le1$.  The new curved boundary
contains the parabolic arc $v=-u^2/4$ ($2<u<\sqrt{24}$).
The exclusion theorem shows that no weak limit of admissible
stationary laws can give this dynamically created arc positive mass.
The plot is an illustration; all claims are proved analytically.}
\label{fig:invcore}
\end{figure}

\section{Attainment regimes and the hexagon phase diagram}\label{sec:regimes}
%=====================================================================

\begin{theorem}[Isotropic window: six states suffice]\label{thm:iso}
Let $V=\max_P|v|$ and $M_u=(\max_Pu)(-\min_Pu)$. If
\[
1<\rho\le\rho_{\rm iso}(P):=\max\Bigl\{1+\frac{R(P)}{V^2},\ \frac{M_u}{(M_u-R(P))_+}\Bigr\}
\]
(with $M_u/0=\infty$), then $s_W(P,\rho)=R(P)$, attained by a design on at most six sites. In particular finite attainment holds near $\rho=1$ on every convex body.
\end{theorem}
\begin{proof}
An optimal measure for $R(P)$ exists by compactness. Its moment vector $m(\mu)\in\R^5$ (the vector of $\E u,\E v,\E uv,\E u^2,\E v^2$) lies in the convex hull of the compact curve $x\mapsto(u,v,uv,u^2,v^2)$, so by Carath\'eodory it is matched by a measure on at most six points of $P$; this measure is optimal too, with $A,B\ge R(P)$. On its sites define either
\[
J_{ij}=p_ip_j\Bigl[1-(\rho-1)\frac{v_iv_j}{B}\Bigr]\qquad\text{or}\qquad J_{ij}=p_ip_j\Bigl[\rho+(\rho-1)\frac{u_iu_j}{A}\Bigr].
\]
For the first, $\sum_jJ_{ij}(x_i-x_j)=p_ix_i-(\rho-1)\frac{p_iv_i}{B}\bigl(\E[v]x_i-\E[vx]\bigr)=p_ix_i+(\rho-1)p_i(0,v_i)=p_i\Lambda x_i$, using $\E v=\E uv=0$. For the second, $\sum_jJ_{ij}(x_i-x_j)=\rho p_ix_i-(\rho-1)p_iu_i(1,0)=p_i\Lambda x_i$. The first is non-negative when $(\rho-1)v_iv_j\le B$, which holds if $(\rho-1)V^2\le R(P)\le B$; the second when $(\rho-1)(-u_iu_j)\le\rho A$, which holds if $(\rho-1)M_u\le\rho R(P)$. Thus a design with $\min(A,B)=R(P)\ge s_W$ exists.
\end{proof}

\begin{theorem}[Six-state attainment for reflection-symmetric bodies]\label{thm:sym}
If $P$ is symmetric under $(u,v)\mapsto(u,-v)$, then $s_W(P,\rho)=R(P)$ for \emph{every} $\rho>1$, and the optimum is attained by a reversible design on at most \emph{six} states. In particular, the capacity does not depend on $\rho$ for every reflection-symmetric triangle, rhombus, kite or other convex body with this symmetry.
\end{theorem}
\begin{proof}
Let $I=[a,b]$ be the horizontal projection of $P$, with $a<0<b$, and write its vertical sections as $[-h(u),h(u)]$, where $h\ge0$ is continuous. For every probability measure on $P$ with $\E u=\E v=\E uv=0$, its horizontal marginal $\nu$ satisfies $\E v^2\le\int_I h(u)^2\,d\nu(u)$ and $\E u^2=\int_Iu^2\,d\nu(u)$. Conversely, given any $\nu$ with $\int_Iu\,d\nu=0$, placing equal conditional masses at $(u,\pm h(u))$ gives a symmetric measure on $P$ with the same $u^2$ moment and vertical moment $\int h^2\,d\nu$. Hence
\[
R(P)=\max_{\nu\in\mathcal P(I),\,\int u\,d\nu=0}
\min\left\{\int u^2\,d\nu,\int h(u)^2\,d\nu\right\}=:S.
\]
By compact convex minimax and linear-programming duality for the single centering constraint, there are $\alpha_*\in[0,1]$ and $\lambda_*\in\R$ such that
\[
\alpha_*u^2+(1-\alpha_*)h(u)^2+\lambda_*u\le S\qquad(u\in I).
\]
Indeed, for each fixed $\alpha$ the minimum over $\lambda$ of the maximum of the left side is the maximum over centred $\nu$ of the corresponding weighted moment; the minimum in $\lambda$ is attained since $a<0<b$. Choose an optimal centred measure $\nu_*$. Its objective is $S$, and integration of the last inequality forces equality $\nu_*$-almost everywhere. Thus the vectors $(u,u^2,h(u)^2)$ in its support lie on the affine plane
$\lambda_*u+\alpha_*u^2+(1-\alpha_*)h(u)^2=S$.
The three moments of $\nu_*$ therefore belong to the convex hull of a compact subset of an affine plane of dimension at most two. By Carath\'eodory, a centred measure on at most three such abscissae has the same two second moments and is optimal. Split the mass at each abscissa equally between $(u,h(u))$ and $(u,-h(u))$ (coincident sites are merged). The resulting law has at most six sites, $\E u=0$ and $\E[v\mid u]=0$, and achieves $R(P)$. Lemma~\ref{lem:vertical} supplies nonnegative symmetric conductances at every $\rho>1$. Since $s_W\le R(P)$, equality follows.
\end{proof}

For strong anisotropy, vertical sections take over. Let $I=\{u:(u,0)\in P\}=[a,b]$ and write the vertical section of $P$ at abscissa $u$ as $[l(u),h(u)]$; put $D(u)=-l(u)h(u)$.

\begin{lemma}[Vertical-section designs]\label{lem:vertical}
If $\E u=0$ and $\E[v\mid u]=0$, then $J_{ij}=p_ip_j\bigl[1+(\rho-1)\mathbf 1\{u_i=u_j\}/m(u_i)\bigr]$, where $m(u)$ is the mass at abscissa $u$, is a design for every $\rho>1$.
\end{lemma}
\begin{proof}
The reset part gives $p_ix_i$; the block part gives $(\rho-1)p_i(x_i-\E[x\mid u_i])=(\rho-1)p_i(0,v_i)$.
\end{proof}

For a polygon let $Z$ consist of $a,0,b$ and the abscissae in $I$ of the vertices of $P$, and
\begin{equation}\label{eq:Rinf}
R_\infty=\max\Bigl\{\min\Bigl(\sum_zw_zz^2,\sum_zw_zD(z)\Bigr):w\ge0,\ \sum_zw_z=1,\ \sum_zzw_z=0,\ z\in Z\Bigr\}.
\end{equation}

\begin{theorem}[Exact polygonal vertical-section compression]\label{thm:knotexact}
For every compact convex polygon $P$ with $0\in\operatorname{int}P$, the full interval problem
\[
R_I=\max_{\nu\in\mathcal P(I),\,\int u\,d\nu=0}
\min\left\{\int u^2\,d\nu,\;\int D(u)\,d\nu\right\}
\]
is \emph{exactly} the finite program \eqref{eq:Rinf}: $R_I=R_\infty$.
More strongly, every centred probability measure $\nu$ on $I$ admits
another centred probability measure $\nu_Z$ supported on $Z$ with
\[
\int u^2\,d\nu_Z=\int u^2\,d\nu,
\qquad \int D(u)\,d\nu_Z\ge\int D(u)\,d\nu.
\]
The maximum $R_\infty$ is attained by a reversible design on at most
six states of $P$ at \emph{every} $\rho>1$ (not necessarily on the
finite dictionary $S(P,\rho)$). In particular,
$R_\infty\le s_6(P,\rho)\le s_W(P,\rho)$ for every $\rho>1$.
\end{theorem}
\begin{proof}
On $I$ write $f=-l\ge0$ and $g=h\ge0$.
The functions $f,g$ are concave and piecewise affine, and $D=fg$.
Between consecutive elements of $Z$ both are affine.
An affine continuation $\widetilde f,\widetilde g$ from any such
interval is a supporting line of the corresponding concave function:
$\widetilde f\ge f\ge0$ and $\widetilde g\ge g\ge0$ on $I$.
Consequently, their product $q=\widetilde f\widetilde g$ is a
quadratic polynomial satisfying $q\ge D$ throughout $I$, with
$q=D$ on the chosen interval.

For $x<0<y$ denote the centred two-point law by
\[
\eta_{x,y}=\frac{y}{y-x}\delta_x+
\frac{-x}{y-x}\delta_y.
\]
Its two moments are
\begin{equation}\label{eq:pair-moments}
A(x,y)=-xy,\qquad
B(x,y)=\frac{yD(x)-xD(y)}{y-x}.
\end{equation}
Fix $y>0$ and let $x$ vary in a negative knot interval on which
$D(x)=q(x)=a_2x^2+a_1x+a_0$.
Set $E=q(y)-D(y)\ge0$.  Direct division in
\eqref{eq:pair-moments} yields
\[
B(x,y)=a_0-a_2xy+\frac{xE}{y-x},\qquad
\frac{\partial^2B}{\partial x^2}
=\frac{2yE}{(y-x)^3}\ge0.
\]
Similarly, fixing $x<0$ and letting $y$ vary in a positive knot
interval with $D(y)=q(y)=a_2y^2+a_1y+a_0$, set
$E=q(x)-D(x)\ge0$. Then
\[
B(x,y)=a_0-a_2xy-\frac{yE}{y-x},\qquad
\frac{\partial^2B}{\partial y^2}
=\frac{-2xE}{(y-x)^3}\ge0.
\]
Both formulas extend continuously when one of the endpoints is zero,
in which case $\eta_{0,y}=\eta_{x,0}=\delta_0$.
Since $A(x,y)=-xy$ is affine in each coordinate separately, splitting
$x$ between the two endpoints of its knot interval preserves $A$
and does not decrease $B$ by convexity.  Next split $y$ between
its own two knot endpoints, with the same properties.  Each resulting
measure is centred; the final pairs have both abscissae in $Z$.

Every centred probability $\nu$ is a mixture of these pair laws and
possibly a point mass at zero.  Indeed, if
$m=\int_{x<0}(-x)\,d\nu=\int_{y>0}y\,d\nu>0$, couple the finite measures
$(-x)\nu|_{(-\infty,0)}$ and $y\nu|_{(0,\infty)}$ by their
normalised product $\gamma$ (each marginal is the stated finite
measure), and set
\[
d\omega(x,y)=\frac{y-x}{(-x)y}\,d\gamma(x,y).
\]
Then $\nu=\nu(\{0\})\delta_0+\int\eta_{x,y}\,d\omega$.
The case $m=0$ is just $\nu=\delta_0$.
Apply the two endpoint-splitting operations to this mixture.
Linearity of the moments produces $\nu_Z$, preserving $A$ and not
reducing $B$. Thus $R_I\le R_\infty$, while $R_I\ge R_\infty$
is immediate.

The finite LP \eqref{eq:Rinf} has an optimal basic solution supported
on at most three abscissae: it has two equalities (mass and centring)
and at most two active moment inequalities for the maximised variable.
For each such abscissa $z$, place the mass $w_z$ at the two endpoints
$(z,l(z))$, $(z,h(z))$, weighted to have conditional mean $v=0$
(and merge them if both endpoints coincide).
This gives at most six sites, $\E u=0$, $\E[v\mid u]=0$, and
$A=\sum w_zz^2$, $B=\sum w_zD(z)$.
Lemma~\ref{lem:vertical} provides nonnegative conductances for
any $\rho>1$, completing the proof.
\end{proof}

\begin{theorem}[Affine vertical midline: exact attainment at every rate]\label{thm:affinemidline}
Let $P\subset\R^2$ be any compact convex body with $0\in\operatorname{int}P$. Suppose its full horizontal projection $H=[a,b]$ is also the interval of abscissae of its intersections with $v=0$, so that $l(u)\le0\le h(u)$ for all $u\in H$, and suppose the midpoint function
$k(u):=l(u)+h(u)=k_0+k_1u$ is affine. Then for every $\rho>1$,
\[
\boxed{s_W(P,\rho)=\max_{\nu\in\mathcal P(H),\ \E_\nu u=0}
\min\{\E_\nu u^2,\E_\nu[-l(u)h(u)]\},\qquad N_{\rm opt}(P,\rho)\le6.}
\]
In particular, the capacity is independent of the relaxation-rate ratio even when $P$ is not symmetric under $v\mapsto-v$.
\end{theorem}
\begin{proof}
For any $l(u)\le v\le h(u)$, expansion of $(v-l(u))(h(u)-v)\ge0$ gives
\begin{equation}\label{eq:midlinebound}
v^2\le D(u)+k(u)v,\qquad D(u)=-l(u)h(u).
\end{equation}
Every design has $\E v=\E uv=0$ by Lemma~\ref{lem:id}. Since $k$ is affine, (\ref{eq:midlinebound}) implies $B\le\E D(u)$; the horizontal marginal is centred and is supported on $H$. This proves the upper bound.

For the converse, the centred measure problem on the right has an optimum by compactness. Minimax and duality for its single centring constraint furnish $\alpha\in[0,1]$ and $\lambda\in\R$ with
$\alpha u^2+(1-\alpha)D(u)+\lambda u\le R$ on $H$, where $R$ denotes its maximum. An optimal law is supported on the contact set; the vectors $(u,u^2,D(u))$ of its support lie on a common affine plane. Carath\'eodory's theorem yields an optimal centred law on at most three abscissae preserving the two moments. Split each of its atoms between the two ends of the vertical section with conditional mean $v=0$. There are at most six sites, and Lemma~\ref{lem:vertical} provides a genuine reversible design on them for every $\rho>1$, attaining $A,B$ of the moment optimum.
\end{proof}

\begin{theorem}[Complete three-regime state-count phase diagram on a hexagon]
\label{thm:hexphase}
Fix $0<b<1$, let $q=1-b^2$, $t=(1+q)^{-1}$, and let
\[
 P_b=\{(u,v):|u|\le1,\ |v|\le1-(1-b)|u|\},
 \qquad b_*=(\sqrt5-1)/2.
\]
For every $\rho>1$ the capacity is $s_W(P_b,\rho)=t$.
The \emph{minimum} number of positive-mass states of an optimal design has the exact phase diagram
\[
N_{\rm opt}(P_b,\rho)=
\begin{cases}
4,&0<b<b_*,\quad 1<\rho\le 1+b^{-1},\\
3,&b=b_*,\quad 1<\rho\le 1+b^{-1},\\
5,&b_*<b<1,\quad 1<\rho\le 1+q^{-1},\\
6,&\text{otherwise}.
\end{cases}
\]
Thus fixed hexagonal confinement can force jumps of $4\to6$, $3\to6$, or $5\to6$, without any change in the optimal capacity. In particular, the general six-state bound for reflection-symmetric bodies is sharp.
\end{theorem}
\begin{proof}
\emph{Capacity and support.} Put $a=1-b$. For $r=|u|\le1$,
\[
 v^2\le(1-ar)^2\le1-(1-b^2)u^2=1-qu^2,
\quad
 [1-qu^2]-(1-ar)^2=2ar(1-r).
\]
Hence $\min(A,B)\le t$. Equality forces $A=B=t$ and all massive sites to belong to the six vertices $(\pm1,\pm b),(0,\pm1)$.
The equations $\E u=\E v=\E uv=0$ hold for every design. Since $A=t$, centring fixes the total mass in each outer vertical section to $t/2$ and the central mass to $qt$. Write $m_\pm\in[-1,1]$ for the conditional mean of the sign of $v$ at $u=\pm1$. The equation $\E uv=0$ gives $m_+=m_-=m$. The outer contribution to $\E v$ is $btm$.
If one outer section uses a single sign, $m=\pm1$, so \emph{both} outer sections use that same single sign. If the central section uses only the sign $\epsilon\in\{\pm1\}$, the equation $\E v=0$ becomes $bm+q\epsilon=0$. These elementary sign constraints classify every optimal support:
\begin{itemize}
\item If $q>b$, the central section needs both signs; the outer sections are either both singleton or both double. Thus the only possible state counts are four and six.
\item If $q=b$, a three-state design has outer signs both opposite the central singleton. Every design with more than three massive vertices must use all six vertices (positive masses on the other signs force $|m|<1$ and both central signs).
\item If $q<b$, both outer sections require both signs, since $|bt|>qt$ rules out $|m|=1$. The centre uses one or two signs: only five or six states are possible.
\end{itemize}

\emph{Two necessary rate bounds.} If the vertex $x=(1,\epsilon b)$ is the only massive point at $u=1$, then for any other massive vertex $x_j$, $|\epsilon b-v_j|\le(1+b)(1-u_j)$. Its horizontal and vertical force balances imply
\[
\rho b\pi_x\le(1+b)\sum_jJ_{xj}(1-u_j)=(1+b)\pi_x,
\qquad \rho\le1+b^{-1}.
\]
The same holds on the section $u=-1$.
Now suppose $q<b$ and an optimal five-state design has only the central point $C=(0,-1)$ (the other sign is symmetric). The mass budgets and $\E v=0$ give the total masses at the outer signs
\[
P_+=\frac t2\left(1+\frac qb\right),\qquad
P_-=\frac t2\left(1-\frac qb\right).
\]
At every outer vertex, horizontal equilibrium implies $J_{iC}\le\pi_i$, since the central edge changes the abscissa by one and every other horizontal contribution has the correct sign. The vertical force equation at $C$ therefore yields
\[
\rho qt=\sum_{i\ne C}J_{iC}(1+v_i)
\le(1+b)P_++(1-b)P_-=t(1+q).
\]
Thus every optimal five-state design satisfies $\rho\le1+q^{-1}$.

\emph{Four and three states when $q\ge b$.} For $1<\rho\le1+b^{-1}$ use
\[
L=(-1,b),\quad R=(1,b),\quad D=(0,-1),\quad U=(0,1),
\]
with
\[
\pi_L=\pi_R=t/2,\quad \pi_D=t(q+b)/2,\quad \pi_U=t(q-b)/2
\]
and nonzero conductances
\[
 J_{LR}=\frac{t(1+b-\rho b)}{4(1+b)},\quad
 J_{LD}=J_{RD}=\frac{t\rho b}{2(1+b)},\quad
 J_{DU}=\frac{\rho t(q-b)}4.
\]
All balances and $A=B=t$ follow by substitution. For $q>b$, all four masses are positive. When $q=b$ the site $U$ has zero mass and all incident conductances vanish; delete it to obtain an optimal three-state design.

\emph{Five states when $q<b$.} Set $\tau=1+q^{-1}$ and $r=q/b\in(0,1)$. At $\rho=\tau$ use
\[
 L_+=(-1,b),\ L_-=(-1,-b),\ R_+=(1,b),\ R_-=(1,-b),\ C=(0,-1),
\]
with masses
\[
\pi_{L_+}=\pi_{R_+}=p_+:=\frac{t(b+q)}{4b},\quad
\pi_{L_-}=\pi_{R_-}=p_-:=\frac{t(b-q)}{4b},\quad
\pi_C=qt.
\]
Join each outer node to $C$ with conductance equal to its own mass:
$J_{L_+C}=J_{R_+C}=p_+$, $J_{L_-C}=J_{R_-C}=p_-$. Add the two vertical edges
\[
 J_{L_+L_-}=J_{R_+R_-}
 =a_0:=\frac{t(b^2-q^2)}{8b^2q}>0.
\]
For each outer node horizontal equilibrium is immediate. At a positive-$v$ outer node the vertical force is $(1+b)p_++2ba_0=\tau b p_+$; at a negative-$v$ outer node it is $(1-b)p_--2ba_0=-\tau b p_-$. At $C$ the vertical force is
$-2(1+b)p_+-2(1-b)p_-=-\tau qt$ and the horizontal forces cancel.
Thus the network realizes rates $(1,\tau)$ and $A=B=t$.
For $1<\rho<\tau$, let $J^0_{ij}=\pi_i\pi_j$ on these five sites and set
\[
 J_{ij}(\rho)=\theta J_{ij}(\tau)+(1-\theta)J^0_{ij},
 \qquad \theta=\frac{\rho-1}{\tau-1}\in(0,1).
\]
Since the law is centred, the reset network $J^0$ gives relaxation matrix $I$, so interpolation gives exactly $\operatorname{diag}(1,\rho)$ and preserves positive masses and conductances.

\emph{Six states for all rates.} Put $\pi_{(\pm1,\pm b)}=t/4$ and $\pi_{(0,\pm1)}=qt/2$. Every vertical block is centred. For distinct sites define
\[
 J_{ij}=\pi_i\pi_j\left[1+(\rho-1)
 \frac{\mathbf1\{u_i=u_j\}}{\sum_{k:u_k=u_i}\pi_k}\right]>0.
\]
The complete-reset force is $\pi_i x_i$; the within-block term is $(\rho-1)\pi_i(0,v_i)$. Therefore $A=B=t$ for every $\rho>1$. The preceding support classification and rate bounds show that no smaller optimal network exists beyond the corresponding thresholds.\qedhere
\end{proof}

\subsection*{A two-parameter geometry--spectrum phase portrait}
Theorem~\ref{thm:hexphase} can also be read with the \emph{rate ratio fixed}
and the shape of the region varied. This reveals an effect invisible in the
single-parameter plots: the smallest optimal state space need not grow
monotonically as the hexagon changes shape. Set
\[
 b_*:=\frac{\sqrt5-1}{2},\qquad
 \rho_c:=1+\frac1{b_*}=\frac{3+\sqrt5}{2}.
\]
The number $b_*$ is the positive solution of $b^2+b=1$, i.e. the
point at which the outer and central mass budgets coincide.

\begin{figure}[htbp]
\centering
\includegraphics[width=.88\textwidth]{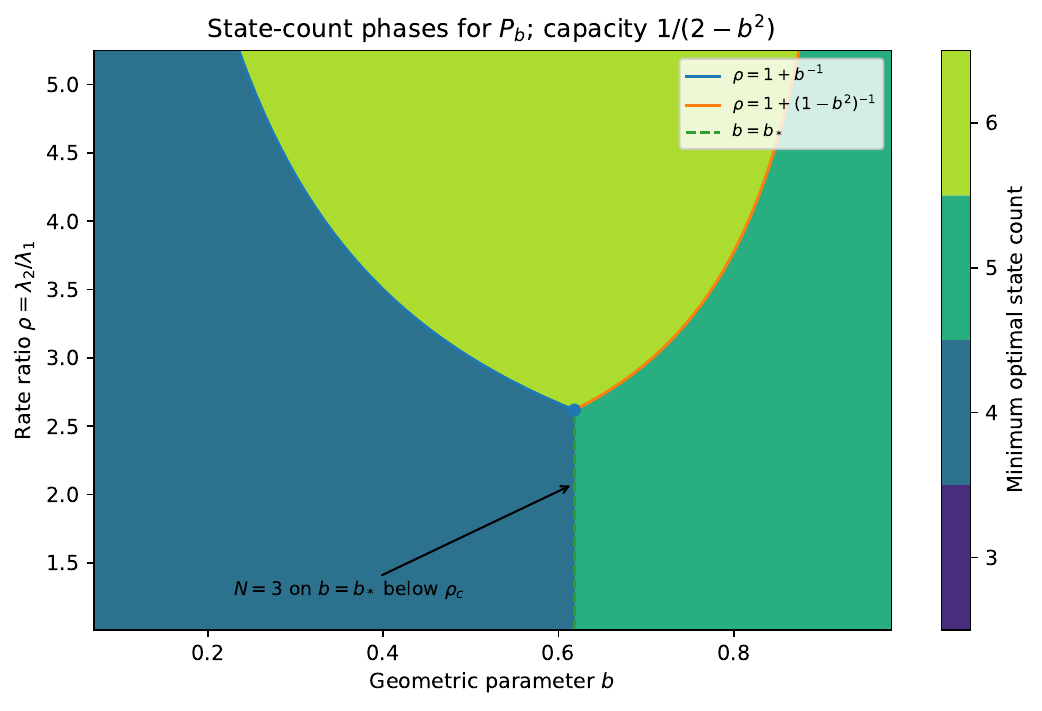}
\caption{Complete geometry--spectrum phase portrait of the family
$P_b$. The colour scale gives the minimum number of positive-mass
states required for an optimal reversible chain. The three-state
regime lies only on the line $b=b_*$ below $\rho_c$, and thus has zero
area in the $(b,\rho)$ plane. The curved boundaries are analytic and
meet at $(b_*,\rho_c)$.}
\label{fig:hex-phase}
\end{figure}

\paragraph{Geometric resonance and its limitations.}
The critical three-state design is not a generic feature of a
three-dimensional parameter region. Its existence requires the exact
balance $q=b$, which is a single algebraic condition on the shape.
For $1<\rho\le\rho_c$, an arbitrarily small perturbation of $b_*$
changes the optimal count to four on one side or five on the other.
The same support cannot therefore represent the optimal model robustly
under arbitrary geometric perturbations. These are \emph{exact}
state-count effects: they do not imply a comparable discontinuity
for approximately optimal designs.

The dictionary $S(P,\rho)$ is defined in Section~\ref{sec:polygons}; let $L_8$ be the supremum over designs on at most eight dictionary sites.

\begin{theorem}[Eight dictionary states at strong anisotropy]\label{thm:eight}
For every convex polygon $P$ with $0\in\operatorname{int}P$ there are $C_P,\rho_P$ such that for $\rho\ge\rho_P$
\[
R_\infty-\frac{C_P}{\rho}\le L_8(P,\rho)\le\LS(P,\rho)\le s_W(P,\rho)\le R_\infty+C_P\frac{1+\log\rho}{\rho}.
\]
\end{theorem}
\noindent The proof is in Appendix~\ref{app:eight}. The lower bound lifts an optimal solution of \eqref{eq:Rinf}, which uses at most three abscissae, onto dictionary points by continuing roots of the leaf equation \eqref{eq:leaf}; the upper bound uses the decay of conditional means $e^{-t\Lambda}x$ over the time $t=(\log\rho)/\rho$.

\section{Polygon certificates and the dictionary reduction}\label{sec:hinges}
On a finite polygonal partition of $P$ let $U_K(x)=c_K+\omega_KRx$ on each open cell $K$, with $R(u,v)=(-v,u)$, and $F_K=q_\alpha+\Lambda x\cdot U_K$.

\begin{proposition}[Exact checkability]\label{prop:pwr}
$F_K\le C$ on every cell iff it holds on all cell edges, and on each edge it is a single one-variable quadratic inequality.
\end{proposition}
\begin{proof}
$\Lambda x\cdot c_K$ is linear and $\Lambda x\cdot Rx=(\rho-1)uv$ is harmonic, so $\Delta F_K=\Delta q_\alpha=2>0$: $F_K$ is subharmonic and attains its maximum on the boundary of the cell, where it is quadratic in the arc parameter.
\end{proof}

\begin{lemma}[Boundary hinges are monotone]\label{lem:hinge}
Let $h\in\partial P$ and let $d$ point from $h$ into $P$ (i.e.\ $n_E\cdot d\le0$ for every side $E$ through $h$). Then for $b\ge0$ the field $U(x)=bR(x-h)$ for $\det(d,x-h)>0$ and $U(x)=0$ otherwise is monotone on $P$, and $\Lambda x\cdot U(x)=b\,K_h(x)$ on its support.
\end{lemma}
\begin{proof}
For $x,y$ on the same side, $(U(x)-U(y))\cdot(x-y)$ is $bR(x-y)\cdot(x-y)=0$ or $0$. For $x$ on the positive side and $y$ on the other, $bR(x-h)\cdot(x-y)=-bR(x-h)\cdot(y-h)=b\det(y-h,x-h)$. All directions from $h$ into $P$ lie in a cone of opening at most $\pi$ containing $d$; $y-h$ lies clockwise and $x-h$ counter-clockwise of $d$ in it, so $\det(y-h,x-h)\ge0$. Finally $\Lambda x\cdot R(x-h)=\det(x-h,\Lambda x)$.
\end{proof}

Sums of a rigid field $c+\omega Rx$ (monotone, with zero work) and boundary hinges are therefore monotone, and Theorem~\ref{thm:gap} applies with \emph{every} trace on cell interfaces.

\subsection{Conditional proof and the structure of optimal designs}

For a piecewise-rigid certificate let
\[
 Z^+=\bigcup_K\{x\in\overline K:F_K(x)=C\}
\]
be the \emph{upper-trace contact set}; it uses all closed-cell traces, not just a particular convention on the interfaces. An exact certificate has $C=s_W(P,\rho)$. A \emph{positive boundary-hinge certificate} has the special form
\[
 U(x)=c+\omega Rx+\sum_{j=1}^m b_j H_{h_j,d_j}(x),\qquad b_j>0,
 \quad H_{h,d}(x)=R(x-h)\,\mathbf1_{\{\det(d,x-h)>0\}},
\]
where each $h_j\in\partial P$, each $d_j$ points into $P$, and all incident traces satisfy $F_K\le C$. A hinge whose cut is entirely on $\partial P$ is absorbed into the rigid field on the interior.

\begin{theorem}[Boundary concentration with an explicit contact condition]\label{thm:boundary}
For an exact bounded piecewise-rigid monotone certificate, every weak limit of mass laws of maximizing designs is supported on $Z^+$. If $Z^+\subset\partial P$ and no boundary segment of positive length has an identically saturated incident quadratic $F_K\equiv C$, then $Z^+$ is finite and a finite optimal design exists. The conclusion is independent of any density theorem for abstract PSD stresses.
\end{theorem}

\begin{lemma}[Zero-work convex-hull locking for positive boundary hinges]\label{lem:locking}
Let $U$ be a positive boundary-hinge certificate and let a positive-mass optimal site $x$ lie in the relative interior of a side $E$ of $P$. Assume that every positive cable from $x$ ends outside $E$. If $x$ has two non-collinear neighbours, then there is a single rigid field $U_0=c_0+\omega_0 Rx$ such that, for the convex hull $T_x$ of $x$ and all its neighbours,
\[
 U_0(z)\text{ is an incident rigid trace of }U\text{ on }T_x,
 \qquad F_0(z):=q_\alpha(z)+\Lambda z\cdot U_0(z)\le C\quad(z\in T_x).
\]
Every one of these sites satisfies $F_0=C$. In particular the unproved convex-hull step for \emph{arbitrary} piecewise-rigid monotone fields is valid for this explicitly restricted positive-hinge class.
\end{lemma}

\begin{theorem}[Conditional finite dictionary for positive hinges]\label{thm:conditional}
Suppose there exists an exact positive boundary-hinge certificate $U$ satisfying:
\begin{enumerate}
\item $Z^+\subset\partial P$ (no interior upper-trace contacts);
\item no positive-length boundary segment is flat at level $C$ for an incident trace;
\item for every two distinct points $x,y\in Z^+$ on a common side, not both vertices, \emph{some} pair of incident traces has $(V_x-V_y)\cdot(x-y)>0$.
\end{enumerate}
Then $s_W(P,\rho)=\LS(P,\rho)$, hence the finite-dictionary conjecture holds for this instance. Every non-vertex site in any optimal finite design is either a coordinate-axis intersection or a leaf attached to a vertex; if the leaf equation vanishes identically on its side, that leaf can be replaced by the endpoints of the side without lowering either objective value. The assertion is a sufficient certificate criterion, \emph{not} a proof of the conjecture for all polygons.
\end{theorem}
\noindent Full proofs, including the exceptional hinge centred at a contact and the hyperbola-level geometry, are in Appendix~\ref{app:boundary}. The restriction to positive boundary hinges and the explicit interior-contact test are essential for what is established here; the corresponding statement for arbitrary monotone piecewise-rigid fields remains unproved.

\subsection{Density, continuity and an arithmetic reduction}

\begin{theorem}[Positive rational networks are dense]\label{thm:dense}
For every design in a compact convex body with $0$ in its interior, its moment pair is a limit of moment pairs of designs at the same $\rho$ on rational interior sites with all off-diagonal conductances strictly positive; masses can be taken rational and, for rational $\rho$, conductances too. Moreover $s_W(P,\rho)$ is right-continuous in $\rho$.
\end{theorem}

\begin{theorem}[Integral witnesses]\label{thm:arith}
If finite attainment together with Conjecture~\ref{conj:sigma} fails for some polygon and some $\rho>1$, then Conjecture~\ref{conj:sigma} fails for an integral polygon with the same number of sides, a rational $\rho$, and a competitor design on distinct integral interior sites with integral masses and conductances (before normalization) and all cables positive. Hence a search over rational data is a semi-decision procedure for a counterexample, and any counterexample has an exactly verifiable finite witness.
\end{theorem}
\noindent Proofs are in Appendix~\ref{app:dense}.

\section{All-rate certificate for the asymmetric quadrilateral}\label{app:pstar}

Write \(r=\rho\), let \(V_0=(-3,-2)\), \(A_0=(2,-2)\),
\(V_2=(4,2)\) and \(B_0=(-1,3)\), and set
\[
 C_r=\left(\frac{2r}{5-2r},-2\right),\qquad
 Z_r=B_0+\frac{6r+5}{10(r-1)}(V_0-B_0).
\]
For \(r\ge r_*=455/118\), these points are on \(V_0A_0\) and
\(B_0V_0\), respectively. The first solves the leaf equation
with hub \(B_0\), and the second is a normal-tangency point.
All rational identities and polynomial signs below can also be
checked by \path{anc/verify_pstar_all_rates.py}.

\paragraph{The moment bound.}
The quadratic
\[
 M(u,v)=\tfrac56u^2+\tfrac16v^2-\tfrac23u+\tfrac1{12}v-\tfrac34uv
\]
has Laplacian \(2\). On the four edges in cyclic order the gaps
\(11/2-M\) are \(\tfrac{125}{6}s(1-s),0,\tfrac{99}{4}s(1-s),0\),
with edge parameter \(s\in[0,1]\). By subharmonicity \(M\le11/2\)
on \(P_*\). The moment identities imply
\(\min(A,B)\le\mathbb E M\le11/2\) for any design.

\paragraph{The intermediate interval \(r_*\le r\le4\).}
Let \(D=106r-179\). Use the sites
\((A_0,V_2,B_0,C_r,Z_r)\) in this order with masses
\begin{align*}
 p_1&=\frac{11(5r+2)(12r-25)}{10(3r-5)D},
 &p_2&=\frac{11(12r-25)}{10D},\\
 p_3&=\frac{44(79r^2+60r-250)}{25(6r+5)D},
 &p_4&=\frac{33(2r-5)^2(118r-455)}{50(3r-5)(4r-15)D},\\
 p_5&=\frac{40(r-1)^2(13r-48)}{(4r-15)(6r+5)D}.
\end{align*}
The nonzero conductance formulas, with any zero value omitted, are
\begin{align*}
 J_{12}&=\frac{(5r+2)(12r-25)}{10D},
 &J_{13}&=\frac{(r+2)(5r+2)(12r-25)}{5(3r-5)D},\\
 J_{23}&=\frac{(4-r)(12r-25)}{5D},
 &J_{34}&=\frac{33r(2r-5)^2(118r-455)}{125(3r-5)(4r-15)D},\\
 J_{35}&=\frac{440r(r-1)^2(13r-48)}{(4r-15)(6r+5)^2D}.
\end{align*}
Substitution gives \(\sum p_i=1\), \eqref{eq:eq} and
\(A=B=L_m=22(103r-110)/(5D)\). All signs follow from
\(r\ge r_*>15/4\), \(r\le4\) and the displayed factors.
At \(r=r_*\) the fourth mass and its cable vanish. The remaining
four sites attain \(11/2\). Interpolating their conductances
by Lemma~\ref{lem:mono} proves the first interval of
\eqref{eq:pstar-all}.

For the intermediate upper bound use the rigid field plus one
positive boundary hinge centred at \(B_0\), with direction \(C_r-B_0\).
In the notation of Section~\ref{sec:hinges}, the parameters are
\[
 \alpha=\frac{27(2r-5)}D,\quad
 c_0=\left(-\frac{436r+385}{5D},-\frac{2(509r-385)}{5rD}\right),
 \quad\omega=-\frac{2(379r-410)}{5(r-1)D},\quad b_B=\frac{135}D.
\]
The hinge is active in the quadrilateral \(C_r A_0 V_2 B_0\).
In the remaining triangle \(V_0 B_0 C_r\), the gap \(L_m-F\)
is \(-\ell N/(5D)\), where
\[
 \ell=5u-2v+11,\qquad
 N=(54r-135)u+(-130r+110)v-206r+220 .
\]
On the three triangle vertices \(\ell\ge0\) and the values of
\(N\) are \(-27(4r-15),-5(130r-137),0\). On the
quadrilateral the four vertex values of \(F\) equal \(L_m\).
There \(F_{uu}=54(2r-5)/D>0\) and its Hessian determinant is
\[
 \frac{273911r^2-915530r+572975}{25D^2}>0 .
\]
For the last sign, shift \(r=r_*+z\), \(z\ge0\), in the numerator
to get positive coefficients. Convexity bounds \(F\) by \(L_m\)
throughout that cell. Each trace on the common chord obeys the bound,
so the monotone work identity certifies \(s_W\le L_m\).

\paragraph{The interval \(r\ge4\).}
Write \(D_h=468r^2-1097r+575\) and
\(Y_r=A_0+\frac{r+2}{2(r-1)}(V_2-A_0)\). Use sites
\((A_0,B_0,C_r,Y_r,Z_r)\) with masses
\begin{align*}
 p_1&=\frac{363r(r-1)(12r-25)}{2(r+2)(3r-5)D_h},
 &p_2&=\frac{66r(4r-5)(13r+5)}{5(6r+5)D_h},\\
 p_3&=\frac{33(2r-5)^2(156r^2-755r+575)}{10(3r-5)(4r-15)D_h},
 &p_4&=\frac{11(r-1)^2(12r-25)}{(r+2)D_h},\\
 p_5&=\frac{300(r-1)^2(4r^2-9r-20)}{(4r-15)(6r+5)D_h}.
\end{align*}
Here \(Y_r\) is a leaf root on \(A_0V_2\) with hub \(A_0\).
The positive conductances are
\begin{align*}
 J_{12}&=\frac{33r(r-1)(12r-25)}{(3r-5)D_h},
 &J_{14}&=\frac{33r(r-1)^2(12r-25)}{(r+2)^2D_h},\\
 J_{23}&=\frac{33r(2r-5)^2(156r^2-755r+575)}{25(3r-5)(4r-15)D_h},
 &J_{25}&=\frac{3300r(r-1)^2(4r^2-9r-20)}{(4r-15)(6r+5)^2D_h}.
\end{align*}
The factors \(D_h\), \(156r^2-755r+575\) and \(4r^2-9r-20\)
have positive coefficients after shifting \(r=4+z\).
Substitution gives normalization, force balance and equal moments
\(A=B=L_h=33r(52r-57)/D_h\).

For the upper bound set \(s=3(r-3)/(8(r-1))\) and
\(q=V_2+s(B_0-V_2)\). Add positive hinges at \(A_0,B_0\) in the
directions \(q-A_0,C_r-B_0\), respectively, with coefficients
\[
 b_A=\frac{11(104r-115)}{2D_h},\qquad b_B=\frac{25(26r-23)}{D_h}.
\]
The rigid part and objective weight are
\begin{align*}
 \alpha&=\frac{5(2r-5)(26r-23)}{D_h},\\
 c_0&=\left(-\frac{r(208r-63)}{D_h},\,\frac{624r-827}{2D_h}\right),
 &\omega&=-\frac{2392r^2-4198r+1725}{2(r-1)D_h}.
\end{align*}
The cuts partition \(P_*\) into a left triangle \(V_0 C_r B_0\)
with only the \(A_0\) hinge active, a middle quadrilateral
\(C_r A_0 q B_0\) with both active, and a right triangle
\(A_0 V_2 q\) with only the \(B_0\) hinge active. This follows
from the order of the chord endpoints along the boundary.

The gap in the left triangle factors as \((5u-2v+11)\)
times an affine form whose three vertex values, in order
\(V_0,C_r,B_0\), are
\[
 \frac{(4r-15)(26r-23)}{D_h},\qquad 0,\qquad
 \frac{5(208r^2-269r+46)}{2D_h}.
\]
Both factors are nonnegative. In the right triangle the gap
factors as \((2u-v-6)\) times an affine form with values
\[
 -\frac{1924r^2-3255r+1150}{2D_h},\quad
 -\frac{780r^2-3279r+2300}{2D_h},\quad
 -\frac{1092r^3-2526r^2-7499r+7475}{16(r-1)D_h}
\]
at \(A_0,V_2,q\). Both factors are nonpositive.
For the middle quadrilateral the gaps vanish at \(A_0,B_0,C_r\),
and at \(q\) the gap is
\[
 \frac{11(5r+1)(260r^3-1661r^2+8839r-5980)}{128(r-1)^2D_h}.
\]
The middle branch of \(F\) has \(F_{uu}=2\alpha>0\) and Hessian determinant
\[
 \frac{862576r^4-3762616r^3+4460279r^2-1084220r-476100}{4D_h^2}.
\]
The sign polynomials in these checks, and \(D_h(1-\alpha)\), have
positive coefficients after shifting \(r=4+z\).
It follows that the middle branch is convex and bounded above
by \(L_h\) throughout its cell. Thus every trace obeys \(F\le L_h\).
Theorem~\ref{thm:gap} applies and gives equality with the
five-site lower design. The two rational formulas agree at \(r=4\),
completing the proof.

\section{Density of finite networks in the continuum stress class}\label{app:closure}
%=====================================================================

All measures are extended by zero to $\R^2$. A matrix measure $T$ is positive semidefinite if $a^{\mathsf T}Ta\ge0$ for all $a$.

\paragraph{Localization.} Let $K=\conv\supp\mu$. For a unit vector $a$ and $b=\max_Ka\cdot x$, test \eqref{eq:weak} with $\varphi=a\,g(a\cdot x)$, $g=0$ on $(-\infty,b]$, $g'>0$ above $b$: this gives $a^{\mathsf T}Ta=0$ on $\{a\cdot x>b\}$; two independent separating normals and positivity give $T=0$ near every point outside $K$. Moreover $e^{-t\Lambda}K\subset K$: for smooth convex $f$, $m(t)=\int f(e^{-t\Lambda}x)d\mu$ has $m'(t)=-\int D^2[f(e^{-t\Lambda}x)]:dT\le0$; apply this to smooth approximations of $(a\cdot x-b)_+^2$.

\paragraph{Shrinking and smoothing.} For $r<1$ the pair $(x\mapsto rx)_\#\mu$, $r^2(x\mapsto rx)_\#T$ satisfies \eqref{eq:weak} and has invariant hull $rK\subset\operatorname{int}P$. Let $\nu$ be a product of smooth even densities $k_1(u)k_2(v)$ of small support and $D$ the diagonal stress $D_{ii}(x)=\lambda_i\bigl(\int_{x_i}^\infty sk_i(s)ds\bigr)\prod_{j\ne i}k_j(x_j)$, so that $-\operatorname{div}D=\Lambda x\,\nu$. Then $(\mu*\nu,T*\nu+\mu*D)$ is smooth and satisfies \eqref{eq:weak}, and tends to the shrunk pair as the support of $\nu$ shrinks, since $\int\tr D=\int x\cdot\Lambda x\,d\nu\to0$. Denote it $(f\,dx,S\,dx)$, supported in the interior of a closed neighbourhood $H$ of $rK$ in $\operatorname{int}P$.

\paragraph{Uniform background designs and the direction of the masses.}
Let
$K=\conv\supp\mu$, and choose $r<1$ and a compact neighbourhood
$H$ of $rK$ contained in $\operatorname{int}P$.  Invariance of $K$
under $e^{-t\Lambda}$ and the positive clearance of $rK$ from the boundary
allow us to choose $\rho'>\rho$, close enough to $\rho$, so that all
$e^{-t\Lambda'}z$ ($z\in H$, $t\ge0$) have positive clearance from
$\partial P$.  Take four fixed anchors $a_{\pm,\pm}=(\pm q,\pm q)$,
where the closed square $Sq=[-q,q]^2$ lies in $\operatorname{int}P$.
Fix $0<\beta<1$ (for instance $\beta=1/2$). Choose $L>0$ so large
that $e^{-L\Lambda'/(1-\beta)}H$ lies in the middle half of $Sq$.
For $m$ large let $h=L/m$, and define
\[
 z_0=z,\qquad z_1=(I-h\Lambda')z_0,\qquad
 z_{i+1}=(I-h\Lambda')z_i+\beta(z_i-z_{i-1})\quad(1\le i<m).
\]
The recurrence is a stable discretization of
$\dot z=-\Lambda'z/(1-\beta)$: uniformly for $z\in H$ and
$0\le i\le m$, its piecewise-linear interpolation differs from
$e^{-ih\Lambda'/(1-\beta)}z$ by $O(h)$.
One way to see this is to set $v_i=z_i-z_{i-1}$, use
$v_{i+1}=\beta v_i-h\Lambda'z_i$, and sum the geometric convolution;
its initial boundary layer contributes $O(h)$ to the positions, while
standard discrete Gronwall controls the accumulated error.
Consequently every $z_i$ lies in $\operatorname{int}P$ and $z_m$
lies in the interior of $Sq$, with clearance uniformly bounded below.

The masses increase along the chain. Set
\[
 w_i=\beta^{-i},\qquad J_{i,i+1}=w_i/h\quad(0\le i<m).
\]
At $z_0$ the force is $w_0\Lambda'z_0$.  For
$1\le i<m$ it is \emph{exactly}
\begin{align*}
 J_{i-1,i}(z_i-z_{i-1})+J_{i,i+1}(z_i-z_{i+1})
 &=\frac{w_i}{h}\bigl[\beta(z_i-z_{i-1})+z_i-z_{i+1}\bigr]\\
 &=w_i\Lambda'z_i,
\end{align*}
using $w_{i-1}=\beta w_i$ and the stated recurrence.
The terminal site has a residual force
\[
 b(z)=w_m\Lambda'z_m-\frac{w_{m-1}}h(z_m-z_{m-1}),
\]
which is uniformly bounded for fixed $m$ and $z\in H$.
Choose a fixed $D_0>0$ sufficiently large that
$z_m-b(z)/D_0$ belongs to the interior of $Sq$ for every $z\in H$.
Write this point in strictly positive barycentric coordinates of the
four anchors and connect $z_m$ to them with total conductance $D_0$.
Then the force at $z_m$ is exactly $w_m\Lambda'z_m$.

It remains to balance the anchors. Start from the four anchors with
masses $M$ and the strictly positive anchor network at rate $\rho'$:
its conductances are $M/4$ on all six pairs, with an additional
$(\rho'-1)M/2$ on each of the two vertical pairs.  Change the
four anchor masses by $\delta M_a$ solving
\[
 \sum_a\delta M_a=0,\quad
 \sum_a(M+\delta M_a)a+\sum_{i=0}^mw_i z_i=0,\quad
 \sum_a(M+\delta M_a)a_u a_v+\sum_{i=0}^mw_i(z_i)_u(z_i)_v=0.
\]
The coefficient matrix has rows $1,a_u,a_v,a_ua_v$ and is invertible
on the square; the corrections are uniformly bounded in $z\in H$
(and independent of the subsequently chosen $M$).  After adding the
path and terminal-to-anchor edges, the residual anchor load has zero
resultant and zero torque.  The six anchor cable columns span exactly
the five-dimensional space of balanced planar loads, since the four
corners form an infinitesimally rigid complete framework.  Thus bounded
signed changes in anchor conductances remove the residual.  Choosing
$M$ sufficiently large makes every anchor mass and all six anchor
conductances strictly positive, uniformly for $z\in H$.
Normalizing the total mass, which is the fixed number
$4M+\sum_iw_i$, gives a rate-$\rho'$ design with mass at $z$
bounded away from zero on $H$.
Finally interpolate its conductances with the complete reset design,
using coefficient $a=(\rho-1)/(\rho'-1)\in(0,1)$ on the
$\rho'$ design.  This is a design at rate $\rho$, and the
reset contribution gives \emph{direct} positive conductances from
$z$ to each of the four anchors (when $z$ coincides with an anchor,
merge the coincident sites).  Their lower bound $\gamma>0$ is uniform
on $H$, as required for the repair step.

\paragraph{Lattice discretization and repair.} Approximate $S$ in $C^1$ by $S_*=\sum_\ell a_\ell(x)v_\ell v_\ell^{\mathsf T}$, $v_\ell\in\mathbb Z^2\setminus\{0\}$, $a_\ell\ge0$ smooth with common compact support (for example, first use matrix-valued Bernstein approximants on a box:
all their matrix coefficients are positive semidefinite and their scalar
Bernstein weights are nonnegative; they converge in $C^1$ for smooth $S$.
Approximate each fixed coefficient matrix by finitely many positive
rank-one matrices with rational directions, clear denominators to obtain
integer $v_\ell$, and multiply by a smooth cutoff equal to one near
$\supp S$ and supported in $H$). On a translated square lattice of mesh $\ell_0$ in $H$ put an edge from $x_i$ to $x_i+\ell_0v_\ell$ with conductance $a_\ell(x_i+\ell_0v_\ell/2)$. The bare forces satisfy $\ell_0^{-2}L_i\to-\operatorname{div}S_*(x_i)$ uniformly, and the chord stresses converge to $S_*dx$ (midpoint sums). With raw masses $\ell_0^2f(x_i)$ choose meshes so that $|\ell_0^2f(x_i)\Lambda x_i-L_i|\le\varepsilon\ell_0^2$, the mass tends to one and the first and mixed moments are $O(\varepsilon)$. Choose the approximation and mesh so that the force residual at each
lattice site is bounded by $C\varepsilon\ell_0^2$, with $C$ independent of
the mesh.  Superpose at every lattice site the background design scaled by
$K\sqrt\varepsilon\ell_0^2$ for a fixed $K>0$, and an anchor-square design of mass $\sqrt\varepsilon$; the added mass and stress trace are $O(\sqrt\varepsilon)$ (uniformly in the mesh). Repair each site's force error by $O(\varepsilon\ell_0^2)$ changes of its four anchor conductances (two differences of anchor positions span the plane; available conductance $\ge K\gamma\sqrt\varepsilon\ell_0^2$; for sufficiently small $\varepsilon$ these signed changes preserve positivity). The resulting anchor residuals sum in magnitude to \(O(\varepsilon)\),
since there are \(O(\ell_0^{-2})\) modified stars.
The background designs are initially centred with zero mixed moment.
Choose \(O(\varepsilon)\) signed changes to the four anchor masses
with zero total change to cancel the first and mixed moment errors
of the lattice masses. This uses the same invertible square moment
matrix as above. All non-anchor force residuals already vanish.
The corrected anchor load has zero resultant and torque, since the
target loads have first and mixed moments zero and cable forces have
zero resultant and torque. Solve on the six anchor pairs with
\(O(\varepsilon)\) signed changes. Their signs and the anchor mass
changes are absorbed by the positive square design of size
\(\sqrt\varepsilon\). Normalize the total mass and all conductances
by the same factor. The resulting exact designs converge to the smooth pair; a diagonal sequence in the smoothing and shrinking parameters proves the forward direction. The converse and compactness were shown in Section~\ref{sec:relax}.

\paragraph{Proof of Corollary~\ref{cor:masslaws}.} Necessity is the edge-work identity. For sufficiency, the hypothesis implies $\int\Lambda x\cdot V\,d\mu\ge0$ for $C^1$ fields with $e(V)\succeq0$ on $P$: $V_\delta=V+\delta x$ is strongly monotone and Lipschitz, $x\mapsto x-tV_\delta(x)$ is non-expansive on $P$ for small $t$, extend it by Kirszbraun and subtract from the identity. On the subspace $\{e(\varphi)|_P\}$ of continuous symmetric matrix fields define $B(e(\varphi))=\int\Lambda x\cdot\varphi\,d\mu$; it is well defined and positive on positive semidefinite fields, the subspace contains $I$, and $\|B\|=B(I)$. A norm-preserving Hahn--Banach extension is positive ($0\preceq G\preceq I$ gives $B(G)=B(I)-B(I-G)\ge0$), and the Riesz theorem represents it by a positive semidefinite measure $T$ with \eqref{eq:weak}.

%=====================================================================
\section{Proof of Theorem~\ref{thm:eight}}\label{app:eight}
%=====================================================================

\paragraph{Lower bound $R_\infty\le s_W$ and the dual inequality.} Mixtures of centred two-endpoint laws on vertical sections are designs at every $\rho$ (Lemma~\ref{lem:vertical}); at a section the largest centred conditional second moment is $D(z)$, attained by its endpoints. LP duality for \eqref{eq:Rinf} gives $0\le\alpha\le1$ and $d\in\R$ with
\begin{equation}\label{eq:Fdual}
F(u):=\alpha u^2+(1-\alpha)D(u)-du\le R_\infty\qquad(u\in I).
\end{equation}
The inequality on the entire interval $I$ follows from Theorem~\ref{thm:knotexact} together with duality for the \emph{continuous} centred-moment problem, not from arbitrary optimal multipliers of the knot LP. More explicitly, compact convex minimax gives
\[
R_I=\min_{0\le\alpha\le1}\max_{\nu\in\mathcal P(I),\,\int u\,d\nu=0}
\int[\alpha u^2+(1-\alpha)D(u)]\,d\nu.
\]
For fixed $\alpha$ the centring-constrained maximum equals
$\min_{d\in\R}\max_{u\in I}[\alpha u^2+(1-\alpha)D(u)-du]$ by elementary moment LP duality. The minimizer in $d$ exists, since $a<0<b$ makes the outer maximum coercive as $|d|\to\infty$; it may be chosen jointly with an optimal $\alpha$ by compactness. Since $R_I=R_\infty$ by Theorem~\ref{thm:knotexact}, a resulting dual pair satisfies \eqref{eq:Fdual} throughout $I$. It is \emph{not} claimed that every optimal dual pair of the finite-knot LP has this global property.

\paragraph{Upper bound.} Let $U=\max_P|u|$, $V=\max_P|v|$ and $(0,\pm r)\in P$. For each state the conditional mean at time $t$ is $(e^{-t}u,e^{-\rho t}v)\in P$. At $t=(\log\rho)/\rho$ mix it with the opposite point $(0,\pm r)$ to reach the axis: every state satisfies $\operatorname{dist}(u,I)\le U(\log\rho)/\rho+V/(r\rho)$. Extend $D$ and $F$ Lipschitz to the full projection. Since $v^2\le(l+h)(u)v+D(u)$ pointwise and $\E u=0$, \eqref{eq:Fdual} bounds the objective by $R_\infty+O((1+\log\rho)/\rho)$ once $|\E[v\,k(u)]|=O(1/\rho)$ for continuous piecewise-affine $k$. Write $k$ as affine plus a combination of $(u-z)_+$; for $g=(u-z)_+$, $Qg\ge\mathbf 1_{u>z}Qu\ge-U$ and $\E Qg=0$, so $\E|Qg|\le2U$. Reversibility gives $\rho|\E[vk(u)]|=|\E[vQk(u)]|\le V\E|Qk(u)|$.

\paragraph{Lifting to eight dictionary sites.} A basic solution of \eqref{eq:Rinf} uses at most three abscissae (four equality constraints after adding slacks, one basis position taken by the objective variable), and a centred law on at most three abscissae is a mixture of at most two laws, each $\delta_0$ or a centred pair $x<0<y$. The section at $0$ is realized by its two axis endpoints. For a pair, with $\epsilon=1/\rho$: a section with $l<0<h$ either has two vertex endpoints (type I) or a vertex $H$ and a non-vertex endpoint $L$ on a non-vertical side (type II); a section with $D=0$ is the axis point (type III). At type II, the leaf equation $(u-H_u)v-\epsilon u(v-H_v)=0$ restricted to the side through $L$ has a simple root $L_\epsilon=L+O(\epsilon)$ (derivative $L_v\ne0$ at $\epsilon=0$), and $L_\epsilon-H$ is a positive multiple of $\Lambda L_\epsilon$. At type III and an opposite vertex $H$, parametrizing the boundary near $(z,0)$ as $u=u(v)$ (piecewise affine), the equation $(u(v)-H_u-\epsilon u(v))v+\epsilon u(v)H_v=0$ has a root $v=O(\epsilon)$ by the intermediate value theorem, since the limiting slope is $z-H_u\ne0$. These are dictionary points.

Balancing: if a type II group meets the rest only at $H$, with external force $f$ there, solve $p_HH_u+p_L(L_\epsilon)_u=f_u$, $p_HH_v+p_L(L_\epsilon)_v=\epsilon f_v$; for bounded $f$ with $f_u/z$ tending to a positive limit this is non-singular with positive solution, and the internal conductance $J(L_\epsilon-H)=p_L\Lambda L_\epsilon$ balances $L_\epsilon$ and then $H$. A type III point attached to $H$ by a cable of weight $w$ is placed at its root $Y$ with mass $w(Y_u-H_u)/Y_u>0$. Two type II groups: join the vertices by a unit cable; types II and III: join vertex and axis root by a unit cable; two type III groups: the centred axis pair; two type I groups: Lemma~\ref{lem:vertical}. One type I group $Z_\pm=(z,v_\pm)$ with a type II or III group: cables of weights $w_\pm$ from $Z_\pm$ to $Y_\pm$ in the other group ($Y_\pm=H$ for type II, the axis roots for type III); horizontal balance fixes $p_\pm=w_\pm(z-(Y_\pm)_u)/z>0$; choose $w_++w_-=1$ with $\sum_\pm w_\pm\frac{z-(Y_\pm)_u}z[v_\pm-\epsilon(v_\pm-(Y_\pm)_v)]=0$ (coefficients of opposite signs, positive solution tending to the centred weights); a vertical cable of conductance $\bigl(\rho p_+v_+-w_+(v_+-(Y_+)_v)\bigr)/(v_+-v_-)>0$ balances $Z_+$ and hence $Z_-$; finish the other group as above. After normalization the limiting groups are the centred endpoint laws, with $O(\epsilon)$ errors. Mixing at most two such designs uses at most eight sites. This proves the lower bound in Theorem~\ref{thm:eight}.

%=====================================================================
\section{Universal parabolic stress identity and proof of Theorem~\ref{thm:parab}(b)}\label{app:parab}
%=====================================================================

The following argument applies to \emph{every} positive semidefinite stress satisfying distributional equilibrium. It does not invoke the stress-density result of Theorem~\ref{thm:closure}.

Write $\Phi(u,v)=(u,t)$ with $t=c+v/b_0$, and let
$\nu=\Phi_\#\mu$ and $S=D\Phi\,(\Phi_\#T)\,D\Phi^{\mathsf T}$,
where $D\Phi=\diag(1,1/b_0)$. Then
\begin{equation}\label{eq:transformed-weak}
 \int \nabla\varphi:dS=\int\bigl(u\varphi_1+\rho(t-c)\varphi_2\bigr)\,d\nu,
 \qquad \E_\nu t=c.
\end{equation}
The transform preserves positive semidefiniteness. Set $d=t-u^2$ and
$B=b_0^2\E_\nu(t-c)^2$.

\begin{lemma}[Universal parabolic stress identity]\label{lem:universal-psd}
Let \((\nu,S)\) be compactly supported and satisfy
\eqref{eq:transformed-weak}, with \(S\succeq0\). Then
\begin{equation}\label{eq:universal-parab}
 3\int \nabla d^{\mathsf T}\,dS\,\nabla d
 =4c(c-B)+\int d\bigl((6\rho-8)t-6\rho c+4d\bigr)\,d\nu.
\end{equation}
If \(\supp\nu\subset\{(u,u^2):u\in\mathbb R\}\) and
\(\supp S\subset\{d\ge0\}\), then \(B\le c\). In case \(B=c\),
the mass is supported on \(-1\le u\le1\) and
\(\nu=(u\mapsto(u,u^2))_\#\beta\), with
\begin{equation}\label{eq:unique-parab-stress}
 S=\bigl(u\mapsto(u,u^2)\bigr)_\#
 \left[\kappa(1-u^2)\begin{pmatrix}1&2u\\2u&4u^2\end{pmatrix}\beta(du)\right],
 \qquad \kappa=\frac{\rho-2}{2},
\end{equation}
where $\beta(du)\propto(1-u^2)^{a_0-1}du$ on $[-1,1]$.
\end{lemma}
\begin{proof}
Test \eqref{eq:transformed-weak} in turn with
$\varphi=(u^3,0)$, $(0,u^2)$, and $(0,t)$. The resulting exact identities are
\[
 \int u^2\,dS_{11}=\tfrac13\E u^4,\quad
 \int u\,dS_{12}=\tfrac\rho2\E[u^2(t-c)],\quad
 S_{22}(\R^2)=\rho\operatorname{Var}(t).
\]
Because $\nabla d=(-2u,1)$,
\[
 \int\nabla d^{\mathsf T}dS\nabla d
 =\tfrac43\E u^4-2\rho\E[u^2t]+2\rho c\E u^2
  +\rho\operatorname{Var}(t).
\]
Substitute $u^2=t-d$, $\E t=c$, and
$\operatorname{Var}(t)=4cB/(3\rho-4)$; direct expansion gives
\eqref{eq:universal-parab}. If $\nu$ lives on the arc, $d=0$
$\nu$-almost everywhere, and the non-negative left side proves $B\le c$.

If $B=c$, the left side vanishes. Positivity of $S$ implies
$S\nabla d=0$ as a vector measure (the Cauchy--Schwarz inequality for
positive matrix measures). Choose any non-negative smooth function
$\chi$ supported in $(0,\infty)$ and test equilibrium with
$\varphi=\chi(d)\nabla d$; a spatial cutoff equal to one on the compact
supports may be inserted. The right side is zero because $\nu$ lives
on $d=0$. On the left, the $\chi'(d)$ term vanishes, while
$D^2d=\diag(-2,0)$, so
\[
 0=-2\int\chi(d)\,dS_{11}.
\]
Therefore $S_{11}$ vanishes on $\{d>0\}$. Together with
$S\nabla d=0$ and positivity, this forces $S$ to vanish there.
By the assumed support condition $\supp S\subset\{d\ge0\}$,
$S$ is supported on the arc. On the arc, $S\nabla d=0$ means that
$S$ is tangent to $(u,u^2)$. Thus, for a compactly supported positive measure \(\tau\)
on the abscissa line,
\[
 S=(u\mapsto(u,u^2))_\#
  \left[\begin{pmatrix}1&2u\\2u&4u^2\end{pmatrix}\tau(du)\right].
\]
Let $\beta_0$ be the $u$-marginal of $\nu$. Applying equilibrium to
$(f(u),0)$ and $(0,g(u))$ yields, for every polynomial $f,g$,
\[
 \int f'(u)\,d\tau=\int uf(u)\,d\beta_0,
 \qquad
 \int2ug'(u)\,d\tau=\rho\int(u^2-c)g(u)\,d\beta_0.
\]
Use $f(u)=ug(u)$ in the first identity and subtract half the second:
\[
 \int g\,d\tau=\tfrac{\rho-2}{2}\int(1-u^2)g\,d\beta_0
 =\kappa\int(1-u^2)g\,d\beta_0.
\]
Both measures are compactly supported. By polynomial approximation,
the displayed identity identifies
\(\tau=\kappa(1-u^2)\beta_0\) as signed measures.
Since \(\tau\ge0\) and \(\kappa>0\), \(\beta_0\) gives no mass to
\(\{|u|>1\}\). We therefore have
\(\int uf\,d\beta_0=\kappa\int(1-u^2)f'\,d\beta_0\)
on \([-1,1]\).
Its moments $m_j=\int u^j\,d\beta_0$ satisfy
$ m_0=1$, $m_1=0$, and
$(1+\kappa j)m_{j+1}=\kappa j m_{j-1}$ for $j\ge1$.
These are exactly the moments of the symmetric beta law $\beta$,
as follows by integration by parts; compactness of $[-1,1]$
ensures uniqueness. Hence $\beta_0=\beta$ and
$\tau=\kappa(1-u^2)\beta$, which proves \eqref{eq:unique-parab-stress}.
\end{proof}

\paragraph{Proof of Theorem~\ref{thm:parab}(b) and necessity.}
Let finite designs have \(\min(A_j,B_j)\to c\). Their mass measures
\(\mu_j\) and chord stresses \(T_j\) have convergent subsequences by
the trace bound \(\tr T_j(P)=\int x\cdot\Lambda x\,d\mu_j\) on the
common compact support. The limits obey equilibrium and are PSD. Since
$0\le\E_jd=c-A_j\le c-\min(A_j,B_j)\to0$, every limiting mass
measure lies on the parabola \(t=u^2\). By the localization argument in
Appendix~\ref{app:closure}, its limiting stress is supported in the
convex hull of its mass support, contained in the parabolic region
$\{d\ge0\}$. The universal identity
\eqref{eq:universal-parab} gives $B_{\rm lim}\le c$; but
$B_j\ge\min(A_j,B_j)\to c$, hence $B_{\rm lim}=c$.
Lemma~\ref{lem:universal-psd} forces the \emph{same} mass and stress
limits for every convergent subsequence. Transforming
\eqref{eq:unique-parab-stress} back to $(u,v)$ gives precisely the
stress in Theorem~\ref{thm:parab}(b). Compactness then gives
convergence of the entire maximizing sequence. Finally, the limiting
mass has full support on $\Gamma$, so the closedness of $P$ implies
$\Gamma\subset P$. This proves both the uniqueness assertion and the
necessity in the equality characterization, without relying on the
density of finite networks in the positive-stress class.

%=====================================================================
\section{Proofs of Theorems~\ref{thm:boundary} and \ref{thm:conditional}}\label{app:boundary}
%=====================================================================

\paragraph{The finite-contact theorem.}
For a bounded piecewise-rigid field choose any measurable trace at interface points, and write $F=q_\alpha+\Lambda x\cdot U(x)$. For every maximizing design $W_n$, the exact gap identity gives
\[
0\le\int_P(C-F)\,d\mu_n\le C-\min(A_n,B_n)\longrightarrow0.
\]
The upper-semicontinuous envelope $F^*(x)=\max_{K:x\in\overline K}F_K(x)$ obeys $F\le F^*\le C$; it is upper semicontinuous because there are finitely many closed cells and each $F_K$ is continuous. Thus $C-F^*$ is nonnegative lower semicontinuous. A weakly convergent subsequence of the masses satisfies, by Portmanteau,
\[
0\le\int (C-F^*)\,d\mu\le\liminf_n\int(C-F^*)\,d\mu_n=0,
\]
so $\operatorname{supp}\mu\subset Z^+$. If the stated contact hypotheses hold, each boundary piece carries a nonzero quadratic $C-F_K\ge0$, hence has at most two roots; $Z^+$ is finite. The finite-support closure theorem then realizes $\mu$ by a genuine finite cable network, preserving $A,B$ and attaining $C$. This proves Theorem~\ref{thm:boundary}.

\paragraph{Saturation of all traces in an optimal finite design.}
Any incident trace $U_K(x)$ is a limit of values of the globally monotone field at interior points of $K$. Replacing the assigned value at one site by any such trace therefore preserves every pairwise monotonicity inequality. The inequality $F_K\le C$ remains valid. Since an optimal design has $\min(A,B)=C$, the gap identity must have equality for \emph{every} such selection. Consequently every positive-mass site saturates every incident trace, and every positive cable has zero work for every selection of endpoint traces. For a positive sum of hinges, each hinge contributes nonnegative work, hence each individual hinge has zero work on every positive cable.

\paragraph{Zero-work geometry of a boundary hinge.}
Fix one hinge $H_{h,d}$, whose active region is the intersection of $P$ with $\det(d,z-h)>0$. Its work on a pair on the same side is zero. If $x,y$ are on opposite open sides, its work equals a positive multiple of $\det(y-h,x-h)$ after the appropriate ordering. The convex cone of rays from the boundary point $h$ to $P$ has aperture at most $\pi$. Therefore opposite-side work is strictly positive unless the two rays are collinear and opposite, in which case $x,y$ lie on the same supporting boundary line through $h$. If one endpoint lies on the cut, the same conclusion holds upon taking either incident trace, except when that endpoint is the centre $h$ (where both traces are zero).

Let now $x$ be in the relative interior of a boundary side $E$ and let $xy$ be a zero-work cable with $y\notin E$. The exceptional collinear-opposite case would put $xy$ along a boundary side through $x$, hence in $E$, and is excluded. If the hinge centre is not $x$, zero work therefore forces $y$ into the same closed half-plane of the hinge as $x$, provided the trace at $x$ has a fixed rigid branch. The hinge centre itself belongs to the boundary of either half-plane and presents no difficulty.

\paragraph{Excluding a hinge centred at a multi-cable boundary contact.}
Suppose $x\in\operatorname{relint}E$ has two non-collinear neighbours, none on $E$. Under the hypotheses of Theorem~\ref{thm:conditional}, any two distinct off-side boundary neighbours are automatically non-collinear with $x$: a ray from $x$ entering $\operatorname{int}P$ meets $\partial P$ at only one further point. For any two incident traces $V,V'$ at $x$, zero work on two non-collinear cables yields $(V-V')\cdot(x-y_j)=0$ for $j=1,2$, hence $V=V'$.

A positive hinge whose cut passes through $x$ but whose centre $h$ is different from $x$ would have a nonzero normal jump at $x$, contradicting this trace equality. Consider hinges centred at $x$ with a cut pointing strictly into $P$. Orient the unit tangent $\tau$ to $E$ so that the interior is on the left and write $n=-R\tau$ for the outward normal. Such a hinge is active on the negative side of the boundary parameter $x+s\tau$ and has contribution $b\,s\Lambda x\cdot R\tau+O(s^2)$ for $s<0$, and zero for $s>0$. Its contribution to the right derivative minus the left derivative of the boundary trace is therefore $b\,n\cdot\Lambda x$. The outward projection of equilibrium gives
\[
p_x\,n\cdot\Lambda x=\sum_y J_{xy}\,n\cdot(x-y)>0,
\]
since no neighbour lies on $E$. Thus the derivative jump produced by all such centred hinges is strictly positive if one is present. Other hinges and the rigid part have continuous boundary derivatives at $x$ (hinges with a cut confined to $E$ are absorbed into a single interior branch). But $F\le C=F(x)$ on both sides of $x$ forces $F'_-(x)\ge0\ge F'_+(x)$, so $F'_+(x)-F'_-(x)\le0$. This contradiction excludes all nontrivial hinges centred at $x$.

Every hinge now has a fixed branch in a small inward neighbourhood of $x$, giving one rigid field $U_0$. By the zero-work half-plane property, every neighbour lies in the \emph{same closed half-plane} for every hinge. These half-planes are convex, so the same rigid branch applies on the convex hull $T_x$ of $x$ and its neighbours (with appropriate one-sided traces at its boundary). Since all branchwise inequalities $F_K\le C$ hold on the cells and by continuity on their closures, $F_0\le C$ throughout $T_x$. All the neighbours and $x$ are saturated for every incident trace, so $F_0=C$ at these points. This proves Lemma~\ref{lem:locking}.

\paragraph{Quadratic geometry: at most two saturated points.}
Let $F_0=q_\alpha+\Lambda z\cdot(c_0+\omega_0 Rz)$ and $x\in\operatorname{relint}E$ be a contact, with $F_0\le C$ in an inward neighbourhood and no flat boundary contact. On the line $E$, the nonzero quadratic $F_0-C$ has an interior local maximum zero, so with $\tau$ tangent to $E$ and $H=D^2F_0$ we have $\nabla F_0(x)\cdot\tau=0$ and $a=\tau^T H\tau<0$. But $\operatorname{tr}H=2$, hence $\det H<0$. Let $n$ be the outward normal. The inward derivative must be strictly negative: if it vanished, then $\nabla F_0(x)=0$ and $n^THn=2-a>0$ would give $F_0(x-tn)>C$ for small $t>0$. Thus $g:=\nabla F_0(x)\cdot n>0$.

Completing the square gives the saddle point $z_c=x-H^{-1}\nabla F_0(x)$ and
\[
 C-F_0(z_c)=\tfrac12\nabla F_0(x)^T H^{-1}\nabla F_0(x)
 =\frac{g^2}{2}\frac{a}{\det H}>0.
\]
After an invertible affine change of coordinates the level $F_0=C$ is therefore $s^2-t^2=k$ with $k>0$, and the sublevel set is $|s|\le\sqrt{k+t^2}$. On either branch $s=\pm\sqrt{k+t^2}$, the chord between any two distinct level points lies strictly outside the sublevel set, because $t\mapsto\sqrt{k+t^2}$ is strictly convex. A convex subset of $\{F_0\le C\}$ can consequently contain at most one contact on each branch, and hence at most two contacts in total. If $x$ had two distinct neighbours, their convex hull would be contained in $\{F_0\le C\}$ with three distinct contact points, impossible. Therefore every non-vertex optimal site is a leaf.

\paragraph{Dictionary reduction and degenerate side roots.}
The strict same-side work hypothesis prevents an optimal positive cable between two contacts on a common side when at least one is not a vertex. Thus the preceding proof applies to every non-vertex optimal site. If such a leaf $x$ is joined to a vertex $h$, its leaf equilibrium gives $K_h(x)=0$. If the polynomial $K_h$ is not identically zero on its side, $x$ is an isolated dictionary point.

If $K_h\equiv0$ on a vertical side $E=\{u=a\}$, necessarily $h=(a(1-1/\rho),0)$ and for every $z=(a,v)\in E$ one has $z-h=(1/\rho)\Lambda z$. Thus the leaf conductance is $J=\rho p$. Splitting an interior leaf $(a,v)$ barycentrically between the two endpoints $(a,v_1),(a,v_2)$, with masses $p_1+p_2=p$ and $p_1v_1+p_2v_2=pv$, and new conductances $\rho p_1,\rho p_2$ to $h$ preserves all force balances, preserves $A$, and weakly increases $B$. On a horizontal side $E=\{v=b\}$, degeneracy forces $h=(0,-(\rho-1)b)$; then $z-h=\Lambda z$, the leaf conductance is $J=p$, and the analogous splitting preserves $B$ and increases $A$. Both operations preserve or improve $\min(A,B)$, so applied to an optimal design they yield another optimal design supported on dictionary points.

If the leaf's neighbour $h$ is not a vertex, it is also a non-vertex optimal site and hence a leaf. The two vertices form a two-state component, whose zero barycentre and $Qu=-u$, $Qv=-\rho v$, $\rho\ne1$, force one coordinate to vanish identically; both sites are intersections of $\partial P$ with a coordinate axis, which belong to the dictionary. This completes Theorem~\ref{thm:conditional}. The theorem does not assert that every optimal design uses dictionary points in the degenerate-root case; only that a dictionary-supported optimum can be constructed.

%=====================================================================
\section{Proofs of Theorems~\ref{thm:dense} and \ref{thm:arith}}\label{app:dense}
%=====================================================================

\paragraph{Theorem~\ref{thm:dense}.} Shrink the design slightly so that its sites have positive clearance from $\partial P$. Let $Q$ be its generator, $b>\max(\rho,\max_i(-Q_{ii}))$ and $K=I+Q/b$. Add one auxiliary state \( [ij] \) for each unordered positive-weight pair, with one loop state when \(K_{ii}>0\): jump from $i$ to $[ij]$ at rate $MK_{ij}$, $M=2b$, and from $[ij]$ to each endpoint at rate $T/2$ (to its endpoint at rate $T$ for a loop). This chain is reversible. For a coordinate mode $f$ of rate $\lambda\in\{1,\rho\}$ keep $f_i$ on the core and set $f_{[ij]}=\frac{T}{T-\eta_\lambda}\frac{f_i+f_j}2$ with $\eta_\lambda(M+T-\eta_\lambda)=T\lambda$ (smaller root); the core equation follows because the averaged midpoint of a $K$-transition is $(1-\lambda/(2b))f_i$. After rescaling time, the fast rate becomes $\rho'=\rho\frac{M+T-\eta_1}{M+T-\eta_\rho}>\rho$; auxiliary sites tend to edge midpoints (inside $P$) and carry vanishing mass. Mix in a tiny four-corner square design, so that the complete reset network $p_ip_j$ is available, and interpolate $J(r)=\frac{\rho'-r}{\rho'-1}p_ip_j+\frac{r-1}{\rho'-1}J'_{ij}>0$ for $1<r<\rho'$; this realizes every rate in an open interval around $\rho$ with fixed sites and masses, and proves right-continuity together with Lemma~\ref{lem:mono}. To rationalize, approximate sites and the masses off the square by rationals and solve the four moment equations ($1,u,v,uv$) on the square; the target loads are balanced, the complete cable matrix has rank $2N-3$ with image the balanced loads, so a basis of cable columns can be solved for the residual, continuously and with rational data when $\rho$ is rational.

\paragraph{Theorem~\ref{thm:arith}.} If the joint assertion fails, Conjecture~\ref{conj:sigma} fails at some $(P,\rho)$ with $s_W>\LS$. By Theorem~\ref{thm:dense} pick a rational positive competitor strictly above $\LS$; its conductances can be re-solved at nearby rational rates. Take rational $r_j\to\rho$ and rational $n$-gons $P_j$ with vertex errors $o(|r_j-\rho|)$ and no horizontal or vertical sides. Then $\limsup_j\LS(P_j,r_j)\le\LS(P,\rho)$: limits of dictionary points are dictionary points (vertices, axis crossings, normal tangencies and isolated leaf roots by continuity; if $K_{h,\rho}\equiv0$ on the limit side, then $K_{h_j,r_j}(x_j)=(r_j-\rho)(u-h_u)v+o(|r_j-\rho|)$ and the side is horizontal or vertical with $h=((1-1/\rho)a,0)$ on $u=a$ or $h=(0,-(\rho-1)b)$ on $v=b$, whose surviving roots are axis crossings), and the finite-support argument of Theorem~\ref{thm:finsupp} passes feasibility to the limit. Hence the competitor beats the dictionary on some rational $(P_j,r_j)$. A common integral dilation and clearing denominators finish the proof.

%=====================================================================
\section{Robust state counting for approximate eigenmodes}\label{app:robust}

\begin{theorem}[State-count bound for approximate relaxation modes]\label{thm:robustmicro}
Let $\rho>2$, $c=(\rho-2)/\rho$, and $b_0^2=(3\rho-4)/(4c)$. Consider a finite stationary reversible chain with $N\ge2$ states and observables $(u,v)$ in $\conv\Gamma$ such that $\E u=\E v=0$, not necessarily exact eigenfunctions. Write $L=-Q$ and
\[
 t=c+v/b_0,\quad e_u=Lu-u,\quad e_t=L(t-c)-\rho(t-c),\quad
 \varepsilon_u=\|e_u\|_{L^2(\pi)},\quad \varepsilon_t=\|e_t\|_{L^2(\pi)}.
\]
With $\delta=c-\min(\operatorname{Var}u,\operatorname{Var}v)\ge0$, one has
\begin{equation}\label{eq:robustmicro}
\boxed{4(c+2)\delta+4\varepsilon_u+9\varepsilon_t\ \ge\ \frac{c^2}{18(N-1)^2}.}
\end{equation}
Consequently, if the denominator is positive,
\[
 N\ge1+\frac{c}{\sqrt{72(c+2)\delta+72\varepsilon_u+162\varepsilon_t}}.
\]
One may use $\varepsilon_t=\|Lv-\rho v\|_{L^2(\pi)}/b_0$. For zero residuals, the sharper constant in Theorem~\ref{thm:microstate} applies.
\end{theorem}

\noindent The residual norms in this theorem are properties of the generator, not merely deviations of fitted two-time correlation curves from exponential shapes. Thus the result makes the state-count bound stable under \emph{quantified dynamical-model errors}, but does not infer an unrestricted physical microstate count from observed relaxation data alone.
%=====================================================================
\begin{proof}[Proof of Theorem~\ref{thm:robustmicro}]
Set $d=t-u^2\ge0$, $A=\E u^2$, $B=b_0^2\E(t-c)^2$, and denote the Dirichlet form by $\mathcal E(f,g)=\E[fLg]=\sum_{i<j}J_{ij}(f_i-f_j)(g_i-g_j)$. Define $W_4=\sum_{i<j}J_{ij}(u_i-u_j)^4$. Since $\E t=c$ and $u^2\le t\le1$, one has $\E d=c-A\in[0,\delta]$ and $|u|,|t-c|,d\le1$.

\emph{An exact identity with residuals.} The edgewise polynomial identity $(a-b)^4=4(a-b)(a^3-b^3)-3(a^2-b^2)^2$ gives
\[
 W_4+3\mathcal E(d,d)=4\mathcal E(u,u^3)-3\mathcal E(t,t)+6\mathcal E(t,d).
\]
Inserting $Lu=u+e_u$ and $Lt=\rho(t-c)+e_t$ and then $u^2=t-d$ yields
\begin{align}\label{eq:robustdefect}
 W_4+3\mathcal E(d,d)={}&4c(c-B)
 +\E\!\left[d\bigl((6\rho-8)t-6\rho c+4d\bigr)\right]\notag\\
 &+4\E[u^3e_u]-3\E[(t-c)e_t]+6\E[d e_t].
\end{align}
This holds for every reversible chain with the stated stationary centring, without assuming any eigenfunction equation. On $\conv\Gamma$ the coefficient of $d$ is at most $8$, whereas Cauchy--Schwarz bounds the three residual terms in absolute value by $4\varepsilon_u+9\varepsilon_t$. Since $B\ge\min(A,B)$, we get
\begin{equation}\label{eq:robustupper}
 W_4+3\mathcal E(d,d)\le4(c+2)\delta+4\varepsilon_u+9\varepsilon_t.
\end{equation}

\emph{Quartic-energy lower bound without an exact eigenfunction.} Let $r_1<\cdots<r_m$ be the distinct values of $u$ ($m\le N$), with gaps $g_k=r_{k+1}-r_k$. Put
\[
 F_k=\sum_{u_i\le r_k<u_j}J_{ij}(u_j-u_i)=-\sum_{u_i\le r_k}\pi_i(Lu)_i\ge0.
\]
Here $\sum_i\pi_i(Lu)_i=0$ by stationarity. Summing contributions of each cable across the crossed gaps gives $\mathcal E(u,u)=\sum_kF_kg_k$ and $W_4\ge\sum_kF_kg_k^3$, while
$\sum_k F_k\le (N-1)\E|Lu|/2$. Weighted H\"older therefore yields
\[
 W_4\ge\frac{4\mathcal E(u,u)^3}{(N-1)^2(\E|Lu|)^2}
\quad\text{whenever }\mathcal E(u,u)>0.
\]
If $\delta<c/2$ and $\varepsilon_u<c/4$, then $A>c/2$, $\varepsilon_u<A/2$, and
\[
 \mathcal E(u,u)=A+\E[ue_u]\ge A/2,\qquad
 \E|Lu|\le\sqrt A+\varepsilon_u\le\tfrac32\sqrt A.
\]
Consequently $W_4\ge2A^2/[9(N-1)^2]\ge c^2/[18(N-1)^2]$, and \eqref{eq:robustmicro} follows from \eqref{eq:robustupper}. In the complementary cases $\delta\ge c/2$ or $\varepsilon_u\ge c/4$, the left side of \eqref{eq:robustmicro} alone exceeds its right side because $0<c<1$ and $N\ge2$. This proves the theorem.
\end{proof}

\section{Proof of strict improvement under a higher cap}\label{app:raisedcap}

Let \(b=\rho-2>2/3\) and choose \(0<\theta<b\) such that
\begin{equation}\label{eq:theta}
 (3\rho-2)(2\theta-1)>3b .
\end{equation}
Such a choice exists because at \(\theta=b\) the difference
between the two sides is \(2(3b-2)(b+1)>0\).
For \(0<\varepsilon<1\), put \(s=1-\varepsilon\) and define
\[
 f_\varepsilon(u)=u^2+\theta(|u|-s)_+^2,\qquad
 H_\varepsilon(u)=u f_\varepsilon'(u)-\rho(f_\varepsilon(u)-c).
\]
The even curve \(t=f_\varepsilon(u)\) agrees with the parabola on
\([-s,s]\). Its tangent diffusivity is
\(a_\varepsilon=H_\varepsilon/f_\varepsilon''\), defined almost
everywhere. In the central interval
\(a_\varepsilon=\tfrac b2(1-u^2)\). On the right tail, writing
\(x=u-s\ge0\), it is
\begin{equation}\label{eq:tail_diffusivity}
 a_\varepsilon(s+x)=
 \frac{b(1-s^2)}{2(1+\theta)}
 -\frac{(b-\theta)s}{1+\theta}x-\frac b2 x^2.
\end{equation}
This quadratic has a positive root \(w_\varepsilon\) and a negative
root \(-L_\varepsilon\). With \(r_\varepsilon=s+w_\varepsilon\),
\[
 w_\varepsilon=\frac{b}{b-\theta}\varepsilon+O(\varepsilon^2),
 \qquad
 r_\varepsilon=1+R\varepsilon+O(\varepsilon^2),
 \qquad R=\frac{\theta}{b-\theta}>0.
\]
These expansions follow by substituting at the simple root \(x=0\)
at \(\varepsilon=0\), where the derivative in \(x\) is nonzero.
The diffusivity is positive on \((-r_\varepsilon,r_\varepsilon)\)
and vanishes linearly at the two ends.
The curve is contained in \(K_T\) as soon as
\(T\ge f_\varepsilon(r_\varepsilon)
      =1+2R\varepsilon+O(\varepsilon^2)\).

\subsection*{An equilibrium supported on the modified curve}

On the interval where \(a_\varepsilon>0\), take
\[
 F(u)\ \propto\ \exp\!\left[-\int_0^u v/a_\varepsilon(v)\,dv\right],
 \qquad p_\varepsilon(u)=F(u)/a_\varepsilon(u),
\]
with normalization \(\int p_\varepsilon=1\).
The flux \(F\) is continuous across the two joins \(u=\pm s\).
It vanishes as a positive power of the distance to each endpoint,
because \(a_\varepsilon\) has a simple root there. Thus
\(p_\varepsilon\) is integrable and \(F'=-u p_\varepsilon\)
holds also across the joins in the weak sense.
Let \(X(u)=(u,f_\varepsilon(u))\) in the \((u,t)\) coordinates.
The pushforwards of \(p_\varepsilon(u)\,du\) and
\(F(u)X'(u)X'(u)^{\mathsf T}\,du\) form a positive semidefinite
equilibrium pair. Indeed, for a smooth vector field \(\varphi\),
\[
 \begin{aligned}
 \int F\,\nabla\varphi(X):X'X'^{\mathsf T}\,du
 &=\int F\big[(\varphi(X)\cdot X')'-\varphi(X)\cdot X''\big]\,du\\
 &=\int \varphi(X)\cdot\big(uX'-a_\varepsilon X''\big)
                        p_\varepsilon\,du\\
 &=\int \varphi(X)\cdot\big(u,\rho(f_\varepsilon-c)\big)
                        p_\varepsilon\,du .
 \end{aligned}
\]
The endpoint terms vanish and \(X'\) is continuous at the joins.
The affine change to \((u,v)\) gives an equilibrium as in
\eqref{eq:weak}. Constant test fields give \(\E t=c\).
Let \(A_\varepsilon,B_\varepsilon\) be its stationary variances
in the original coordinates.

\subsection*{The gain in vertical variance}

Put \(d=f_\varepsilon(u)-u^2\) and let \(\E_d\) denote expectation
under the probability law with density proportional to
\(d p_\varepsilon\).
The scalar operator \(\mathcal G=a_\varepsilon\partial_{uu}
-u\partial_u\) has Dirichlet form
\(\mathcal E(h,g)=\int a_\varepsilon h'g' p_\varepsilon\,du\)
by the same integration by parts. It gives
\(\mathcal G u=-u\), \(\mathcal G f_\varepsilon=-\rho(f_\varepsilon-c)\)
and \(\mathcal G(u^2)=-2u^2+2a_\varepsilon\).
The derivation of \eqref{eq:slackidentity} applies to this form with
\(S=2a_\varepsilon\) and \(\mathcal W_4=0\): the latter identity follows
here from \(4\mathcal E(u,u^3)=3\mathcal E(u^2,u^2)\) by the chain
rule. In particular
\begin{equation}\label{eq:cap_ratio}
 \frac{4c(B_\varepsilon-c)}{c-A_\varepsilon}
 =4+(3\rho-2)\E_d[f_\varepsilon-1]
        -2\E_d[d]-6\E_d[a_\varepsilon].
\end{equation}
The denominator is \(\E d>0\). Because the weighted law is even and
supported on the two tails, all three expectations in
\eqref{eq:cap_ratio} can be computed from its conditional law on
the positive tail.

Write \(w=w_\varepsilon\), \(L=L_\varepsilon\) and set \(\xi=x/w\).
By \eqref{eq:tail_diffusivity},
\(a_\varepsilon(s+x)=\tfrac b2(w-x)(x+L)\).
A partial fraction integration of \(F'/F\) shows that \(p_\varepsilon\,du\)
on the right tail, up to a constant, is
\[
 (1-\xi)^{\gamma_\varepsilon-1}
 (1+w\xi/L)^{-\lambda_\varepsilon-1}\,d\xi,\qquad 0<\xi<1 ,
\]
where
\(\gamma_\varepsilon=r_\varepsilon/[\tfrac b2(w+L)]\)
and \(\lambda_\varepsilon=(s-L)/[\tfrac b2(w+L)]\).
Indeed
\((s+x)/a_\varepsilon(s+x)=\gamma_\varepsilon/(w-x)
+\lambda_\varepsilon/(x+L)\).
Weighting by \(d=\theta w^2\xi^2\) gives the limiting beta density
proportional to \(\xi^2(1-\xi)^{\gamma-1}\) with
\(\gamma=(1+\theta)/(b-\theta)>0\).
For this convergence, \(w/L\to0\), \(\lambda_\varepsilon\) is bounded
and \(\gamma_\varepsilon\to\gamma\), so the integrands for normalization
and any bounded moment have an integrable power bound near \(1\).

Uniformly for \(0\le\xi\le1\) one has, with \(u=s+w\xi\),
\[
 \frac{u-1}{\varepsilon}\longrightarrow -1+(1+R)\xi,\qquad
 \frac{a_\varepsilon(u)}{\varepsilon}\longrightarrow
                        \frac{b}{1+\theta}(1-\xi),
 \qquad d=O(\varepsilon^2).
\]
The limiting beta mean is \(3/(3+\gamma)\).
It gives
\[
 \frac{\E_d[u-1]}{\varepsilon}\to\frac{2\theta-1}{3b+1-2\theta},
 \qquad
 \frac{\E_d[a_\varepsilon]}{\varepsilon}\to\frac{b}{3b+1-2\theta}.
\]
Since \(f_\varepsilon-1=2(u-1)+O(\varepsilon^2)\) on the tail,
\eqref{eq:cap_ratio} becomes
\begin{equation}\label{eq:cap_slope}
 \frac{4c(B_\varepsilon-c)}{c-A_\varepsilon}
 =4+\ell_\rho\varepsilon+o(\varepsilon),\qquad
 \ell_\rho=
 \frac{2[(3\rho-2)(2\theta-1)-3b]}{3b+1-2\theta}>0.
\end{equation}

\subsection*{From a positive stress to the capped upper bound}

For each fixed sufficiently small \(\varepsilon\) with
\(f_\varepsilon(r_\varepsilon)\le T\), use Theorem~\ref{thm:closure}
on \(K_T\) to approximate the equilibrium by finite reversible
designs. The moment ratio is continuous at this pair since
\(c-A_\varepsilon>0\). Thus a finite design \(W\) can be chosen
with \(a=c-A(W)>0\), \(z=B(W)-c>0\) and
\(4cz/a\) as close to \eqref{eq:cap_slope} as desired.
Let \(m\) be its number of sites and mix it with an
\((N-m)\)-site on-arc design of deficit \(\delta_\Gamma\).
Choose the weight \(\delta_\Gamma/(a+z+\delta_\Gamma)\) on \(W\).
Both coordinate variances of the mixture are then
\(c-a\delta_\Gamma/(a+z+\delta_\Gamma)\).
The mixture has at most \(N\) states. By
Theorem~\ref{thm:arcsharp}, with \(W\) fixed,
\((N-1)^2\delta_\Gamma\to I_\rho^3/(4c)\).
Thus every such \(W\) yields
\(\limsup_N F_N(T)\le I_\rho^3/(4c+4cz/a)\).
Taking the infimum of these bounds over the approximating designs gives
\[
 \limsup_{N\to\infty}F_N(T)
 \le \frac{I_\rho^3}{4c+\mathcal R_\varepsilon},
 \qquad
 \mathcal R_\varepsilon=\frac{4c(B_\varepsilon-c)}{c-A_\varepsilon}.
\]
By \eqref{eq:cap_slope} this is a strict improvement for every
\(T>1\) upon choosing \(\varepsilon\) small enough.

For \(T\) close to \(1\), put \(h=T-1\) and choose \(\varepsilon\)
so \(f_\varepsilon(r_\varepsilon)=T\). The local positive root is
analytic in \(\varepsilon\) and the height has derivative \(2R>0\)
at zero, giving \(\varepsilon=h/(2R)+O(h^2)\).
The stress remains supported in \(K_T\) and is approximable by
Theorem~\ref{thm:closure}, so the same bound holds. In particular
\[
 \limsup_{N\to\infty}F_N(1+h)
 \le \frac{I_\rho^3}{4(c+1)+(\ell_\rho/(2R))h+o(h)}
 \qquad(h\downarrow0).
\]
This gives \eqref{eq:cap_strict}, for example with
\(m_\rho=\ell_\rho/(4R)\) for sufficiently small \(h\).
For \(\rho=4\), choosing \(\theta=1\) gives \(R=1\) and
\(\ell_\rho/(2R)=4/5\). Together with \eqref{eq:cap_bounds},
this proves a first-order change in the optimal scaled deficit.

\end{document}